\let\savedKernelLabel\label
\documentclass[aps,prb,superscriptaddress,reprint,floatfix]{revtex4-2}
\let\label\savedKernelLabel

\usepackage[dvipsnames]{xcolor}
\usepackage{graphicx}
\usepackage{rotating}

\usepackage{bm,amssymb}
\usepackage{amsfonts}
\usepackage{mathbbol}
\usepackage{physics}
\usepackage{slashed}
\usepackage{upgreek}

\usepackage{footmisc}
\usepackage{setspace}
\usepackage{comment}
\usepackage[normalem]{ulem}
\usepackage{enumerate}
\usepackage{tikz}

\usepackage[
    pagebackref=false,
    colorlinks=true,
    linkcolor=blue,
    urlcolor=blue,
    filecolor=black,
    citecolor=red,
    pdfstartview=FitV,
    pdftitle={},
    pdfauthor={},
    pdfsubject={},
    pdfkeywords={},
    pdfpagemode=UseNone,
    bookmarksopen=true
]{hyperref}

\usepackage{etoolbox}
\apptocmd{\sloppy}{\hbadness 10000\relax}{}{}

\makeatletter
\AtBeginDocument{
    \renewcommand{\p@subsection}{\thesection.}
}
\makeatother

\renewcommand{\theequation}{\arabic{equation}}

\renewcommand{\bar}[1]{\overline{#1}}

\newcommand{\vA}{\bm{A}}
\renewcommand{\vb}{\bm{b}}
\newcommand{\vc}{\bm{c}}
\newcommand{\ve}{\bm{e}}

\newcommand{\vJ}{\bm{J}}
\newcommand{\vk}{\bm{k}}
\newcommand{\vK}{\bm{K}}
\newcommand{\vL}{\bm{L}}
\newcommand{\vM}{\bm{M}}
\newcommand{\vp}{\bm{p}}
\newcommand{\vP}{\bm{P}}
\newcommand{\vq}{\bm{q}}
\newcommand{\vR}{\bm{R}}
\newcommand{\vs}{\bm{s}}
\newcommand{\vS}{\bm{S}}
\newcommand{\vw}{\bm{w}}
\newcommand{\vx}{\bm{x}}
\newcommand{\vz}{\bm{z}}
\newcommand{\vsigma}{\bm{\sigma}}
\newcommand{\vepsilon}{\bm{\epsilon}}
\renewcommand{\vnabla}{\bm{\nabla}}

\newcommand{\fE}{\mathcal{E}}
\newcommand{\fJ}{\mathcal{J}}
\newcommand{\fK}{\mathcal{K}}
\newcommand{\fL}{\mathcal{L}}
\newcommand{\fM}{\mathcal{M}}
\newcommand{\fO}{\mathcal{O}}
\newcommand{\fP}{\mathcal{P}}
\newcommand{\fQ}{\mathcal{Q}}
\newcommand{\fQA}{\mathcal{Q}_\text{A}}
\newcommand{\fQX}{\mathcal{Q}_\text{X}}
\newcommand{\fS}{\mathcal{S}}
\newcommand{\fW}{\mathcal{W}}
\newcommand{\fX}{\mathcal{X}}
\newcommand{\fY}{\mathcal{Y}}

\newcommand{\Sint}{S_\text{int}}
\newcommand{\SD}{S_\text{D}}
\newcommand{\SX}{S_\text{X}}
\newcommand{\SA}{S_\text{A}}
\newcommand{\SP}{S_\text{P}}
\newcommand{\SB}{S_\text{B}}
\renewcommand{\SS}{S_\text{S}}
\newcommand{\PiA}{\Pi_\text{A}}
\newcommand{\PiX}{\Pi_\text{X}}
\newcommand{\Psib}{\Psi_\text{b}}
\newcommand{\Psis}{\Psi_\text{s}}
\newcommand{\Psibb}{\bar{\Psi}_\text{b}}
\newcommand{\Psisb}{\bar{\Psi}_\text{s}}
\newcommand{\Lambdac}{\Lambda_\text{c}}
\newcommand{\epsilonS}{\epsilon_S}
\newcommand{\Mh}{\hat{\vM}}

\begin{document}
\title{Quantum theory of electron energy-loss spectroscopy}

\author{Michael J. Landry}
\email{mjlandry@mit.edu}
\affiliation{Quantum Measurement Group, Massachusetts Institute of Technology, Cambridge, MA 02139, USA}
\affiliation{Department of Nuclear Science and Engineering, Massachusetts Institute of Technology, Cambridge, MA 02139, USA}
\affiliation{Department of Physics, Massachusetts Institute of Technology, Cambridge, MA 02139, USA}

\author{Abhijatmedhi Chotrattanapituk}
\affiliation{Quantum Measurement Group, Massachusetts Institute of Technology, Cambridge, MA 02139, USA}
\affiliation{Department of Electrical Engineering and Computer Science, Massachusetts Institute of Technology, Cambridge, MA 02139, USA}

\author{Ashley S. Dale}
\affiliation{Quantum Measurement Group, Massachusetts Institute of Technology, Cambridge, MA 02139, USA}
\affiliation{Laboratory for Information and Decision-making Systems, Massachusetts Institute of Technology, Cambridge, MA 02139, USA}

\author{Zack West}
\affiliation{Quantum Measurement Group, Massachusetts Institute of Technology, Cambridge, MA 02139, USA}
\affiliation{Department of Electrical Engineering and Computer Science, Massachusetts Institute of Technology, Cambridge, MA 02139, USA}

\author{L. Dagny Davis}
\affiliation{Quantum Measurement Group, Massachusetts Institute of Technology, Cambridge, MA 02139, USA}
\affiliation{Department of Nuclear Science and Engineering, Massachusetts Institute of Technology, Cambridge, MA 02139, USA}

\author{Lijun Wu}
\affiliation{Condensed Matter Physics and Materials Science Department, Brookhaven National Laboratory, Upton, NY 11973, USA}

\author{Ray F. Egerton}
\affiliation{Physics Department, University of Alberta, Edmonton, AB T6G 2E1, Canada}

\author{Yimei Zhu}
\affiliation{Condensed Matter Physics and Materials Science Department, Brookhaven National Laboratory, Upton, NY 11973, USA}

\author{Mingda Li}
\email{mingda@mit.edu}
\affiliation{Quantum Measurement Group, Massachusetts Institute of Technology, Cambridge, MA 02139, USA}
\affiliation{Department of Nuclear Science and Engineering, Massachusetts Institute of Technology, Cambridge, MA 02139, USA}

\begin{abstract}
Electron energy-loss spectroscopy (EELS) is a powerful technique for probing quantum excitations with increasingly high spatial and energy resolution. Standard EELS theory treats beam electrons as distinguishable probes interacting electromagnetically with the sample (direct channel), neglecting fermionic exchange. While exchange has been studied in nonrelativistic surface spectroscopy and relativistic scattering from free electrons, atoms, and ions, a unified treatment for general many-body samples across EELS beam energies is lacking. Here we develop a fully relativistic quantum mechanical framework by decomposing the electron field into beam and sample modes. We identify the exchange coefficient as the relativistic generalization of the Ochkur kernel and show that its interference with the direct channel changes sign at a beam kinetic energy of 511 keV, owing to magnetic coupling of the exchange photon. The framework extends to positron scattering, where the same terms yield annihilation and electron-positron pair production. Additional channels vanish at leading order and enter only through loop corrections. We recover textbook results and establish the conditions under which exchange and annihilation reduce to local interactions. We derive a scaling law quantifying when exchange matters: for unpolarized beams at 60--300 keV and momentum transfers below $3\,\text{\AA}^{-1}$, exchange corrections are below $5\times 10^{-4}$, justifying conventional EELS theory. At 10--100 eV, however, corrections reach tens of percent. For magnon creation, the direct and exchange channels leave the outgoing electron in orthogonal spin states, strongly suppressing exchange for unpolarized beams while allowing significant contributions for polarized beams. Our framework provides a foundation for fully quantum mechanical EELS, enabling more precise measurements and motivating spin-sensitive experiments.
\end{abstract}

\maketitle

\section{Introduction}
When a beam of electrons passes through a thin sample, some of the electrons create excitations in the material, thereby losing energy. The distribution of energy lost provides the spectrum of those excitations. 
This technique is known as electron energy-loss spectroscopy (EELS). 

In a scanning transmission electron microscope, the beam can be focused to atomic dimension and, with the addition of a monochromator the energy resolution, now reaches a few millielectronvolts. Consequently, phonons \cite{krivanek2014} and, very recently, magnons \cite{kepaptsoglou2025,mm2025} can be measured with momentum resolution in samples a few nanometers thick. At the opposite end of the energy scale, momentum-resolved reflection EELS (M-EELS) operates at primary energies of 10--100 eV. This technique measures the density response $\chi(\vq,\omega)$ of quantum materials with meV resolution and controlled momentum transfer across several Brillouin zones \cite{vig2017}. Its results are increasingly compared with resonant inelastic x-ray scattering (RIXS), optical spectroscopy, and monochromated microscope EELS on the same materials. Such complementary spectroscopic measurements are particularly valuable for correlated quantum materials like kagomé metals~\cite{nguyen2022kagome}. Both techniques are described by the same theory, so the questions we address here are relevant to both.

In almost all cases, the theory used to interpret such spectra relies on treating the beam electron as a moving charge whose electromagnetic field perturbs the sample. Then first-order perturbation theory yields the double-differential cross section as a Coulomb kernel multiplied by the sample's dynamic structure factor. This is Bethe's theory \cite{bethe1930,inokuti1971}. Its relativistic extension, which accounts for the beam electron's current as well as its charge and includes retardation, has appeared in the literature in several forms \cite{moller1932,schattschneider2005,sorini2008,iatrakis2026a,iatrakis2026b} and is the basis of the standard treatment in microscopy \cite{egerton2011}. In this picture, the beam electron is a distinguishable probe that comes in, electromagnetically interacts with the sample, and goes out.

There is, however, an important quantum mechanical subtlety that this picture neglects. The beam electron is identical to the electrons in the sample, which can have important quantum effects. In particular, as we do not know which electron leaves the sample the different possible histories can interfere. The consequences of this indistinguishability are familiar at low energies but are neglected at high energies. In a process known as \emph{exchange}, the scattering electron can take the place of one of the sample electrons, which then leaves in its stead. We will see that exchange is mediated by a contact interaction, so the Pauli exclusion principle demands that the target and sample electrons have opposite spin. Thus at leading order, exchange always flips spins. At the few-electronvolt beam energies of surface spectroscopy, exchange is the dominant way to excite a spin wave, a fact on which spin-polarized electron energy-loss spectroscopy has been built for decades \cite{kirschner1985,plihal1999,vollmer2003}. In this regime, exchange is treated non-relativistically.

Relativistic exchange has also been studied. In 1954, Rohrlich and Carlson used the M{\o}ller and Bhabha cross sections to compute the difference in energy loss between electrons and positrons passing through matter \cite{rohrlich1954} and found that exchange enters the relativistic Born theory of atomic ionization as well. At the 30--300 keV energies of the microscope, however, the object of interest is the spectrum of a many-body sample rather than its stopping power. In this setting, exchange has been neglected as a low-energy correction \cite{iatrakis2026a,iatrakis2026b}, but its size has never been established. Further, there are two other consequences of indistinguishability for relativistic probes: (1) positron beams, which have no exchange with electrons, can annihilate with them, and (2) in principle, for any beam, a pair of sample electrons can be converted into a pair of beam electrons. No framework has yet contained all of them for a general many-body sample across the full range of beam energies in use. Thus, whether any of these processes matter at microscope energies has not been determined. 

This paper constructs such a framework and uses it to answer several key questions: How does exchange scale with beam energy and momentum transfer, and when must it be considered? What changes when the beam consists of positrons? Can the conversion of sample electrons into beam electrons ever contribute to a measured spectrum?

To this end, we split the electron field into modes that belong to the beam and modes that belong to the sample. The single interaction of quantum electrodynamics (QED), in which one photon is exchanged between two currents, then separates, exactly and without double counting, into a small number of terms labeled by which fields they contain. There is a term in which the beam current couples to the sample current (the direct channel), a term in which the beam electron becomes a sample electron and vice versa (the exchange channel), and a term in which two sample electrons become two beam electrons (the pair-production channel). In addition, there are beam-beam and sample-sample terms that do not involve energy transfer to the sample. As the probe is treated fully relativistically, we obtain results for positronic beams for free. In this case, the exchange term in the action is reinterpreted as annihilation followed by pair creation. We then treat the beam exactly, to all orders in its velocity, while expanding in the ratios of sample momenta to beam momenta, which are small at every energy the microscope uses. It turns out that each channel's photon lives in a distinct kinematic region, with low momentum for the direct channel and high momentum for exchange and annihilation. Integrating out this photon then yields a local operator with a coefficient that we can compute. The result is an effective field theory (EFT), constructed by the method of regions \cite{beneke1998}, in which the heavy scale is the beam momentum, the light scales are those of the sample, and the small parameter is their ratio. 

Using this framework, we arrive at the following results. The direct channel reproduces the known relativistic cross section, which we present in a form that fixes conventions for everything else. The exchange channel is a contact interaction whose coefficient is the relativistic generalization of the Ochkur kernel of atomic collision theory \cite{ochkur1964,rudge1968}. Its interference with the direct channel is destructive at low energy, as in M{\o}ller scattering. Curiously, however, the interference reverses sign at beam kinetic energy of $m$ ($\sim$511 keV) for a reason we explain in Sec.~\ref{sec:e2m}. For a positron beam, the same operator describes annihilation, with a coefficient smaller than the exchange coefficient by $T/(T+2m)$, where $T$ is the beam's kinetic energy. In the non-relativistic limit, this operator becomes the contact interaction responsible for the virtual-annihilation contribution to the positronium hyperfine splitting, which we recover exactly. The pair channel vanishes identically at leading order and first appears in a one-loop diagram whose intermediate state is off shell by twice the beam energy. We compute this diagram in an exactly solvable model and bound it at $10^{-13}$ of the direct amplitude at 100 keV. Lastly, we establish the hierarchy of two-photon corrections, calibrating the second Born approximation on an exactly solvable potential.

Throughout, we pay particular attention to why exchange is small in the microscope, how it scales with beam energy and momentum transfer, and when it must be included. Table~\ref{tab:when} and Fig.~\ref{fig4} collect these rules of thumb, and Sec.~\ref{sec:meels} applies them to low-energy reflection spectroscopy, where exchange is not small. As an application, we derive the cross section for creating a magnon in a ferromagnet with both the direct and the exchange channel present, which is the calculation behind the companion work \cite{companion}. A particularly interesting feature of this calculation is that exchange enters the intensity only at second order in a small ratio, as the two channels leave the beam electron in orthogonal spin states, whereas it enters the beam electron's polarization at first order.

\section{Notation}
\label{sec:notation}
We use natural units $\hbar=c=1$. The electron mass is denoted by $m$, and $e$ denotes the magnitude of its charge, so the electron charge is $-e$ and $e^2=4\pi\alpha$, where $\alpha$ is the fine-structure constant. 

We specify space-time components with Greek indices that run over 0--3, while spatial components use Roman indices which run over 1--3. We also use the mostly-minus space-time metric $\eta_{\mu\nu} = \text{diag}(1,-1,-1,-1)$. Repeated space-time component indices are summed, while repeated state, spin, and operator labels require explicit summation. When a parameter contains both component indices and identifying labels, they are separated by a comma. Lorentz contractions can be written with the space-time metric explicitly or with raised and lowered indices, while spatial dot and cross products use $\cdot$ and $\times$, respectively. The Euclidean metric is $\delta_{ij}$, and the antisymmetric symbols in three and four dimensions are defined by $\epsilon_{123}=1$ and $\epsilon^{0123}=1$, respectively.

We use calligraphic letters for four-momenta, four-currents, the beam tensor, and the sample current-response tensors. Spatial vectors are bold, and their magnitudes are written in plain type. For example, a four-momentum $\fQ=(\omega, \vq)$ which has energy $\omega$, spatial momentum $\vq$ with magnitude $q=\abs{\vq}$, and invariant
\begin{equation}
    \fQ^2\equiv\eta_{\mu\nu}\fQ^\mu\fQ^\nu=\fQ^\mu\fQ_\mu=\omega^2-q^2\,.
\end{equation}
We also denote symmetrization by placing indices in parentheses, for example,
\begin{equation}
    \fS^{(\mu\nu)}\equiv\dfrac{\fS^{\mu\nu}+\fS^{\nu\mu}}{2}\,.
\end{equation}

The superscripts $\ast$, $\dagger$, and $\top$ denote complex conjugation, Hermitian conjugation, and transposition, respectively. For cleanliness, we use ``h.c." for Hermitian conjugate of the entire preceding expressions. We also define the 4 by 4 Dirac matrices $\gamma^\mu$ which satisfy the algebra
\begin{equation}
    \gamma^\mu\gamma^\nu + \gamma^\nu\gamma^\mu = 2\eta^{\mu\nu}\mathbb{1}_4\,,
\end{equation}
where $\mathbb{1}_d$ is the $d$-dimensional identity matrix. We further define the Dirac adjoint $\bar{\Psi} = \Psi^\dagger\gamma^0$, the matrix $\gamma^5=i\gamma^0\gamma^1\gamma^2\gamma^3$, and the slash notation $\slashed{\fQ}=\fQ^\mu\gamma_\mu$.

A tilde marks a Fourier-transformed kernel or operator. In particular, the Fourier representation of a space-time kernel is
\begin{equation}
    \Pi(\fX-\fY)=\int\dfrac{\text{d}^4\fQ}{(2\pi)^4}\,e^{-i\fQ^\mu(\fX_\mu-\fY_\mu)}\widetilde{\Pi}(\fQ)\,,
    \label{eq:photon-fourier}
\end{equation}
where $\fX$ and $\fY$ are space-time coordinates, while the spatial Fourier transform of an operator $\fO$ is
\begin{equation}
    \widetilde{\fO}(\vq)=\int \text{d}^3\vx\,e^{i\vq\cdot\vx}\fO(0,\vx)\,.
\end{equation}
The transition matrix element $\widetilde{\fO}_{nm} \equiv \bra{n}\widetilde{\fO}\ket{m}$ connects the sample states $\ket{m}$ to $\ket{n}$.

The operator $\text{T}$ denotes chronological ordering of fields, placing later times to the left and including a minus sign for each interchange of fermion fields.

\section{Channel decomposition of the beam-sample interaction}
\label{sec:framework}
\subsection{Photon-mediated interaction}
Consider an electron described by a Dirac field $\Psi(\fX)$ with four-current $\fJ^\mu=\bar{\Psi}\gamma^\mu\Psi$. The photon field $A^\mu$ enters the QED Lagrangian quadratically through kinetic term and linearly through coupling $e\fJ^\mu A_\mu$. Consequently, the photon field can be integrated out exactly via a Gaussian path integral. In the Feynman gauge, this yields an effective interaction action for the fermion field containing a non-local current-current interaction mediated by the photon propagator:
\begin{equation}
    \Sint = -\dfrac{e^2}{2}\iint \Pi(\fX-\fY)\fJ^\mu(\fX)\fJ_\mu(\fY)\,,
    \label{eq:Sint}
\end{equation}
where double integral denotes integration over $\fX$ and $\fY$,
\begin{equation}
    \widetilde{\Pi}(\fQ) = -\frac{1}{\fQ^2+i0} = \frac{1}{q^2-\omega^2-i0}\,,
    \label{eq:photon-propagator}
\end{equation}
and $\fQ = (\omega,\vq)$ is the photon four-momentum. 

It should be emphasized that Eq.~\eqref{eq:Sint} is exact, i.e., it is the covariant form of the Coulomb interaction together with magnetic and retarded corrections. When evaluated to all orders, this action generates all virtual-photon processes with no external photons---radiative losses to the vacuum require an external photon and are suppressed by an additional factor of $\alpha$, while radiation into the medium, such as Cherenkov loss, is a collective mode of the sample electrons and is therefore already contained in the sample response. 

Although the photon integration is exact, we approximate the scattering amplitudes by truncating the interaction expansion. In particular, through Born approximation, we only keeps single-insertion contributions inside $\Sint$ which correspond to amplitude factor of order $e^2$ while selected two-insertion contributions of order $e^4$ are discussed in Sec.~\ref{sec:beyond}. The scattering calculations in this work neglect explicit vacuum radiative corrections since these omitted corrections are negligible at the momenta of interest.

\subsection{Beam and sample modes}
\label{subsec:beam-sample-modes}
In an EELS experiment performed on a sample with Fermi surface energy $E_\text{F}$ by an electron beam with kinetic energy $T=E-m$, it is generally the case that $T$ can be high enough for relativistic effect to be non-trivial while $E_\text{F}$ can be considered non-relativistic. Furthermore, the energy that the beam actually loses to the sample can be considered small compare to $T-E_\text{F}$. Hence, we can apply EFT by letting $\Lambdac\ll T-E_\text{F}$ be the EFT cutoff which can be considered the bound on energy loss of beam electrons. We, then, can split the field into two distinct parts, the potentially relativistic \emph{beam field} $(\Psib)$ and the non-relativistic \emph{sample field} ($\Psis$) by energy,
\begin{equation}
    \Psi = \Psib + \Psis\,,
\end{equation}
where $\Psib$ and $\Psis$ contain the modes within energy $\Lambdac$ from $T$ and $E_\text{F}$, respectively. Following the method of regions~\cite{beneke1998}, we expand each channel's photon in the region where its momentum lies, and extend the domains of $\Psib$ and $\Psis$ to all momenta. The overlap between regions do not contribute at any finite order in perturbation theory and are hence dropped. 

We will treat the beam field $\Psib$ exactly with Dirac spinors, while the non-relativistic sample field $\Psis$ can be written in terms of a two-component field $\psi$, whose components $\psi_s$ are labeled by spin states $s\in\{\uparrow, \downarrow\}$, as
\begin{equation}
    \Psis(t,\vx) = \dfrac{e^{-imt}}{\sqrt{2m}}\sum_s u^s(0)\psi_s(\vx)+\order{p/m}\,,
    \label{eq:sample-field}
\end{equation}
where $t$ is time, $u^s$ are the rest-frame Dirac spinors, $p$ is the sample characteristic electron momentum with assumption that $p/m$ is small.

Notice that the free Hamiltonian conserves the number of beam electrons $N_\text{b}$ and the number of sample electrons $N_\text{s}$ separately, because the two sets of modes are disjoint. This property is of crucial importance. It means that every term of the interaction can be labeled by the change $(\Delta N_\text{b},\Delta N_\text{s})$ it produces, and terms with different labels cannot interfere.

\subsection{Channel decomposition}
\begin{figure}[!tb]
    \includegraphics[width=\columnwidth]{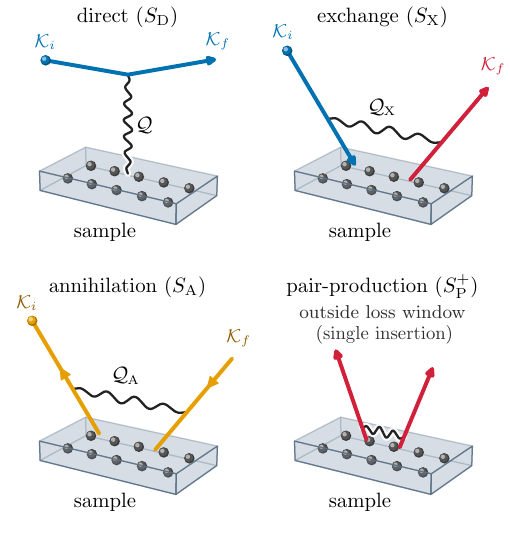}
    \caption{Selected beam-sample processes at one insertion of $\Sint$. Wavy lines represent virtual photons. Arrow heads on fermion lines indicate fermion flow and point opposite to the direction of motion for positrons. In the direct channel, the beam electron is deflected and transfers four-momentum $\fQ$ to the sample. In the exchange channel, the photon carries the remaining $\fQX$ after creating a sample electron. For a positron beam, the same exchange term describes annihilation and recreation of the electron-positron pair. The photon carries the combine $\fQA$ from annihilating with a sample electron. In pair production, two sample electrons enter beam states, increasing the beam kinetic energy by about $2T$ which falls outside the retained loss window $\Lambdac$.}
    \label{fig:channels}
\end{figure}

Substituting $\Psi=\Psib+\Psis$ into the current, we find 
\begin{equation}
    \fJ^\mu = \sum_{i,j\in\{\text{b},\text{s}\}}\fJ^\mu_{ij}\,,
\end{equation}
where $\fJ^\mu_{ij}=\bar{\Psi}_i\gamma^\mu\Psi_j$ represents the transition current from $j$ field to $i$ field. Hence, the product $\fJ^\mu(\fX)\fJ_\mu(\fY)$ in Eq.~\eqref{eq:Sint} contains sixteen terms. For notational ease, define 
\begin{equation}
    S_{ij,kl} = -\dfrac{e^2}{2}\iint\Pi(\fX-\fY)\fJ^\mu_{ij}(\fX)\fJ_{kl,\mu}(\fY)\,,
\end{equation}
where $i,j,k,l\in\{\text{b},\text{s}\}$. The symmetry of the kernel under interchange of space-time coordinates gives equal contributions from the same pair of distinct currents, i.e., 4 terms with repeated currents, and 6 pairs of terms with distinct currents as shown in Table~\ref{tab:channels}. Figure~\ref{fig:channels} depicts some of these processes that are important to EELS, and would be discussed in greater detail in this work.
\begin{table}[!htb]
    \caption{Different contributing channels in the beam-sample electron system interaction. Channels with distinct current terms have extra factor of 2 from its symmetric counterparts.}
    \label{tab:channels}
    \begin{ruledtabular}
    \begin{tabular}{lcccccccc}
    Channel & Symbol & $ij$ & $kl$ & Multiplicity & $\Delta N_\text{b}$ & $\Delta N_\text{s}$\\
    \hline
    beam-beam & $\SB$ & bb & bb & 1 & 0 & 0\\
    sample-sample & $\SS$ & ss & ss & 1 & 0 & 0\\
    direct & $\SD$ & bb & ss & 2 & 0 & 0\\
    exchange & $\SX$ & bs & sb & 2 & 0 & 0\\
    pair-production & $\SP^+$ & bs & bs & 1 & +2 & $-2$\\
    pair-absorption & $\SP^-$ & sb & sb & 1 & $-2$ & +2\\
    beam-ionize & $\SB^+$ & bb & bs & 2 & +1 & $-1$\\
    sample-ionize & $\SS^+$ & bs & ss & 2 & +1 & $-1$\\
    beam-absorb & $\SB^-$ & bb & sb & 2 & $-1$ & +1\\
    sample-absorb & $\SS^-$ & sb & ss & 2 & $-1$ & +1
    \end{tabular}
    \end{ruledtabular}
\end{table}

In the beam-beam channel, $\SB$ only contains beam fields. For a single beam electron, it is the electron's self-interaction and is absorbed into its physical mass. For a beam of many electrons, it describes M{\o}ller scattering within the beam, which is a matter of beam optics rather than spectroscopy. On the other hand, $\SS$ only contains sample fields and represents the interaction part of the sample's Hamiltonian, which we treat it exactly rather than perturbatively. In particular, it determines the sample's eigenstates, and the sample response functions.

We refer to $\SD$ and $\SX$ as the \emph{direct} and \emph{exchange} channels, respectively. In $\SD$, the beam current at $\fX$ couples through the photon to the sample current at $\fY$. In this process, the beam electron scatters from momentum $\fK_i$ to $\fK_f$, and the photon carries the momentum transfer $\fQ = \fK_i-\fK_f$ into the sample. In $\SX$, however, a beam electron is annihilated at $\fY$ and a sample electron is created there, while the reverse occurs at $\fX$. In other words, the beam electron has gone into the sample and a sample electron has come out. In this case, photon connects a fast electron to a slow one at each vertex, so it carries a momentum of order $\fK_i$ rather than $\fQ$. This difference in photon momentum is the physical distinction between the two channels, and, as we will see, it is what makes exchange a contact interaction.

The remaining terms involve the change in number of beam and sample electrons. The unnamed channels correspond to single-electron ionization or absorption while the pair-production channel $\SP^+$ converts two sample electrons into two beam electrons, and pair-absorption $\SP^-$ performs the reverse. The pair term is $\SP\equiv\SP^++\SP^-$.

However, under our assumption, the first Born approximation containing one $\Sint$ insertion of order $e^2$ would have energy transfer of size $|\Delta N_\text{b}||T|$ which is larger than $\Lambdac$ if $|\Delta N_\text{b}|\neq 0$. Furthermore, the energy loss would never occur if the term does not involve both beam and sample fields. These implies that only direct ($\SD$) and exchange ($\SX$) channels can contribute to this order.

With the relativistic treatment of beam field, we can seamlessly extend the results toward the case for positron beam interaction with sample electrons. In this case, exchange term describes a positron annihilating with a sample electron at one vertex while a positron and a sample electron are created at the other, that is, annihilation into a virtual photon and back. We refer to this process as the \emph{annihilation} ($\SA$) channel and treat it in Sec.~\ref{sec:positron}. Thus the annihilation channel is simply the exchange term read with positron modes.

Beyond first order approximation, multiple $\Sint$ insertions occur which allowed the prohibited modes in the first order to temporary exist as the intermediate processes. For instance, at order $e^4$, two insertions of $\Sint$ can contribute to the energy-loss amplitude through a virtual intermediate state with different beam and sample populations such as the pair-production can be followed by the pair-absorption restoring the initial electron numbers in both beam and sample as shown in Sec.~\ref{sec:pair}.

\subsection{Small parameters and validity of the approximations}
\label{sec:power}
Four momenta enter the problem, namely the beam momentum $k$, the momentum transfer $q$, the sample momentum $p$, and the mass $m$. There are also two velocities: the beam velocity $v=k/E$ with $E=\sqrt{k^2+m^2}=m+T$, and the sample velocity $v_t=p/m$. Our framework is an expansion in the small quantities
\begin{equation}
    \lambda_q\equiv\frac{q}{k},\qquad\lambda_p\equiv\frac{p}{k},\qquad v_t=\frac{p}{m},\qquad\frac{\alpha}{v},
    \label{eq:smallparams}
\end{equation}
with the beam velocity $v$ kept to all orders. To get a sense of their size, notice that at 100~keV we have $\lambda_q\sim\lambda_p\sim v_t\sim10^{-2}$ for a transfer of $1$~\AA$^{-1}$ and $\alpha/v\sim10^{-2}$, while at 1~keV we have $\lambda_q\sim0.06$. At the primary energies of reflection spectroscopy, 10--100 eV, $\lambda_q$ reaches $0.2$--$0.6$ for transfers of one to three inverse \aa ngstr\"oms, and $\alpha/v\simeq0.5$. In this regime the expansion is at the limit of its validity. As a result, the leading-order results of this paper should be understood as estimates of the size of each channel rather than as controlled expansions, and the corrections of Secs.~\ref{sec:deltaX} and \ref{sec:beyond} are of comparable size. We will return to what survives in that regime in Sec.~\ref{sec:meels}. Finally, energy transfers to sample excitations satisfy $\omega\lesssim qv_t$, so $\omega/q$ is itself of order $v_t$.

Keeping $v$ exact is the point of our construction. The coefficients that multiply powers of the small parameters are functions of $v$ that contain $m/T$, and they vary by orders of magnitude between 1 and 300~keV. An expansion in $v$ would therefore fail at 100~keV, where $v=0.55$. An expansion in $\lambda_q$ and $\lambda_p$, however, does not fail, because these parameters stay small at every energy the microscope uses. Thus, when we say that a channel is of order $\lambda_q^2$, we mean that its amplitude, relative to the direct amplitude, is $\lambda_q^2$ times a function of $v$ that we give exactly.

\section{Direct channel and its cross section}
\label{sec:direct}
The direct channel corresponds to the standard theory, and its cross section is well known \cite{moller1932,schattschneider2005,iatrakis2026a}. We nevertheless derive it here briefly, for two reasons. First, the same objects, namely the beam vertex, the sample spectral tensor, and the normalization of the cross section, recur in every later channel. Second, the exchange and annihilation amplitudes will be compared to this one.

\subsection{Direct amplitude}
\label{subsec:direct-amplitude}

Let the beam electron have four-momentum $\fK_i=(E_i,\vk_i)$ before and $\fK_f=(E_f,\vk_f)$ after scattering, with spinors $u_i$ and $u_f$ normalized to $\bar{u} u=2m$, and let the sample go from its ground state $\ket{0}$ to an excited state $\ket{n}$ of energy $\omega_{n0}$ above it. The four-momentum transfer is $\fQ=\fK_i-\fK_f=(\omega,\vq)$ with $\omega=E_i-E_f=\omega_{n0}$. The beam current has the matrix element $\bra{\fK_f}\fJ^\mu_\text{bb}(\fX)\ket{\fK_i}=\bar{u}_f\gamma^\mu u_i\,e^{-i\fQ^\rho\fX_\rho}$, and the integrals over $\fX$ and $\fY$ in $\SD$ then produce the photon kernel at momentum $\fQ$ and a delta function of energy and momentum conservation. Stripping the delta function, we define the amplitude $\fM$ by $\bra{f}iS\ket{i}=i\fM\,(2\pi)^4\delta^{(4)}(\cdots)$ and find
\begin{equation}
    \fM_D=-e^2\,\widetilde{\Pi}(\fQ)\;\bar{u}_f\gamma^\mu u_i\;\bra{n}\widetilde{\fJ}_{\text{ss},\mu}(\vq)\ket{0},
    \label{eq:MD}
\end{equation}
where $\fJ_{\text{ss},\mu}=\Psisb\gamma_\mu\Psis$ is the sample current. The amplitude consists of three factors, namely the photon kernel, the beam vertex, and the sample transition current. As we will see, this structure is the same in every channel; only the form of each of the three factors changes.

\subsection{Spin sums and sample response}
\label{subsec:spin-response}

For an unpolarized beam we average over the initial spin and sum over the final spin. The beam vertex then enters through the beam tensor
\begin{align}
    \fL^{\mu\nu}&\equiv\dfrac{1}{2}\sum_{\rm spins}(\bar{u}_f\gamma^\mu u_i)(\bar{u}_f\gamma^\nu u_i)^*\nonumber\\
    &=\dfrac{1}{2}\Tr\big[\gamma^\mu(\slashed{\fK}_f+m)\gamma^\nu(\slashed{\fK}_i+m)\big]\nonumber\\
    &=2\big[\fK_f^\mu \fK_i^\nu+\fK_i^\mu \fK_f^\nu-\eta^{\mu\nu}(\fK_i^\rho\fK_{f,\rho}-m^2)\big].
    \label{eq:lepton}
\end{align}
The sample enters through its spectral tensor. With $\widetilde{\fO}(\vq)\equiv\int\text{d}^3\vx\,e^{i\vq\cdot\vx}\fO(0,\vx)$ the Fourier component of an operator that injects momentum $\vq$, define
\begin{equation}
    S^{ab}(\vq,\omega)\equiv\frac1V\sum_n\bra{0}\widetilde{\fO}_b^{\dagger}(\vq)\ket{n}\bra{n}\widetilde{\fO}_a(\vq)\ket{0}\,\delta(\omega-\omega_{n0}),
    \label{eq:Sab}
\end{equation}
the spectral function of the pair of operators $(\fO_a,\fO_b)$ per unit volume. For any amplitude of the form $\fM=\sum_a c_a\bra{n}\widetilde{\fO}_a(\vq)\ket{0}$, Fermi's golden rule with the beam flux and phase space gives the cross section per unit sample volume
\begin{equation}
    \frac1V\frac{\text{d}^2\sigma}{\text{d}\Omega\,\text{d}\omega}=\frac{k_f}{16\pi^2k_i}\;\overline{\sum_{\rm spins}}\;\sum_{a,b}c_ac_b^*\,S^{ab}(\vq,\omega).
    \label{eq:master0}
\end{equation}
The factor $k_f/16\pi^2k_i$ collects the relativistic normalization of the spinors, the phase space $k_fE_f\,\text{d}\Omega\,\text{d}\omega/(2\pi)^3$, and the flux $k_i/E_i$. Since $e^2=4\pi\alpha$, an amplitude with two powers of $e$ turns this factor into $\alpha^2k_f/k_i$.

For the direct channel, the operators are the four components of the current, and $\fS_{JJ}^{(\mu\nu)}$ is the current--current spectral function. Current conservation, $\omega\widetilde{\fJ}^0_\text{ss}(\vq)=\vq\cdot\widetilde{\vJ}_\text{ss}(\vq)$, fixes its time components in terms of the density response. Moreover, for an isotropic sample the tensor has only two independent functions, longitudinal and transverse:
\begin{equation}
    \begin{aligned}
    \fS_{JJ}^{00}&=R_L,\qquad \fS_{JJ}^{(0i)}=\frac{\omega\fQ^i}{q^2}R_L,\\
    \fS_{JJ}^{(ij)}&=\frac{\omega^2}{q^2}\hat q^i\hat q^jR_L+\dfrac{1}{2}\big(\delta^{ij}-\hat q^i\hat q^j\big)R_T,
    \end{aligned}
    \label{eq:LT}
\end{equation}
where $R_L(\vq,\omega)$ is the dynamic structure factor of the density and $R_T(\vq,\omega)$ is the trace of the current spectral function over the two directions transverse to $\vq$. Contracting Eq.~\eqref{eq:lepton} with Eq.~\eqref{eq:LT} gives
\begin{equation}
    \fL_{\mu\nu}\fS_{JJ}^{\mu\nu}=D_L\,R_L+D_T\,R_T,
\end{equation}
where
\begin{equation}
    \begin{aligned}
    D_L&=\frac{(4E_iE_f+\fQ^2)(\fQ^2)^2}{q^4},\\
    D_T&=\frac{2|\vk_i\times\vk_f|^2}{q^2}-\fQ^2.
    \end{aligned}
    \label{eq:DDT}
\end{equation}

\subsection{Relativistic Bethe cross section}

Inserting Eq.~\eqref{eq:MD} into Eq.~\eqref{eq:master0} with $\widetilde{\Pi}(\fQ)=-1/\fQ^2$, we obtain
\begin{equation}
    \begin{aligned}
    \frac1V\frac{\text{d}^2\sigma_D}{\text{d}\Omega\,\text{d}\omega}=&\alpha^2\frac{k_f}{k_i}\bigg[\frac{4E_iE_f+\fQ^2}{q^4}\,R_L(\vq,\omega)\\
    &+\frac{2|\vk_i\times\vk_f|^2/q^2-\fQ^2}{(\fQ^2)^2}\,R_T(\vq,\omega)\bigg].
    \end{aligned}
    \label{eq:master}
\end{equation}
Equation~\eqref{eq:master} is the relativistic Bethe formula. Notice that its longitudinal term reproduces Bethe's non-relativistic result \cite{bethe1930,inokuti1971} when $E_i,E_f\to m$; in particular, with the Bohr radius $a_0=1/m\alpha$ it becomes $(4/a_0^2q^4)(k_f/k_i)R_L$. At relativistic energies, the factor $4E_iE_f\to4m^2\gamma^2$ is the $\gamma^2$ of the standard microscopy formula \cite{egerton2011}. Moreover, the longitudinal coefficient is identical to that of Iatrakis and Brudanin \cite{iatrakis2026a}, whose $(E_i+E_f)^2-q^2$ equals $4E_iE_f+\fQ^2$ identically, and the longitudinal--transverse structure is that of de Forest and Walecka \cite{deforest1966} and Schattschneider \emph{et al.} \cite{schattschneider2005}. The transverse term describes the coupling of the beam's current to that of the sample. As we will see in Sec.~\ref{sec:magnon}, it is this term that excites magnons in the direct channel.

\section{Exchange channel}
\label{sec:exchange}

\subsection{Exchange amplitude}

Consider the matrix element of $\SX$ between the same states. At $\fY$, the beam field annihilates the incoming electron and the sample field creates a sample electron of momentum $\fP'$. At $\fX$, the sample field annihilates a sample electron of momentum $\fP$ and the beam field creates the outgoing beam electron. In order to bring the two beam operators and the two sample operators together, the fermion operators must be reordered, which produces one minus sign relative to the direct channel. This is just the familiar relative sign of M{\o}ller scattering. Writing the beam spinor indices explicitly, we have
\begin{equation}
    \begin{aligned}
    \bra{f}i\SX\ket{i}=+ie^2\!\sum_{a,b,c,d}\iint\Pi(\fX-\fY)\,e^{i\fK_f^\rho\fX_\rho-i\fK_i^\rho\fY_\rho}\\
    \times\bar{u}_{f,a}\gamma^\mu_{ab}\,\gamma_{\mu,cd}\,u_{i,d}\;\bra{n}\bar\Psi_{{\rm s},c}(\fY)\Psi_{{\rm s},b}(\fX)\ket{0}.
    \end{aligned}
    \label{eq:SXme}
\end{equation}
The integrals fix the momentum flowing through the photon. The sample electron gains $\fP'-\fP=\fQ$, as it must, while the photon carries $\fQX=\fK_i-\fP'$, which is the difference between the incoming beam momentum and an outgoing \emph{sample} momentum. Using $\fP^{\prime 0}=\sqrt{m^2+p'^2}$ with $p'=|\vp'|$, we find that its virtuality is
\begin{equation}
    \begin{aligned}
    -\fQX^2&=-(\fK_i-\fP')^2\\
    &=2m(E-m)+\frac{E}{m}p'^2-2\vk_i\!\cdot\!\vp'+O(Ep'^4/m^3).
    \end{aligned}
\label{eq:Q2}
\end{equation}
The first term is $2mT$, which is hundreds of kiloelectronvolts squared, while the others are smaller by $p'^2/mT$ and $2kp'/mT$, i.e.\ by $\lambda_p$. As a result, the photon is far off shell and spacelike, and its kernel is approximately a constant,
\begin{equation}
    \widetilde{\Pi}(\fQX)\;\to\;\PiX\equiv\frac{1}{2m(E-m)}=\frac{1}{2mT}.
\label{eq:PiX}
\end{equation}
Non-relativistically, $2mT=k^2$ and $\PiX=1/k^2$, which is the Ochkur approximation of atomic collision theory \cite{ochkur1964,rudge1968}; that is, the exchange amplitude at high energy is the Born amplitude with $1/q^2$ replaced by $1/k^2$. Equation~\eqref{eq:PiX} is therefore the relativistic generalization of the Ochkur \emph{kernel}, and it is exact in the beam energy. The replacement $k^2\to2m(E-m)$ is the entire relativistic correction to the kernel. It should be noted, however, that the vertex structure that multiplies the kernel, given in Eqs.~\eqref{eq:MX}--\eqref{eq:MXlead} below, goes beyond Ochkur's approximation and reduces to it only in the non-relativistic limit. Since a constant kernel in momentum space is a delta function in position space, the exchange interaction is a contact interaction. The two vertices in Eq.~\eqref{eq:SXme} collapse to a single point, and the sample enters only through operators evaluated at that point.

\subsection{Local approximation and operator basis}
\label{subsec:exchange-operators}

Even with the vertices at one point, the beam spinors and the sample fields in Eq.~\eqref{eq:SXme} remain tangled together, since the beam spinor $\bar{u}_f$ sits next to one sample field and the beam spinor $u_i$ sits next to the other. To untangle them, we use the Fierz identity. For the product of two Dirac matrices with indices arranged as in Eq.~\eqref{eq:SXme}, we have
\begin{equation}
    \begin{aligned}
    \gamma^\mu_{ab}\,\gamma_{\mu,cd}&=\sum_A c_A\,\Gamma^A_{ad}\,\Gamma_{A,cb},\\
    \Gamma^A&=(\mathbb{1}_4,\gamma^5,\gamma^\mu,\gamma^5\gamma^\mu),\\
    c_A&=\Big(1,\,-1,\,-\dfrac{1}{2},\,-\dfrac{1}{2}\Big),
    \end{aligned}
    \label{eq:fierz}
\end{equation}
where $\Gamma_A$ denotes the same matrices with lowered index (see Appendix~\ref{app:fierz} for the derivation and a numerical check). The tensor structure $\sigma^{\mu\nu}$ does not appear, because it is absent from the Fierz expansion of $\gamma^\mu\otimes\gamma_\mu$. Inserting Eq.~\eqref{eq:fierz} into Eq.~\eqref{eq:SXme} and stripping the delta function, we find
\begin{equation}
    \fM_X=+e^2\PiX\sum_Ac_A\,(\bar{u}_f\Gamma^Au_i)\,\bra{n}\widetilde{\Psisb\Gamma_A\Psis}(\vq)\ket{0}.
    \label{eq:MX}
\end{equation}
This amplitude has the same structure as the direct amplitude \eqref{eq:MD}, namely a kernel, a beam vertex, and a sample matrix element. Now, however, there are four beam vertices instead of one, four sample operators instead of the current, and the hard kernel in place of $1/q^2$. In the language of effective field theory, Eq.~\eqref{eq:MX} is the leading local operator produced by integrating out the hard photon, and the Fierz coefficients $c_A$ are its Wilson coefficients, which are exact at order $\alpha$.

We must now express the sample operators in terms of the non-relativistic field. To leading order in $v_t$, we have
\begin{equation}
    \begin{aligned}
    \Psisb\Psis&\to\rho,\\
    \Psisb\gamma^0\Psis&\to\rho,\\
    \Psisb\gamma^5\gamma^i\Psis&\to-s^i,
    \end{aligned}
    \label{eq:NRops}
\end{equation}
while $\Psisb\gamma^i\Psis$ is the current $\fJ^i_\text{ss}$, of order $v_t$, and $\Psisb\gamma^5\Psis$ and $\Psisb\gamma^5\gamma^0\Psis$ are of order $v_t$ as well. Collecting the leading terms, we find
\begin{equation}
    \begin{aligned}
    \fM_X=\\
    e^2\PiX\Big\{&\big[\bar{u}_f\big(\mathbb{1}_4-\tfrac12\gamma^0\big)u_i\big]\widetilde{\rho}_{n0}+\dfrac{1}{2}(\bar{u}_f\gamma^iu_i)\,\widetilde{\fJ}^i_{\text{ss},n0}\\
    &-\dfrac{1}{2}(\bar{u}_f\gamma^5\gamma^iu_i)\,\widetilde{s}^i_{n0}\Big\}+O(v_t),
    \end{aligned}
    \label{eq:MXlead}
\end{equation}
where $\widetilde{\rho}_{n0}=\bra{n}\widetilde{\rho}(\vq)\ket{0}$ and so on. We see that exchange couples to the sample's density through the scalar and time-vector vertices, which combine into the single bracket shown, and to its spin density through the axial vertex. The spatial current term is of order $v_t$, but we keep it because it interferes with the direct current at the same order as other terms we keep below.

We can now understand the essential physics of the two channels. In the direct channel, the beam couples to the sample's charge and current with strength $e^2/q^2$, which is large at small transfer. In the exchange channel, it couples to the sample's density and spin density with strength $e^2/2mT$, which is independent of the transfer because the interaction is local. The ratio of the two, $q^2/2mT$, is of order $\lambda_q^2$, and it is in this sense that exchange is small at microscope energies. Nevertheless, the exchange channel can do something that the direct channel cannot do at leading order, namely couple to the spin density, and it does so with a coefficient that does not fall with $q$.

\subsection{Density channel}
\label{sec:density}
\label{subsec:exchange-density}

Suppose first that the sample has no magnetic order. Then density and spin fluctuations are uncorrelated, and only the density terms of Eq.~\eqref{eq:MXlead} interfere with the direct amplitude. The interference contributes $2\,\Re(\fM_D^*\fM_X)$ to the cross section. To compute it, we need two beam spin sums in addition to Eq.~\eqref{eq:lepton},
\begin{equation}
    \begin{aligned}
    \dfrac{1}{2}\sum_{\rm spins}(\bar{u}_f\gamma^\nu u_i)^*&(\bar{u}_fu_i)\\
    &=\dfrac{1}{2}\Tr\big[\gamma^\nu(\slashed{\fK}_f+m)(\slashed{\fK}_i+m)\big]\\
    &=2m\,(\fK_i+\fK_f)^\nu ,
    \end{aligned}
    \label{eq:trace1}
\end{equation}
and the sample spectral functions of Eq.~\eqref{eq:LT}. Contracting the scalar-vertex term with the density--current spectral function, and using $\vq\!\cdot\!(\vk_i+\vk_f)=k_i^2-k_f^2=\omega(E_i+E_f)$, we find $(\fK_i+\fK_f)^\nu\eta_{\nu\rho}\fS_{JJ}^{0\rho}=-(E_i+E_f)(\fQ^2/q^2)R_L$, while the vector-vertex term contracts to $\fL_{\mu\nu}\fS_{JJ}^{\mu\nu}$ of Eq.~\eqref{eq:DDT}. Letting $\mathcal D\equiv e^4\widetilde{\Pi}(\fQ)^2(D_LR_L+D_TR_T)$ denote the direct contribution to Eq.~\eqref{eq:master0}, the result is
\begin{equation}
    \begin{aligned}
    \mathcal I_X&\equiv2\,\Re\,\overline{\sum}\,\frac1V\sum_n\fM_D^*\fM_X\,\delta(\omega-\omega_{n0})\\
    &=e^4\,\widetilde{\Pi}\,\PiX\Big[(D_LR_L+D_TR_T)\\
    &\quad\qquad\qquad+4m(E_i+E_f)\frac{\fQ^2}{q^2}R_L\Big].
    \end{aligned}
    \label{eq:IX}
\end{equation}
It is convenient to define the kinematic ratio
\begin{equation}
    \Lambda\equiv\frac{4m(E_i+E_f)}{4E_iE_f+\fQ^2}\,\frac{q^2}{\kappa}\,\frac{D_LR_L}{D_LR_L+D_TR_T},
    \label{eq:Lambda}
\end{equation}
where $\kappa\equiv q^2-\omega^2$. The ratio $\Lambda$ tends to $2$ in the non-relativistic limit and to $2m/E$ in the ultra-relativistic one, and reduces to $8mE/(4E^2-\kappa)$ for quasi-elastic scattering with $R_T=0$. We then find
\begin{equation}
    \frac{\mathcal I_X}{\mathcal D}=\frac{\kappa\,(1-\Lambda)}{2m(E-m)}.
    \label{eq:IXratio}
\end{equation}
Non-relativistically, $\Lambda=2$ and $\mathcal I_X/\mathcal D\to-q^2/k^2$. In this limit, exchange reduces the cross section by the M{\o}ller factor, which for an unpolarized sample is twice the average of the spin-exchange operator, $2\langle P_{\rm ex}\rangle q^2/k^2$ with $\langle P_{\rm ex}\rangle=\dfrac{1}{2}$. Ultra-relativistically, by contrast, $\Lambda\to0$ and $\mathcal I_X/\mathcal D\to+q^2/2mE$, meaning that the interference is constructive. The two limits have opposite signs. The crossing occurs where $\Lambda=1$, which for quasi-elastic scattering is $E=2m$, corresponding to a beam kinetic energy of 511~keV.

\subsection{Sign reversal of exchange interference}
\label{sec:e2m}

Why should the sign of the interference reverse? As we will now see, the reversal has a simple origin. In the exchange diagram, the hard photon connects a fast electron to a slow one at each vertex. The transition current between a beam spinor of momentum $k$ and a sample spinor at rest has time component $\sqrt{2m(E+m)}$ and space components $\sqrt{2m(E+m)}\,k/(E+m)$. Consequently, its magnetic part is smaller than its electric part by
\begin{equation}
    \frac{|\vJ|}{\fJ^0}=\frac{k}{E+m}=\sqrt{\frac{E-m}{E+m}}\,.
\end{equation}
\begin{figure}[!htb]
\includegraphics[width=\columnwidth]{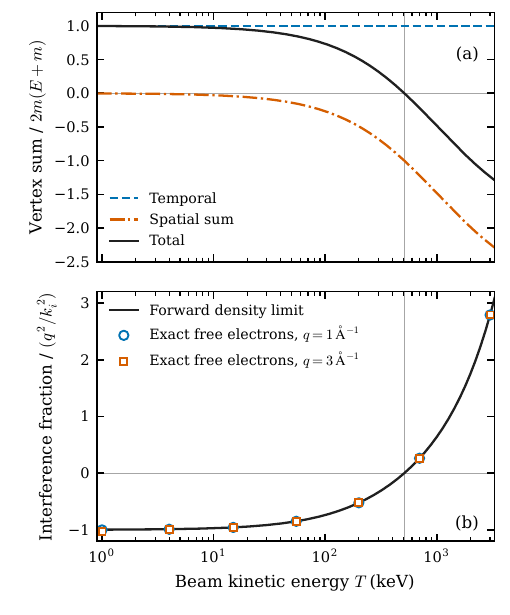}
\caption{Origin and free-electron check of the leading density-interference sign reversal. (a) The coefficient of $\xi_f^\dagger\xi_i$ in the sum of Eq.~\eqref{eq:4m2mE}, divided by $2m(E+m)$. The temporal contribution is $1$, the summed spatial contribution is $-3T/(T+2m)$ and their total vanishes at $T=m$. (b) Interference divided by the direct weight, in units of $q^2/k_i^2$. The solid line is the retained density/current fraction $\mathcal I_X/\mathcal D$ in the forward, small-loss density limit of Eq.~\eqref{eq:IXratio}, with $\Lambda=2m/E$. Markers show the full direct--exchange interference divided by the direct weight for a free electron pair, calculated from the single-photon amplitudes with an initially stationary target, recoil and $q=1$ or $3$~\AA$^{-1}$. Initial beam and target spins are averaged independently with total weight $1/4$, and the spins of both final electrons are summed. Gray vertical lines mark $T=m\simeq511$~keV. For an isotropic sample, response corrections can shift the retained density/current zero through $\Lambda$ in Eq.~\eqref{eq:Lambda}.}
\label{fig:exchange-sign}
\end{figure}
Now, the exchange amplitude is the product of two such currents contracted with the metric, $\fJ^0\fJ'{}^0-\vJ\cdot\vJ'$. When we sum over the sample spin, the three spatial components contribute equally, so the magnetic term is $3(E-m)/(E+m)$ times the electric one. As a result, the forward exchange amplitude summed over sample spin is
\begin{equation}
    2m(E+m)\Big[1-3\,\frac{E-m}{E+m}\Big]=4m\,(2m-E).
\label{eq:4m2mE}
\end{equation}
The electric coupling of the hard photon dominates at low energy and gives the M{\o}ller sign. The magnetic coupling, however, grows with the transition velocity and overtakes the electric coupling when that velocity reaches $1/\sqrt3$, which corresponds to $E=2m$. Above this energy, the exchange interference is constructive. We verified Eq.~\eqref{eq:4m2mE} with explicit Dirac spinors at six energies (see Appendix~\ref{app:numerics}). Figure~\ref{fig:exchange-sign} shows the two contributions and compares the resulting interference with exact scattering off a free electron. The same bookkeeping, with three spatial components against one temporal component, will reappear in the correction to the kernel below.

\subsection{Spin channel}
\label{sec:spinX}

Now suppose that the sample is magnetic. Then the axial term of Eq.~\eqref{eq:MXlead} couples the beam to the spin density, giving $\fM_X^{\rm spin}=-\dfrac{1}{2} e^2\PiX(\bar{u}_f\gamma^5\gamma^iu_i)\widetilde{s}^i_{n0}$. Fortunately, the vertex takes an exact and simple form in the basis of canonical (rest-frame) spin. For $\vk_i$ along $\hat{\vz}$ and small scattering angle, we have
\begin{equation}
    \begin{aligned}
    \bar{u}_f\gamma^5\gamma^{x,y}u_i&=-2m\,\xi_f^\dagger\sigma^{x,y}\xi_i,\\
    \bar{u}_f\gamma^5\gamma^{z}u_i&=-2E\,\xi_f^\dagger\sigma^{z}\xi_i ,
    \end{aligned}
    \label{eq:axialvertex}
\end{equation}
where $\xi$ are the two-component spinors of the beam electron. Notice that the components transverse to the beam carry the mass, while the component along the beam carries the energy; the latter is the familiar statement that $\gamma^5\gamma^z$ measures helicity. As a result, the exchange spin amplitude does not grow with beam energy for transverse spin components, whereas the direct magnetic amplitude, which couples through the beam's convective current $\sim2k$, does. This difference is the origin of the crossover between the two channels in magnon scattering, which we discuss in Sec.~\ref{sec:magnon}. Finally, since the spin density $\vs$ is a pseudovector and $\gamma^5\gamma^i$ is an axial vertex, the operator is parity-even, as it must be.

\subsection{Corrections to the contact kernel}
\label{sec:deltaX}
\label{subsec:exchange-kernel}

The kernel \eqref{eq:PiX} is only the leading term in the expansion of Eq.~\eqref{eq:Q2}. Writing $\widetilde{\Pi}(\fQX)=\PiX[1+\delta(\vp')]$, we find
\begin{equation}
    \delta(\vp')=\dfrac{2\vk_i\!\cdot\!\vp'}{2mT}-\dfrac{E\,p'^2}{2m^2T}+\dfrac{(2\vk_i\!\cdot\!\vp')^2}{(2mT)^2}+\cdots .
    \label{eq:exchange-kernel-expansion-main}
\end{equation}
The linear term becomes an operator with one derivative acting on the sample creation field $\Psisb$ via $\vp'\mapsto i\vnabla$. In the free-electron momentum average of Appendix~\ref{app:exchange-kernel}, it enters at relative order $\vk_i\!\cdot\!\langle\vp'\rangle/mT=O(\lambda_p)$. For an isotropic benchmark weight, however, it averages to zero, and the first correction is quadratic,
\begin{equation}
    \delta_X=\frac{2m-E}{3m}\;\frac{\langle p'^2\rangle_{\rm iso}}{2mT}.
    \label{eq:deltaX}
\end{equation}
The coefficient arises in the same way as in Eq.~\eqref{eq:4m2mE}. In particular, with the common denominator $(2mT)^2$, the angular average of $(2\vk_i\!\cdot\!\vp')^2$ gives $4k^2p'^2/3$, to be compared with $2E(E-m)p'^2$ from the energy, and $4k^2/3-2E(E-m)\propto2m-E$. We verified Eq.~\eqref{eq:deltaX} against the exact angular average of $1/(-\fQX^2)$ to five digits at six energies. Suppose, for example, that the relevant sample electrons are $3d$ electrons with $\langle p'^2\rangle_{\rm iso}\approx25$~\AA$^{-2}$. The benchmark correction is then $3\times10^{-4}$ at 100~keV, $0.3\%$ at 10~keV, and $1\%$ at 3~keV. It vanishes at $E=2m$ and is negative above. The corrections from the sample side are the $O(v_t)$ operators omitted in Eq.~\eqref{eq:NRops}.

\subsection{Magnitude and scaling of exchange}
\label{sec:magnitudes}

Table~\ref{tab:magnitudes} gives the size of the exchange channel across the microscope range. We see that the density interference $|\mathcal I_X/\mathcal D|$ is of order $\lambda_q^2$, lying between $10^{-5}$ and $10^{-3}$ at 10--300~keV. Thus, for unpolarized samples and spin-blind detection, exchange is a negligible correction to the direct cross section throughout the microscope range, which justifies the standard practice of ignoring it. The spin amplitude, however, behaves quite differently. Its natural point of comparison is the direct \emph{magnetic} amplitude, which is itself small. We defer this comparison to Sec.~\ref{sec:magnon}, where we will find that the ratio is $\epsilonS\simeq qm/(2Tk)$ and that it reaches $1/\sqrt{2}$ near $T_0\simeq8$~keV for $q=1$~\AA$^{-1}$.

\begin{figure}[!htb]
\centering
\begin{tikzpicture}
\node[anchor=south west,inner sep=0] (map) at (0,0)
  {\includegraphics[width=\columnwidth]{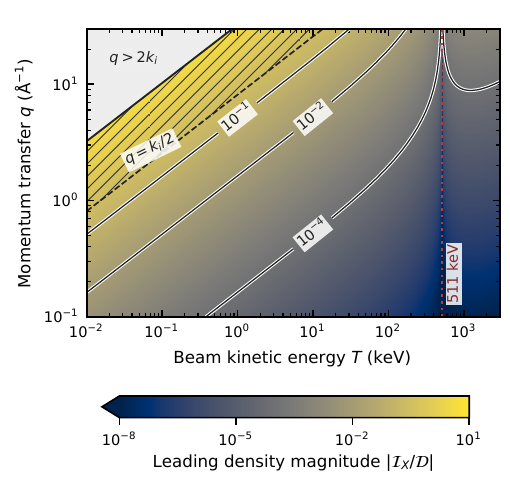}};
\begin{scope}[x={(map.south east)},y={(map.north west)}]
\def\EELSbox#1#2#3#4#5{%
  \pgfmathsetmacro{\xl}{0.17+0.81*(ln(#1)-ln(0.01))/(ln(3000)-ln(0.01))}%
  \pgfmathsetmacro{\xr}{0.17+0.81*(ln(#2)-ln(0.01))/(ln(3000)-ln(0.01))}%
  \pgfmathsetmacro{\yb}{0.34+0.60*ln(#3/0.1)/ln(300)}%
  \pgfmathsetmacro{\yt}{0.34+0.60*ln(#4/0.1)/ln(300)}%
  \draw[red!80!black,line width=0.7pt]
    (\xl,\yb) rectangle (\xr,\yt);
  \node[anchor=north,font=\scriptsize,fill=white,
    fill opacity=0.9,text opacity=1,inner sep=1pt]
    at ({(\xl+\xr)/2},{\yt-0.012}) {#5};
}
\EELSbox{0.01}{0.1}{0.1}{3}{M-EELS}
\EELSbox{1}{30}{0.1}{10}{SEM/LV}
\EELSbox{60}{300}{0.1}{10}{STEM}
\end{scope}
\end{tikzpicture}
\caption{Leading density-interference magnitude from Eq.~\eqref{eq:IXratio} for quasi-elastic kinematics. Color and labeled contours show $|\mathcal I_X/\mathcal D|$. Values below $10^{-8}$ use the lowest color, including the leading zero near $511$~keV marked by the vertical dotted line. Gray excludes $q>2k_i$, which is inaccessible at zero loss. Hatching marks $k_i/2\leq q\leq2k_i$, where the required $q\ll k_i$ is not satisfied. Red boxes mark illustrative windows for momentum-resolved reflection EELS (M-EELS), scanning or low-voltage electron microscopy (SEM/LV) and scanning transmission electron microscopy (STEM). Their energy ranges are $10$--$100$~eV, $1$--$30$~keV and $60$--$300$~keV, respectively. The M-EELS box spans $q=0.1$--$3$~\AA$^{-1}$ and the other boxes span $0.1$--$10$~\AA$^{-1}$. These windows are guides, not instrument specifications or validity bounds. The plotted $q$ is the total spatial transfer and need not equal the reported in-plane transfer in reflection. All colored regions remain subject to the loss, response and local-kernel conditions in the text.}
\label{fig4}
\end{figure}

When, then, must exchange be included? To answer this question, we must understand why exchange is small and how it scales, both of which can be read off from Eq.~\eqref{eq:IXratio}. Exchange is small because the exchange interaction is a contact interaction, with coefficient $1/2mT$ set by the hard photon, while the direct interaction is a Coulomb interaction with coefficient $1/q^2$. The ratio of the two is therefore $q^2/2mT$. Non-relativistically, $|\mathcal I_X/\mathcal D|\simeq q^2/k^2$, and in the quasi-elastic limit it is $q^2/k^2$ times $|1-\Lambda|\,(E+m)/2m$, a factor that lies between $0.34$ and $1$ from 300~keV down to the non-relativistic limit. As a result, exchange grows as the square of the momentum transfer and falls as the inverse of the beam's kinetic energy, and the same ratio $q^2/2mT$ is the parameter that determines when exchange must be included. Table~\ref{tab:when} tabulates $|\mathcal I_X/\mathcal D|$ on a grid of representative beam energies and momentum transfers, together with the transfer at which exchange reaches one and ten percent of the direct cross section, and Fig.~\ref{fig4} plots it.

From these results, we can identify three regimes. First, for monochromated microscope spectroscopy at 60--300~keV, exchange is below $10^{-4}$ for every transfer inside the first few Brillouin zones and reaches one percent only above $14$~\AA$^{-1}$. In this regime, the direct-channel theory is quantitatively complete for spin-blind measurements, and this provides a quantitative basis for standard practice. Second, at 1--20~keV, the range of low-voltage microscopes and of scanning electron microscopes, exchange reaches one percent at $q\simeq1.6$--$7.5$~\AA$^{-1}$, which is within reach of core-loss and large-angle measurements. Third, at 10--100~eV, exchange is tens of percent at one to three inverse \aa ngstr\"oms. In this regime, exchange must be included in any quantitative treatment of the cross section.

Two further rules of thumb follow from the operator structure itself, independent of the numbers. First, the leading exchange correction multiplies the density spectral function $R_L(\vq,\omega)$ by $1+\mathcal I_X/\mathcal D$ at fixed $\vq$. Because both channels are linear in the same sample operator, this correction rescales the weight of the spectrum as a function of momentum transfer but does not change its shape in energy. As a result, a measurement that compares line shapes at fixed $\vq$ is unaffected, whereas a measurement that extracts the absolute response, or compares intensities across momentum transfers, is affected. Second, the exchange amplitude reverses the beam electron's spin while the direct amplitude does not. Consequently, a spin-polarized source with a spin analyzer separates the two channels regardless of their relative size. This is the principle behind spin-polarized surface spectroscopy \cite{kirschner1985}, and in Sec.~\ref{sec:magnon} we will see how a polarized source alone can recover it in the microscope.

\begin{table}[t]
\caption{Magnitude of the leading density interference from Eq.~\eqref{eq:IXratio} for quasi-elastic kinematics. The first column gives the beam kinetic energy in keV. The next four columns give $|\mathcal I_X/\mathcal D|$ at the indicated spatial transfers in \AA$^{-1}$; entries $a\,(-n)$ mean $a\times10^{-n}$. The signed interference is negative at every listed energy. The last two columns give the formal transfers $q_{1\%}$ and $q_{10\%}$ in \AA$^{-1}$ at which this leading expression reaches the indicated magnitude. They are extrapolations of that expression, not controlled thresholds for the full exchange correction. Every entry remains subject to the low-loss, density-dominated and small-transfer conditions and the local-kernel conditions of Sec.~\ref{subsec:exchange-operators}. In particular, the larger transfers at $100$~eV and the percentage thresholds need separate applicability checks.}
\label{tab:when}
\begin{ruledtabular}
\begin{tabular}{lcccccc}
$T$ [keV] & $q=0.3$ & $q=1$ & $q=2$ & $q=3$ & $q_{1\%}$ & $q_{10\%}$\\
\hline
0.1 & $3.4\,(-3)$ & $3.8\,(-2)$ & $0.15$ & $0.34$ & 0.5 & 1.6\\
1 & $3.4\,(-4)$ & $3.8\,(-3)$ & $1.5\,(-2)$ & $3.4\,(-2)$ & 1.6 & 5.1\\
20 & $1.6\,(-5)$ & $1.8\,(-4)$ & $7.0\,(-4)$ & $1.6\,(-3)$ & 7.5 & 24\\
60 & $4.5\,(-6)$ & $5.0\,(-5)$ & $2.0\,(-4)$ & $4.5\,(-4)$ & 14 & 45\\
100 & $2.3\,(-6)$ & $2.6\,(-5)$ & $1.0\,(-4)$ & $2.3\,(-4)$ & 20 & 62\\
300 & $3.0\,(-7)$ & $3.3\,(-6)$ & $1.3\,(-5)$ & $3.0\,(-5)$ & 55 & 170\\
\end{tabular}
\end{ruledtabular}
\end{table}

\section{Annihilation channel for positron beams}
\label{sec:positron}

A positron beam is described by the same field $\Psib$, which now contains positron modes, $\Psib\supset b^\dagger_{\vk}v(\fK)e^{+i\fK^\rho\fX_\rho}$, where $b^\dagger$ creates a positron and $v(\fK)$ is the positron spinor. The channel decomposition is therefore unchanged, and only the interpretation of its terms is different.

\subsection{Charge independence of the direct channel}
\label{subsec:positron-direct}

For an incoming positron $\fK_i$ and outgoing positron $\fK_f$, the connected matrix element of the beam current is
\begin{equation}
    \bra{e^+(\fK_f)}\fJ^\mu_\text{bb}(\fX)\ket{e^+(\fK_i)}=-\,\bar v_i\gamma^\mu v_f\,e^{-i\fQ^\rho\fX_\rho},
\end{equation}
which has the same phase as for an electron, so that the sample receives the same $\fQ$, and which carries a minus sign from the anticommutation of $b_{\vk_i}b^\dagger_{\vk_f}$, reflecting the positron's opposite charge. Hence $\fM_D^{(+)}=+e^2\widetilde{\Pi}(\fQ)\,\bar v_i\gamma^\mu v_f\,\bra{n}\widetilde{\fJ}_{\text{ss},\mu}(\vq)\ket{0}$. The positron beam tensor is $\dfrac{1}{2}\Tr[\gamma^\mu(\slashed{\fK}_f-m)\gamma^\nu(\slashed{\fK}_i-m)]$, in which the mass enters squared, so it is equal to Eq.~\eqref{eq:lepton}. Therefore, we find
\begin{equation}
    \sigma_D^{(+)}=\sigma_D^{(-)}
\end{equation}
exactly, at order $e^2$, for any sample. The intuition behind this result is that a cross section at order $e^4$ cannot know the sign of $e$. Indeed, the sample only learns the sign of the beam's charge when two photons are exchanged, as we discuss in Sec.~\ref{sec:beyond}.

\subsection{Annihilation from the exchange operator}
\label{subsec:annihilation-amplitude}

Consider now $\SX$ with positron modes. In this case, the operator $\Psibb$ at $\fX$ \emph{annihilates} the incoming positron while $\Psis$ annihilates a sample electron, so that the pair disappears into the photon. At $\fY$, the pair is recreated. The photon carries $\fQA=\fK_i+\fP$, the sum of the positron and electron momenta, and
\begin{equation}
    \fQA^2=(\fK_i+\fP)^2\simeq2m(E+m),
\end{equation}
which is timelike and far from the pole. As a result, the kernel is again a constant,
\begin{equation}
    \PiA\equiv-\frac{1}{2m(E+m)},
    \label{eq:annihilation-kernel}
\end{equation}
which is smaller in magnitude than $\PiX$ by $(E-m)/(E+m)=T/(T+2m)$ and opposite in sign. Moreover, the reordering of fermion operators now produces no net minus sign, since two signs cancel where for electrons they combined. The Fierz rearrangement then proceeds as before, giving
\begin{equation}
    \fM_A^{(+)}=-e^2\PiA\sum_Ac_A\,(\bar v_i\Gamma^Av_f)\,\bra{n}\widetilde{\Psisb\Gamma_A\Psis}(\vq)\ket{0}.
    \label{eq:MA}
\end{equation}
This is just the exchange operator read with positron modes, and we refer to it as the annihilation channel. As a check, suppose that the sample electron is free. Then the Fierz identity undoes itself on c-number spinors, and $\fM_D^{(+)}+\fM_A^{(+)}$ becomes
\begin{equation}
    -e^2\left[\frac{(\bar v_i\gamma^\mu v_f)(\bar{u}'\gamma_\mu u)}{t}-\frac{(\bar v_i\gamma^\mu u)(\bar{u}'\gamma_\mu v_f)}{s}\right],
\end{equation}
which is the Bhabha amplitude with its correct relative sign \cite{bhabha1936}, just as $\fM_D+\fM_X$ gives the M{\o}ller amplitude.

\subsection{Non-relativistic limit}

For slow positrons, the positron vertices reduce to $\bar v_iv_f\to-2m\,\eta_i^\dagger\eta_f$, $\bar v_i\gamma^0v_f\to+2m\,\eta_i^\dagger\eta_f$, and $\bar v_i\gamma^5\gamma^kv_f\to-2m\,\eta_i^\dagger\sigma^k\eta_f$, where $\eta$ is the two-spinor in $v$. The physical spin of the positron is $\xi$ with $\eta=i\sigma^2\xi^*$, which gives $\eta_i^\dagger\eta_f=\xi_f^\dagger\xi_i$ and $\eta_i^\dagger\sigma^k\eta_f=-\xi_f^\dagger\sigma^k\xi_i$. Using Eq.~\eqref{eq:NRops}, we then find
\begin{equation}
    \fM_A^{(+)}\;\to\;-\frac{e^2}{4m}\,\xi_f^\dagger\big(3\widetilde{\rho}_{n0}+\vsigma\!\cdot\!\widetilde{\vs}_{n0}\big)\xi_i .
    \label{eq:annihilation-nonrel-amplitude}
\end{equation}
For a single sample electron, $3+\vsigma_{e^+}\!\cdot\vsigma_{e^-}=4P_{S=1}$, where $P_{S=1}$ is the projector on the spin-triplet pair. The annihilation channel is therefore a contact interaction that acts only on triplet pairs. This is exactly as it must be, since a single virtual photon has spin one. The sign is repulsive, opposite to that of the attractive Coulomb term. Converting to a potential with $\widetilde{\rho}_{n0}\to2m$ and the Born relation $\fM=-(2m)^2\widetilde{V}_{\rm ann}(\vq)$, we obtain
\begin{equation}
    V_{\rm ann}(\bm{r})=\frac{2\pi\alpha}{m^2}\,P_{S=1}\,\delta^{(3)}(\bm{r}),
\end{equation}
which is the virtual-annihilation potential of positronium responsible for the term $\alpha^4m/4$ in its ground-state hyperfine splitting \cite{blp}. Remarkably, our many-body scattering formalism, when reduced to a single electron at rest, reproduces this textbook potential down to the coefficient. This agreement is the most stringent check we have of the signs and normalizations in Eqs.~\eqref{eq:MX} and \eqref{eq:MA}. We also confirmed numerically that the threshold amplitude has eigenvalues $\{0,-2,-2,-2\}\times e^2$ in the pair-spin space (see Appendix~\ref{app:numerics}).

\subsection{Annihilation interference and the electron--positron ratio}

The calculation of Sec.~\ref{sec:density} goes through as before, with $\PiX\to\PiA$, $u\to v$, and the trace \eqref{eq:trace1} replaced by $\dfrac{1}{2}\Tr[\gamma^\nu(\slashed{\fK}_i-m)(\slashed{\fK}_f-m)]=-2m(\fK_i+\fK_f)^\nu$. Here the sign of the mass term flips relative to the vector term, and the result is
\begin{equation}
    \begin{aligned}
    \mathcal I_A&=e^4\,\widetilde{\Pi}\,\PiA\Big[(D_LR_L+D_TR_T)\\
    &\quad\qquad\qquad-4m(E_i+E_f)\frac{\fQ^2}{q^2}R_L\Big],\\
    \frac{\mathcal I_A}{\mathcal D}&=-\frac{\kappa\,(1+\Lambda)}{2m(E+m)}.
    \end{aligned}
    \label{eq:IAratio}
\end{equation}
Non-relativistically, this ratio is $-3q^2/4m^2$, which reflects the competition between the repulsive triplet term, acting on the three-quarters of pairs that are triplets, and the attractive Coulomb term. The annihilation interference is destructive at all energies. Since the direct cross sections are equal, the fractional contrast of electron to positron cross sections at the same kinematics is, to first order in the small quantities,
\begin{equation}
    \begin{aligned}
    \Delta_{e^\pm}\equiv\frac{\sigma^{(-)}}{\sigma^{(+)}}-1&\simeq\frac{\mathcal I_X-\mathcal I_A}{\mathcal D}\\
    &=\frac{\kappa}{2m}\left[\frac{1-\Lambda}{E-m}+\frac{1+\Lambda}{E+m}\right]\\
    &\;\xrightarrow{\ \text{quasi-elastic}\ }\;\frac{q^2}{k^2}\,\frac{E^2-2m^2}{mE}.
    \end{aligned}
    \label{eq:Rratio}
\end{equation}
Non-relativistically, $\Delta_{e^\pm}\simeq-q^2/k^2$, which is just the M{\o}ller deficit, while ultra-relativistically it is $+q^2/mE$. It crosses zero at $E=\sqrt2\,m$, corresponding to a beam energy of 212~keV. At $q=1$~\AA$^{-1}$, it is $-3.8\times10^{-3}$ at 1~keV, $-1.7\times10^{-5}$ at 100~keV, and $+3.2\times10^{-6}$ at 300~keV. We verified both interference terms against brute-force spin sums with explicit Dirac spinors (see Appendix~\ref{app:numerics}).

Since this contrast is free of sample parameters at leading order, one might hope to use it as a null test of exchange. Unfortunately, however, we will show in Sec.~\ref{sec:beyond} that the two-photon distortion of the beam by the sample's mean potential produces a charge-odd asymmetry with the same energy scaling and a larger coefficient for every solid. As a result, the density-channel contrast does not isolate exchange. The spin channel, however, does isolate it. In particular, the positron's annihilation amplitude to the spin density is smaller than the electron's exchange amplitude by
\begin{equation}
    \frac{|\PiA|}{\PiX}=\frac{T}{T+2m},
\end{equation}
which is $10^{-3}$ at 1~keV and $0.09$ at 100~keV, while the direct magnetic amplitude is the same for both beams. Thus, below $\sim10$~keV, a positron beam is effectively unable to flip a sample spin except through the direct magnetic coupling. Comparing the ratio of spin-flip to non-spin-flip intensities between electron and positron beams therefore provides a clean test in which the distortion cancels.

\section{Exchange in low-energy reflection spectroscopy}
\label{sec:meels}

Momentum-resolved reflection EELS measures the loss spectrum of a surface at primary energies of 10--100~eV as a function of a momentum transfer parallel to the surface, which is scanned across several Brillouin zones. The spectrum is then interpreted, through the direct Coulomb coupling, as the density response $\chi(\vq,\omega)$ of the material \cite{vig2017}. At a primary energy of 50~eV, the beam momentum is $k=3.6$~\AA$^{-1}$. In this case, Eq.~\eqref{eq:IXratio} in its non-relativistic form gives an exchange correction of $-q^2/k^2=-7.6\%$ at $q=1$~\AA$^{-1}$ and $-31\%$ at $q=2$~\AA$^{-1}$. At 100~eV, the corresponding numbers are $-3.8\%$ and $-15\%$. At these kinematics, exchange must be included as part of the cross section instead of being estimated as a correction.

What does our framework tell us about the exchange contribution in this regime? There are three important points.

First, consider its form. To leading order, the density-channel cross section is the direct-channel result multiplied by $1-q^2/k^2+O(q^4/k^4)$. This factor depends on $\vq$ and on the beam energy but not on $\omega$, because the exchange operator is the same density $\rho$ that the direct channel couples to, evaluated at a point. As a result, the extracted $\chi''(\vq,\omega)$ is correct in its energy dependence at each $\vq$, but its momentum dependence is wrong by a smooth, calculable factor that suppresses large transfers, which is precisely the direction in which the spectrum is compared across zones.

Second, at order $q^4/k^4$ the spin channel enters. In particular, the square of the axial term of Eq.~\eqref{eq:MXlead} adds to the density spectrum a contribution proportional to the spin-fluctuation spectrum of the sample. For a paramagnet with uncorrelated spins, this contribution is $\frac34(q^2/k^2)^2$ times the density spectrum. At $q=2$~\AA$^{-1}$ and 50~eV, this amounts to a seven-percent admixture, so a paramagnon or a spin-fluctuation continuum can appear in a nominal charge spectrum at large transfer. A spin-polarized source removes this ambiguity, as discussed in Sec.~\ref{sec:spinX}.

Third, consider the limits of our approach. At 50~eV, the Born parameter $\alpha/v\simeq0.5$ and the ratio $\lambda_p\simeq0.3$ are not small. As a result, the one-photon amplitudes of this paper are only the leading terms of series whose next terms are tens of percent, and the standard treatment of impact scattering at these energies is the multiple-scattering theory of low-energy electron diffraction extended to inelastic events \cite{ibach1982}. It should be noted, however, that the exchange-to-direct \emph{ratio} is more robust than either amplitude, because both share the same sample operator and the same distortion of the beam wave. The numbers above should be read as the size of the effect, to be built into a distorted-wave analysis rather than applied as a multiplicative correction. Although the relativistic content of our framework is irrelevant at these energies, the framework still supplies the operator content of exchange, namely density and spin with the coefficients of Eq.~\eqref{eq:fierz}, together with a single formula that connects the surface and microscope regimes.

\section{Pair-excitation channel}
\label{sec:pair}

\subsection{Leading order selection rule}

Recall that the pair term $\SP$ converts two sample electrons into two beam electrons or the reverse, thereby changing $(N_{\rm b},N_{\rm s})$ by $(\pm2,\mp2)$. As a result, any process with the same number of beam electrons before and after, whether single-electron energy loss, two-electron coincidence, or any other, has zero matrix element of $\SP$ at first order. The processes that $\SP$ does connect, namely the capture of two beam electrons into the sample or the emission of two from it, would require the sample to absorb or supply an energy $2T$, which is not available in the non-relativistic sector. Thus at order $e^2$ the pair channel vanishes identically for every on-shell process of the theory. We emphasize that this vanishing is exact, as it follows from a selection rule imposed by the mode split rather than from a small suppression factor.

\subsection{One-loop pair excitation}

At order $e^4$, however, the combination $\SP^-\SP^+$ does contribute to single-electron energy loss. In this process, the sample emits a beam pair $(\vk',\vk_f)$ from two of its electrons $(\vp_1,\vp_2)$, and then the incoming beam electron $\vk_i$ together with the virtual $\vk'$ are absorbed as sample electrons $(\vp_1',\vp_2)$. Momentum conservation at the emission vertex fixes $\vk'=\vp_1+\vp_2-\vk_f\simeq-\vk_f$, meaning that the virtual beam electron is fast and travels backward. The intermediate state contains three beam electrons, and its energy exceeds the initial energy by
\begin{equation}
    \fE_I-\fE_i=T'+T_f-\epsilon_1-\epsilon_2\simeq2T ,
\end{equation}
where $\epsilon_{1,2}$ are the sample electrons' non-relativistic energies. Notice that this denominator is set by the beam energy rather than by any sample scale; it is this feature that makes the channel negligible. With hard kernels at both vertices, second-order perturbation theory gives, for one partner electron $\vp_2$ in a volume $\mathcal V$ and non-relativistic Coulomb vertices,
\begin{equation}
    W_{PP}^{\rm path}=-\frac{e^4}{\mathcal V^2}\;\frac{1}{|\vk_f-\vp_1|^2\,|\vk_i-\vp_1'|^2\,(T'+T_f)} ,
    \label{eq:MPP}
\end{equation}
which should be compared with the direct matrix element $W_D^{\rm model}=e^2/(q^2\mathcal V)$ in the same normalization.

We verified Eq.~\eqref{eq:MPP} against an exact second-order calculation in a 220-state three-electron Fock space (see Appendix~\ref{app:numerics}). The closed form is reproduced to seven figures, including its sign, and every intermediate state is off shell by $2T$ to within $0.4\%$. The same structure, namely a virtual conversion between the beam and sample sectors with a denominator set by the beam energy, also governs the second-order contributions of the $(\pm1,\mp1)$ terms of Sec.~\ref{sec:framework}. These contributions, in which a single beam electron is virtually captured and a sample electron is re-emitted, are bounded in the same way.

Summing coherently over the $n\mathcal V$ possible partners (the sign is the same for each) and restoring the relativistic vertex factors, we find, for the resulting matrix element $W_{PP}^{\rm model}$,
\begin{equation}
    R_{PP}^{\rm model}\equiv\frac{|W_{PP}^{\rm model}|}{|W_D^{\rm model}|}
    \leq\frac{n\,e^2q^2}{2k^4T}\left[\frac{(E+m)^2}{4mE}\right]^2c_s^2
    \label{eq:RPP}
\end{equation}
where $n$ is the sample's electron density, $c_s\leq1+3(E-m)/(E+m)$ is a spin factor of order one, and the squared bracket is $1.016$ at 100~keV. In the power counting of Sec.~\ref{sec:power}, this ratio is $(\alpha/v)\lambda_p^3\lambda_q^2$, which is four powers of the small parameters below the leading two-photon correction. For a dense metal with $n=0.1$~\AA$^{-3}$ and for $q=1$~\AA$^{-1}$, $R_{PP}^{\rm model}$ is $10^{-7}$ at 1~keV, $10^{-10}$ at 10~keV, and $10^{-13}$ at 100~keV (see Table~\ref{tab:magnitudes}). Relative to the Born exchange amplitude, it is $3\times10^{-5}$, $3\times10^{-7}$, and $3\times10^{-9}$, respectively.

\begin{table*}[!htb]
\caption{Selected dimensionless diagnostics at beam kinetic energy $T$ and spatial transfer $q$. The signed density-interference fractions use Eqs.~\eqref{eq:IXratio} and \eqref{eq:IAratio} with $\kappa=q^2$ and $\Lambda=2m/E$, omitting loss and response corrections. The circular-magnon ratio $\epsilonS=|b_0|/|a_0|$ uses Eq.~\eqref{eq:2x2} with elastic beam coefficients and $q_\perp=q$ in the leading small-transfer geometry. Its current denominator includes direct and exchange contributions. The last column uses the independent-partner Coulomb estimate in Eq.~\eqref{eq:RPP} with $n=0.1$~\AA$^{-3}$ and $k_i^2=2mT$. It assumes partner factorization and coherent path addition and is not a general pair-channel bound. Entries with $T$ comparable with $m$ are algebraic extrapolations of this non-relativistic pair model. The columns compare selected interference fractions and model amplitude ratios, rather than parts of a common cross section. Each retains the sample and local-kernel conditions of its source calculation.}
\label{tab:magnitudes}
\begin{ruledtabular}
\begin{tabular}{rrrrrr}
$T$ & $q$ & $\mathcal I_X/\mathcal D$ & $\mathcal I_A/\mathcal D$ & $\epsilonS$ & $R_{PP}^{\rm model}$\\
{[keV]} & {[\AA$^{-1}$]} & & & & \\
\hline
1 & 1 & $-3.8\times10^{-3}$ & $-1.1\times10^{-5}$ & 16 & $10^{-7}$\\
3 & 1 & $-1.3\times10^{-3}$ & $-1.1\times10^{-5}$ & 3.0 & $5\times10^{-9}$\\
10 & 1 & $-3.7\times10^{-4}$ & $-1.1\times10^{-5}$ & 0.50 & $10^{-10}$\\
30 & 1 & $-1.1\times10^{-4}$ & $-1.0\times10^{-5}$ & 0.095 & $5\times10^{-12}$\\
100 & 1 & $-2.6\times10^{-5}$ & $-9.1\times10^{-6}$ & 0.015 & $10^{-13}$\\
300 & 1 & $-3.3\times10^{-6}$ & $-6.5\times10^{-6}$ & 0.0027 & $3\times10^{-15}$\\
100 & 0.3 & $-2.3\times10^{-6}$ & $-8.2\times10^{-7}$ & 0.0045 & $10^{-14}$\\
100 & 3 & $-2.3\times10^{-4}$ & $-8.2\times10^{-5}$ & 0.045 & $10^{-12}$\\
\end{tabular}
\end{ruledtabular}
\end{table*}

\subsection{Superconductors and coincidence}

One might wonder whether a pairing instability in the sample could enhance the pair channel. This is not the case, because the intermediate state is off shell by $2T\sim10^5$~eV regardless of the sample's spectrum, so no meV structure can produce a resonance. Moreover, BCS coherence factors $u_pv_p\leq\dfrac{1}{2}$ only reduce the pair-removal matrix elements. Detecting two beam electrons in coincidence does not reveal the pair channel either. The Cooper box has the same structure in a coincidence measurement, while the leading two-electron amplitude carries two potential-region kernels and a sample-scale denominator. As a result, the ratio falls to $\sim\lambda_q^4\,\Delta E_{\rm sample}/2T\sim10^{-16}$. We therefore conclude that the pair channel is a purely virtual channel of the beam--sample coupling, which our framework accounts for but which has no measurable effect.

\section{Two-photon corrections}
\label{sec:beyond}

\subsection{Structure of the two-photon terms}

At order $e^4$, the amplitude contains products of pairs of channels, which can be ordered by their energy denominators. Consider first $DD$, in which both photons are in the potential region. In this case, the intermediate beam electron is nearly on shell and the denominator is a sample excitation energy. This is ordinary second-Born scattering, that is, the distortion of the beam wave by the sample's potential. In $DX$ and $XX$, one or two of the photons are hard. The intermediate beam electron, however, still has $T'\simeq T$, because the momentum it exchanged was a sample momentum. The denominator is therefore again a sample scale, and these terms are suppressed relative to $DD$ by one or two powers of $\PiX/\widetilde{\Pi}\sim\lambda_q^2$. Finally, in $PP$ the denominator is $2T$, which produces an additional suppression by $E_{\rm sample}/2T$. Thus we have the hierarchy $DD\gg DX\gg XX\gg PP$.

\subsection{Screened-potential second Born amplitude}
\label{subsec:second-born}

How large is $DD$? This is the standard question of the Born series, and the standard answer \cite{byron1977} is that the second Born amplitude has an imaginary part of relative order $\alpha/v$ (the eikonal phase, which does not change the cross section) and a real part smaller by a further factor of the sample's momentum scale over the beam momentum. This real part is the only two-photon term that changes cross sections at leading order, and it is odd in the beam's charge. Its size therefore determines both the validity of the Born treatment and the viability of the electron--positron test. We established this scaling directly by computing exact scattering off a screened Coulomb potential $V=g\,e^{-\mu r}/r$ with the variable-phase method and isolating the second-Born term as the part of the amplitude even in $g$. We then checked its imaginary part against the generalized unitarity relation $\Im f_2(\vk_f,\vk_i)=\frac{k}{4\pi}\int \text{d}\Omega'f_1(\vk_f,\vk')f_1(\vk',\vk_i)$ to five significant figures (see Appendix~\ref{app:numerics}). We find that, at fixed $q$ and $k=5$--$40\mu$,
\begin{equation}
    \begin{aligned}
    \frac{\Im f_2}{f_1}&=-c_I\,\frac{mg}{k},\\
    \frac{\Re f_2}{f_1}&=-c_R\,\frac{mg}{k}\cdot\frac{\mu}{k},
    \label{eq:born2scaling}
    \end{aligned}
\end{equation}
where $c_I$ and $c_R$ are constant to four digits over this range. These are, however, the coefficients for elastic scattering off the screened potential. The distortion of an inelastic amplitude has the same denominator structure, and we will suppose that its coefficients are of the same order. This assumption is all that the bounds below require. Notice that the real part carries the factor $\mu/k$, which is just $\lambda_p$ with $\mu$ the sample's inverse screening length. The coefficients depend on $q/\mu$; for $q/\mu=0.5,\,1,\,2,\,4,\,8$, we find $c_R=0.29,\,0.40,\,0.62,\,0.85,\,0.95$ and $c_I=0.60,\,0.86,\,1.56,\,2.74,\,4.13$. Thus $c_R$ saturates near unity at large $q/\mu$, while $c_I$ grows logarithmically as the screened remnant of the Coulomb phase.

\subsection{Effect on the electron--positron ratio}

Now suppose that the potential reverses with the beam's charge. Then $|f_1\pm f_2|^2$ gives a charge-odd asymmetry of magnitude $|\Delta_{\rm dist}|=4\,|\Re f_2/f_1|=4c_R\,Z\alpha m\mu/k^2$, where $Z\alpha$ is the strength of the sample's mean potential. In the electron--positron contrast $\Delta_{e^\pm}$ this asymmetry is positive, since the attractive potential focuses the electron beam onto the sample, whereas the first-Born contribution $\Delta_{e^\pm}^{\rm Born}$ in Eq.~\eqref{eq:Rratio} is negative in the non-relativistic limit, $\Delta_{e^\pm}^{\rm Born}\simeq-q^2/k^2$. Comparing their magnitudes, we find
\begin{equation}
    \dfrac{|\Delta_{e^\pm}^{\rm Born}|}{|\Delta_{\rm dist}|}\simeq\dfrac{q^2a_0}{4c_R\,Z\,\mu}\,,
    \label{eq:XvsDist}
\end{equation}
which, for a Thomas--Fermi inverse screening length $\mu\simeq Z^{1/3}/a_0$, is $0.23$ for hydrogen and $0.02$ for carbon at $q=1$~\AA$^{-1}$, and $1.1$ and $0.15$ at $3$~\AA$^{-1}$. Distortion thus dominates the density-channel asymmetry for every solid. As a result, the electron--positron test of exchange must be made in the spin channel, where distortion, being spin-blind, cancels.

\section{Magnon creation with direct and exchange channels}
\label{sec:magnon}

The magnon provides the natural test of our framework, because it is the one excitation for which two channels contribute at comparable size in an accessible regime. The direct channel couples to a magnon only through the beam's current and the sample's magnetization current, which is small. The exchange channel couples to the magnon through the spin density, which is not small. In this section, we derive the cross section with both channels present, and we then investigate the spin structure of the amplitude that underlies the companion work \cite{companion}.

\subsection{Spin-density matrix element of a magnon}
\label{subsec:magnon-matrix}

Suppose that the sample is a ferromagnet with spin $S$ on each of $N_{\rm m}$ sites $\vR_j$, with $n_{\rm m}=N_{\rm m}/V$ sites per unit volume and all spins ordered along a unit vector $\Mh$. We define $\Mh$ as the direction of the ordered electron \emph{spins}. Because the electron's moment is antiparallel to its spin, the magnetization points along $-\Mh$; this distinction affects only the sign of a polarization predicted below \footnote{All signs in this section and in the companion work refer to $\Mh$ as the spin direction. For a sample magnetized along the beam by a lens field, the ordered spins point against the beam.}. The electron spin density of the magnetic electrons is then
\begin{equation}
    \vs(\vx)=2\sum_j\vS_j\,w(\vx-\vR_j),\qquad\int \text{d}^3\vx\,w(\vx)=1,
\end{equation}
where $w$ is the density of the magnetic orbital, and its Fourier component is $\widetilde{\vs}(\vq)=2F(\vq)\sum_j\vS_je^{i\vq\cdot\vR_j}$ with $F(\vq)=\int \text{d}^3\vx\,e^{i\vq\cdot\vx}w(\vx)$ the magnetic form factor. For spin-only moments, the magnetic electrons carry no orbital current, and a one-magnon state has no density matrix element. As a result, the sample enters both channels through $\vs$ alone. In particular, it enters the direct channel through the magnetization current $\vJ_\text{ss}=\vnabla\times\vs/2m$ and the exchange channel through the axial term of Eq.~\eqref{eq:MXlead}.

In the Holstein--Primakoff description \cite{holstein1940}, to leading order in $1/S$, the operator $S_j^-=\sqrt{2S}\,a_j^\dagger$ creates a spin deviation at site $j$, and a magnon of wavevector $\vq$ is the state $\ket{\vq}=N_{\rm m}^{-1/2}\sum_je^{i\vq\cdot\vR_j}a_j^\dagger\ket{0}$ with energy $\omega_{\rm m}(\vq)$. We then have $\bra{\vq}S_j^-\ket{0}=\sqrt{2S/N_{\rm m}}\,e^{-i\vq\cdot\vR_j}$, and with $S^\pm=S^x\pm iS^y$ we find
\begin{equation}
    \begin{aligned}
    \bra{\vq}\widetilde{\vs}(\vq)\ket{0}&=\sqrt{2SN_{\rm m}}\,F(\vq)\,\vepsilon,\\
    \vepsilon&\equiv\hat{\ve}_1+i\hat{\ve}_2,\\
    \hat{\ve}_1\times\hat{\ve}_2&=\Mh,
    \end{aligned}
    \label{eq:magnonME}
\end{equation}
where $(\hat{\ve}_1,\hat{\ve}_2,\Mh)$ is a right-handed triad. The complex vector $\vepsilon$ is the circular polarization of the precessing spins. The spectral tensor of the spin density, Eq.~\eqref{eq:Sab} with $\fO_a=s^k$, is then
\begin{equation}
    \begin{aligned}
    S_{ss}^{kl}(\vq,\omega)&=2Sn_{\rm m}\,|F(\vq)|^2\,\big(\delta^\perp_{kl}-i\epsilon_{klm}\Mh_m\big)\\
    &\qquad\quad\delta(\omega-\omega_{\rm m}(\vq)),\\
    \delta^\perp_{kl}&\equiv\delta_{kl}-\Mh_k\Mh_l ,
    \end{aligned}
    \label{eq:Skl}
\end{equation}

whose antisymmetric imaginary part encodes the sense of precession.

\subsection{Direct and exchange amplitudes}
\label{subsec:magnon-amplitudes}

Consider first the direct channel. In Eq.~\eqref{eq:MD}, the time component of the current vanishes for a magnon, and $\fJ_{\text{ss},\mu}\gamma^\mu=\fJ^0_\text{ss}\gamma^0-\vJ_\text{ss}\cdot\bm\gamma$ leaves $\fM_D=e^2\widetilde{\Pi}(\fQ)\,\bm\Gamma_V\!\cdot\!\widetilde{\vJ}_{\text{ss},n0}$ with $\bm\Gamma_V\equiv\bar{u}_f\bm\gamma u_i$. With the injecting-Fourier convention of Eq.~\eqref{eq:Sab}, the curl becomes $-i\vq\times$, so $\widetilde{\vJ}_\text{ss}(\vq)=-\frac{i}{2m}\vq\times\widetilde{\vs}(\vq)$, and
\begin{equation}
    \begin{aligned}
    \fM_D&=-\frac{ie^2\widetilde{\Pi}}{2m}\,(\bm\Gamma_V\times\vq)\cdot\widetilde{\vs}_{n0}\equiv\vc^{\rm D}\!\cdot\widetilde{\vs}_{n0},\\
    c^{\rm D}_k&=-\frac{ie^2\widetilde{\Pi}}{2m}\,\epsilon_{ijk}\,\Gamma_V^i\,q_j .
    \end{aligned}
    \label{eq:Kmag}
\end{equation}
This amplitude has the structure of the beam's vector vertex crossed into $\vq$. Physically, up to a phase, $(\bm\Gamma_V\times\vq)/2mq^2$ is the Fourier component at $\vq$ of the magnetic field of the moving beam charge (i.e.\ the Biot--Savart field), projected on the precessing spin. Notice the factor $1/2m$ that comes from the magnetization current. It is this factor, together with the $1/q^2$ of the kernel, that makes the direct magnon amplitude small.

Next consider the exchange channel. From the axial term of Eq.~\eqref{eq:MXlead}, we have
\begin{equation}
    \begin{aligned}
    \fM_X^{\rm ax}&=-\frac{e^2\PiX}{2}\,\bm\Gamma_A\!\cdot\!\widetilde{\vs}_{n0}\equiv\vc^{\rm X,ax}\!\cdot\widetilde{\vs}_{n0},\\
    c^{\rm X,ax}_k&=-\frac{e^2\PiX}{2}\,\Gamma_A^k,\qquad\bm\Gamma_A\equiv\bar{u}_f\gamma^5\bm\gamma u_i .
    \end{aligned}
    \label{eq:KX}
\end{equation}
The current term of Eq.~\eqref{eq:MXlead}, $\dfrac{1}{2}(\bar{u}_f\gamma^iu_i)\widetilde{\fJ}_{\text{ss},n0}^{\,i}$, also has a magnon matrix element through the magnetization current. It has the same structure as the direct amplitude and simply replaces $\widetilde{\Pi}$ in Eq.~\eqref{eq:Kmag} by
\begin{equation}
    \Pi_{\rm eff}\equiv\widetilde{\Pi}+\dfrac{1}{2}\PiX ,
    \label{eq:Pieff}
\end{equation}
a correction of relative size $q^2/4mT$, which we keep because it enters the interference below at the same order as the spin terms. (The pseudo-scalar vertex, which couples to $\vnabla\cdot\vs/2m$, contributes at relative order $q^2/m^2$ and is retained in Appendix~\ref{app:magnon-derivation}.) All three amplitudes are linear in the same sample operator. As a result, all carry the same form factor $F(\vq)$, and their ratios are properties of the beam alone. Writing $\vc=\vc^{\rm D}+\vc^{\rm X,ax}$ with $\widetilde{\Pi}\to\Pi_{\rm eff}$ in $\vc^{\rm D}$, we find that the cross section is given by Eq.~\eqref{eq:master0} with $S^{ab}\to S_{ss}^{kl}$ of Eq.~\eqref{eq:Skl}.

\subsection{Magnon cross section}
\label{subsec:magnon-cross-section}

To evaluate this cross section, we need three spin sums (see Appendix~\ref{app:spinsums}). These are the spatial part of the beam tensor $\fL^{ij}$ of Eq.~\eqref{eq:lepton}; the axial tensor
\begin{equation}
    \begin{aligned}
    A^{ij}&\equiv\dfrac{1}{2}\sum_{s_i,s_f}\Gamma_A^i\Gamma_A^{j*}\\
    &=2\big[\fK_i^i\fK_f^j+\fK_i^j\fK_f^i+\delta^{ij}(\eta_{\mu\nu}\fK_i^\mu\fK_f^\nu+m^2)\big];
    \end{aligned}
\end{equation}
and the mixed tensor
\begin{equation}
    \begin{aligned}
    C^{ij}&\equiv\dfrac{1}{2}\sum_{s_i,s_f}\Gamma_V^i\Gamma_A^{j*}=2i\,\epsilon_{ijl}\,w_l,\\
    \vw&\equiv E_f\vk_i-E_i\vk_f ,
    \end{aligned}
\end{equation}
which is purely imaginary and antisymmetric, as the product of a vector and an axial vertex must be. Its mass terms vanish because a trace with one $\gamma^5$ requires four other $\gamma$'s. We now contract $\overline{\sum_{s_i,s_f}}c_kc_l^*$ with Eq.~\eqref{eq:Skl}. The direct and exchange terms are symmetric in $(k,l)$ and see only $\delta^\perp_{kl}$. The interference tensor $\frac{ie^4\widetilde{\Pi}\PiX}{4m}\epsilon_{ijk}q_jC^{il}=-\frac{e^4\widetilde{\Pi}\PiX}{2m}(q_lw_k-\delta_{kl}\,\vq\!\cdot\!\vw)$, however, is real, so its contraction with the imaginary antisymmetric part of $S_{ss}^{kl}$ drops out of the cross section. The result is
\begin{equation}
    \begin{aligned}
    \frac1V\frac{\text{d}^2\sigma}{\text{d}\Omega\,\text{d}\omega}&=\alpha^2\frac{k_f}{k_i}\;2Sn_{\rm m}|F(\vq)|^2\,\delta(\omega-\omega_{\rm m}(\vq))\\
    &\quad\times\underbrace{\left[\frac{\Pi_{\rm eff}^2}{4m^2}\,\mathcal G_{\rm V}+\frac{\Pi_{\rm eff}\PiX}{m}\,\mathcal G_{\rm VA}+\frac{\PiX^2}{4}\,\mathcal A\right]}_{\mathcal B_{\rm m}},
    \end{aligned}
    \label{eq:magnonxsec}
\end{equation}
with
\begin{align}
    \mathcal G_{\rm V}&=q^2\delta_{ij}\fL^{ij}-q_i\fL^{ij}q_j-(\Mh\times\vq)_i\fL^{ij}(\Mh\times\vq)_j,\label{eq:calD}\\
    \mathcal G_{\rm VA}&=\vq\!\cdot\!\vw+(\vq\!\cdot\!\Mh)(\vw\!\cdot\!\Mh),\label{eq:calI}\\
    \mathcal A&=\Tr A-\Mh\!\cdot\!A\!\cdot\!\Mh .\label{eq:calA}
\end{align}
Here $\mathcal G_{\rm V}$ projects the beam's vector vertex onto the directions transverse to $\vq$ (the curl) and to $\Mh\times\vq$, while $\mathcal A$ projects the axial vertex onto the plane transverse to $\Mh$. We checked Eq.~\eqref{eq:magnonxsec} against direct summation over explicit Dirac spinors for non-relativistic and relativistic kinematics and several orientations of $\Mh$, and found exact agreement (see Appendix~\ref{app:numerics}). The interference is constructive for quasi-elastic kinematics.

Consider first the geometry in which the ordered spins lie along the beam, $\Mh\parallel\vk_i$, and suppose that the scattering angle is small. Then $\mathcal G_{\rm V}\simeq4k^2q^2$, $\mathcal G_{\rm VA}\simeq Eq^2$, and $\mathcal A\simeq8m^2$, so with $\widetilde{\Pi}=1/q^2$ and $\PiX=1/2mT$ the bracket becomes
\begin{equation}
    \begin{aligned}
    \Big[\cdots\Big]&\simeq\frac{k^2}{m^2q^2}\Big(1+\frac{q^2}{2mT}\Big)+\frac{E}{2m^2T}+\frac{1}{2T^2},\\
    \frac{\sigma_{\rm int}}{\sigma_D}&=\frac{q^2E}{2Tk^2}+\frac{q^2}{2mT},\\
    \frac{\sigma_X}{\sigma_D}&=\frac{q^2m^2}{2T^2k^2},
    \end{aligned}
    \label{eq:magnonratios}
\end{equation}
where $\sigma_{\rm int}$ collects the terms linear in $\PiX$, namely the interference of the direct amplitude with the spin term of the exchange channel and with its current term. The three terms have different $q$ dependence. The dipole term falls as $1/q^2$ because it is a Coulomb-type field, the exchange term is flat because it is a contact interaction, and the interference lies in between. At $q=1$~\AA$^{-1}$, the pure exchange term is $4.5\times10^{-4}$ of the dipole term at 100~keV and $2\%$ at 30~keV, and to leading order it is equal to the dipole term at
\begin{equation}
    T_0=\Big(\frac{q^2m}{4}\Big)^{1/3}\approx7.9\ \text{keV}\quad(q=1\ \text{\AA}^{-1}),
\end{equation}
which scales as $q^{2/3}$. The interference term is smaller than the pure exchange term by $ET/m^2+Tk^2/m^3$, which is $0.32$ at 100~keV. Thus, below $T_0$, exchange dominates magnon creation, whereas at microscope energies it is a correction below a part in a thousand. Existing treatments of magnon excitation in the microscope contain the direct channel, namely the beam's charge and current and, in some cases, its spin-magnetic coupling, but they do not contain exchange \cite{lyon2021,mendis2021,mendis2022,castellanos2023,nascimento2024,nascimento2025,castellanos2025}. These treatments are therefore complete to this accuracy for unpolarized beams and spin-blind detection. Moreover, our dipole term $\mathcal G_{\rm V}\simeq4k^2q^2\cos^2\phi$ reproduces the quadratic growth with beam momentum and the angular dependence found for the charge coupling in Ref.~\cite{nascimento2025}. The amplification of the transverse magnetic coupling over the Coulomb coupling discussed in Ref.~\cite{iatrakis2026b} is, in our language, the statement that the direct spin-flip channel is carried by the beam current rather than by the beam charge.

Next suppose that the spins are in-plane, $\Mh\perp\vk_i$, with $\vq$ at azimuth $\phi$ from $\Mh$. Then the same reductions give $\mathcal G_{\rm V}\simeq4k^2q^2\cos^2\phi+q^4$, $\mathcal G_{\rm VA}\simeq Eq^2(1+\cos^2\phi)$, and $\mathcal A\simeq8m^2+4k^2$. The factor $\cos^2\phi$ is the dipole selection rule; that is, the beam current couples to the magnetization current only if $\vk\times\vq$ has a component transverse to $\Mh$. As a result, at $\phi=90^\circ$ the dipole channel is reduced to the tiny $q^4$ term, while the exchange channel is not. In this corner of the geometry, exchange dominates by $(2m^2+k^2)/T^2\approx60$ at 100~keV, and the outgoing electrons are polarized to $\sim0.95$ along $\Mh$ (see Appendix~\ref{app:numerics}). The rate, however, is only $\sim5\times10^{-4}$ of that at the allowed azimuth. This geometry provides a cleaner but much slower signature of exchange, and it requires a spin analyzer rather than a polarized source.

\subsection{Spin structure and polarization}
\label{sec:magnonspin}
\label{subsec:magnon-spin}

Why is the interference so much smaller than the amplitude ratio suggests? Since $\sigma_X/\sigma_D=4.5\times10^{-4}$ at 100~keV, the amplitude ratio is $\epsilonS\simeq1.5\times10^{-2}$, and one would naively expect interference of order $2\epsilonS=3\times10^{-2}$, which is about two hundred times larger than what Eq.~\eqref{eq:magnonratios} gives. The answer lies in the spin structure of the two vertices, which take exact and simple forms in the basis of canonical spin. For elastic kinematics, we have
\begin{equation}
    \bm\Gamma_V=\xi_f^\dagger\big[\vK+i\,\vq\times\vsigma\big]\xi_i,\qquad\vK\equiv\vk_i+\vk_f,
    \label{eq:GVexact}
\end{equation}
exactly, while $\bm\Gamma_A$ is given by Eq.~\eqref{eq:axialvertex}. With $\Mh$ along the beam, the magnon polarization $\vepsilon$ is transverse, and the leading amplitude is a $2\times2$ matrix in the beam spin,
\begin{equation}
    \begin{aligned}
    \mathsf T^{(0)}&=a_0\,\mathbb{1}_2+\vb_0\cdot\vsigma,\\
    a_0&=-\frac{ie^2\Pi_{\rm eff}}{2m}\,(\vK\times\vq)\cdot\vepsilon,\\
    \vb_0&=b_0\,\vepsilon,\qquad b_0=e^2\PiX\,m,
    \end{aligned}
    \label{eq:2x2}
\end{equation}
where $\Pi_{\rm eff}\simeq1/q^2$, and where we have dropped the common factor $\sqrt{2SN_{\rm m}}F(\vq)$ and the small spin-dependent part of Eq.~\eqref{eq:GVexact}. Notice that the dipole amplitude is proportional to the unit matrix. Physically, this is because the Biot--Savart field couples to the beam's orbital motion and leaves its spin alone. The exchange amplitude, by contrast, is proportional to $\vsigma\cdot\vepsilon$, which is twice the raising operator about $\Mh$, because the electron that leaves is a sample electron of opposite spin. For an unpolarized beam with the final spin summed, we find $\dfrac{1}{2}\Tr[\mathsf T^{(0)}\mathsf T^{(0)\dagger}]=|a_0|^2+2|b_0|^2$, and the cross term vanishes because $\Tr[\vsigma]=0$. In other words, the two channels leave the beam electron in orthogonal spin states, and orthogonal final states do not interfere. The residual interference in Eq.~\eqref{eq:magnonratios} therefore comes from two small terms: the $i\vq\times\vsigma$ term of Eq.~\eqref{eq:GVexact}, which describes the beam's own spin-magnetization current and is smaller than $\vK$ by $q/2k$, and the exchange current of Eq.~\eqref{eq:Pieff}, which is smaller than the direct current by $q^2/4mT$. With $\epsilonS\equiv|b_0|/|a_0|\simeq qm/(2Tk)$, both are of order $\epsilonS\,q/k$, which is the origin of $\sigma_{\rm int}/\sigma_D\sim\epsilonS\,q/k$.

This suppression, however, is merely an artifact of the unpolarized observable. Indeed, the beam electron's spin has recorded which channel acted, and since the spin is a two-state system, this record can be read in any basis. To see this, consider the polarization of the scattered electrons. Averaging over the incident spin, we find
\begin{equation}
    \vP_f=\frac{\Tr[\vsigma\,\mathsf T^{(0)}\mathsf T^{(0)\dagger}]}{\Tr[\mathsf T^{(0)}\mathsf T^{(0)\dagger}]}=\frac{2\,\Re(a_0\,\vb_0^*)+2|b_0|^2\Mh}{|a_0|^2+2|b_0|^2},
    \label{eq:Pf}
\end{equation}
whose transverse component is first order in $\epsilonS$. Writing $\vb_0=b_0\,\vepsilon$ and evaluating with $\vK\times\vq$ transverse, we obtain
\begin{equation}
    \vP_{f,\perp}=\frac{2\zeta_M\epsilonS}{1+2\epsilonS^2}\,\hat{\vq}_\perp,
    \label{eq:Pperp}
\end{equation}
where $\zeta_M=\Mh\cdot\hat{\vk}_i$. The radial projection $P_{f,\perp}=\hat{\vq}_\perp\cdot\vP_f$ reverses with $\Mh$. For $q=1$~\AA$^{-1}$, this polarization is $3\%$ at 100~keV and $19\%$ at 30~keV, and its magnitude peaks at $1/\sqrt2$ for $\epsilonS=1/\sqrt2$ near $T_0$. Moreover, the incident analyzing vector $\vA_{\rm in}$ for a polarized beam has the same leading transverse part.
\begin{figure}[!htb]
\includegraphics[width=\columnwidth]{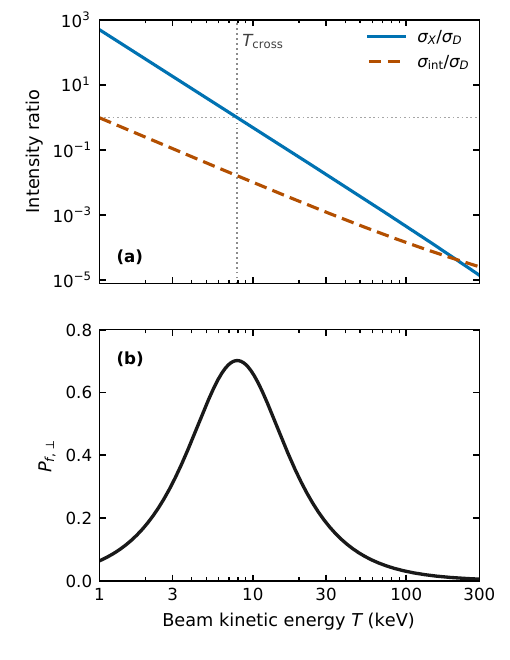}
\caption{Retained local-exchange model for a circular magnon with $\Mh\parallel\vk_i$ and $q=1$~\AA$^{-1}$, using elastic beam coefficients. (a) Full exchange square $\sigma_X/\sigma_D$ and direct--exchange interference $\sigma_{\rm int}/\sigma_D$ from Eq.~\eqref{eq:magnonratios-app}. The horizontal guide marks unity and the vertical guide marks $T_{\rm cross}=7.90$~keV, where $\sigma_X=\sigma_D$. (b) Radial outgoing polarization $P_{f,\perp}=\hat{\vq}_\perp\cdot\vP_f$ from Eqs.~\eqref{eq:Pf-app} and \eqref{eq:2x2-app}. Both panels include all retained current, axial and pseudo-scalar terms. The sample and kernel corrections discussed in Sec.~\ref{subsec:magnon-corrections} are omitted.}
\label{fig3}
\end{figure}

More generally, for any spin excitation with matrix element $\bm{\fE}=\bra{n}\widetilde{\vs}(\vq)\ket{0}$, the amplitudes are $a\propto(\vK\times\vq)\cdot\bm{\fE}$ and $\vb\propto\bm{\fE}$. Since $\vK\times\vq$ is real, we then have
\begin{equation}
    \begin{aligned}
    2\,\Re(a\,\vb^*)&=\frac{e^4\PiX}{m\,q^2}\,\mathsf G_0\big[(\vK\times\vq)\times\vL_\fE\big],\\
    \vL_\fE&\equiv-\frac i2\,\bm{\fE}^*\!\times\bm{\fE},\\
    \mathsf G_0&={\rm diag}(m,m,E),
    \end{aligned}
    \label{eq:law}
\end{equation}
where $\mathsf G_0$ carries the vertex factors of Eq.~\eqref{eq:axialvertex}. The first-order signal is proportional to the spin angular momentum $\vL_\fE$ carried by the mode. Consequently, it vanishes for linearly polarized modes and for degenerate antiferromagnetic pairs, and its component along any beam polarization axis is selected by geometry. Symmetry-breaking antiferromagnets are particularly interesting in this context, given that substantial spin polarization can emerge even in materials with nearly vanishing net magnetization~\cite{chou2024spin}. These are the results used in the work. We verified Eqs.~\eqref{eq:Pf}--\eqref{eq:law} against exact Dirac-spinor sums and against exact M{\o}ller scattering off a free electron with a spin flip, which contains neither the Fierz identity nor the non-relativistic reduction, and found agreement to $4\times10^{-5}$ at 100~keV (see Appendix~\ref{app:numerics}).

\subsection{Count rate and measurement time}
\label{subsec:magnon-count}

We now estimate how long such a measurement would take. The probability that a beam electron creates a magnon into a window $\Delta^2q_\perp$ of transverse momentum around $\vq$, in a film of thickness $t$, follows from Eq.~\eqref{eq:magnonxsec} by integrating the delta function over $\omega$ and converting solid angle to transverse momentum, $\text{d}\Omega=\text{d}^2\vq_\perp/k^2$. For the dipole term in the along-beam geometry, we find
\begin{equation}
    \begin{aligned}
    p&=t\int_{\Delta\Omega}\text{d}\Omega\int \text{d}\omega\,\frac1V\frac{\text{d}^2\sigma}{\text{d}\Omega\,\text{d}\omega}\\
    &=2Sn_{\rm m}\,|F(\vq)|^2\,r_e^2\,t\,\frac{\Delta^2q_\perp}{q^2},\qquad r_e=\frac{\alpha}{m},
    \end{aligned}
    \label{eq:countrate}
\end{equation}
where the $k^2$ of $\mathcal G_{\rm V}$ has cancelled against the $1/k^2$ of the solid angle, meaning that the magnon probability per electron per momentum window is independent of beam energy and is set by the classical electron radius. Consider yttrium iron garnet, which has a cubic cell with $a=12.376$~\AA\ containing eight formula units and a net spin of $5/2$ per formula unit \cite{cherepanov1993,princep2017}. Taking $|F(\vq)|^2\approx0.85$, we find from Eq.~\eqref{eq:countrate} that $p=5.7\times10^{-11}$ per electron for $t=10$~nm and $\Delta q_\perp=0.2$~\AA$^{-1}$. A measurement of the asymmetry $A_{\rm m}\simeq(2/\pi)P_bA_{{\rm in},\perp}$ to $\nu\sigma$ requires $\nu^2/A_{\rm m}^2$ counts, where $A_{{\rm in},\perp}=\hat{\vq}_\perp\cdot\vA_{\rm in}$ and the factor $2/\pi$ comes from summing the $\cos\varphi$ modulation over half-planes, with $\varphi$ measured from the transverse source polarization. Moreover, if only a fraction $f$ of the counts in the chosen energy window are magnons, with the rest being phonons, which carry no spin asymmetry, the time grows as $1/f$. Strained-superlattice photocathodes deliver a source polarization $P_b=0.85$ \cite{nishitani2005,maruyama2004}, which is close to the $82\%$ measured for a spin-polarized microscope beam at 30~keV \cite{kuwahara2014}. With this polarization, a $5\sigma$ measurement at 30~keV on a 10~nm film requires $2.5\times10^3$ magnon counts and takes
\begin{equation}
    \tau_{5\sigma}\simeq\frac{2\ {\rm h}}{f}\times\frac{1\ {\rm nA}}{I_b}\qquad(30~\text{keV}),
\end{equation}
where $I_b$ is the beam current. At 100~keV on a 30~nm film, the count requirement is $9\times10^4$ and the time is twelve times longer. The beam current is the decisive parameter. The magnon measurement of Ref.~\cite{kepaptsoglou2025} used about 1~pA in a 1.3~nm probe at 7~meV resolution. The present measurement, however, does not require a focused probe, and a parallel beam illuminating a larger area at the same brightness and energy width carries far more current. With such a beam and a magnon-clean energy window, the 30~keV measurement takes as little as hours.

\begin{figure}[!htb]
\includegraphics[width=\columnwidth]{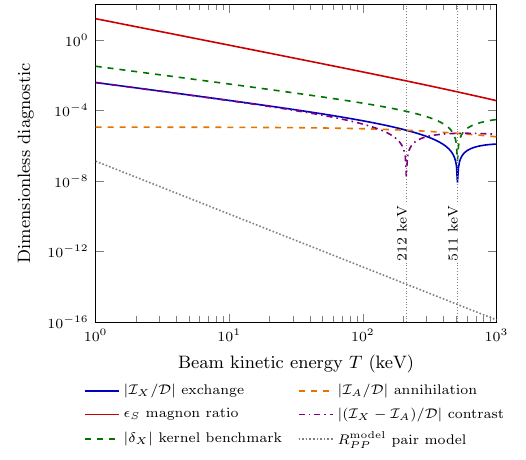}
\caption{Beam-energy dependence of dimensionless diagnostics at $q=1$~\AA$^{-1}$. The electron exchange and positron annihilation curves are magnitudes of the leading density-interference fractions from Eqs.~\eqref{eq:IXratio} and \eqref{eq:IAratio}, with $\kappa=q^2$ and $\Lambda=2m/E$. The contrast curve is the magnitude of their signed difference, which gives the magnitude of the leading fractional contrast $\Delta_{e^\pm}$ only under the additional squared-amplitude conditions in Eq.~\eqref{eq:Rratio}. The vertical guides mark its leading zero near $212$~keV and the exchange-interference zero near $511$~keV. The magnon curve is the spin/current amplitude ratio $\epsilonS$ of Eq.~\eqref{eq:2x2}, with elastic beam coefficients and $q_\perp=q$ in the leading small-transfer circular-mode geometry. Its current amplitude includes direct and exchange terms. The quadratic free-electron kernel benchmark $|\delta_X|$ uses Eq.~\eqref{eq:deltaX} with $\langle p'^2\rangle_{\rm iso}=25$~\AA$^{-2}$. It is not a material-specific correction, and higher orders remain at its leading zero. The pair curve uses the independent-partner Coulomb estimate $R_{PP}^{\rm model}$ of Eq.~\eqref{eq:RPP}, with $n=0.1$~\AA$^{-3}$ and non-relativistic beam momentum $k_i^2=2mT$. It assumes partner factorization and coherent path addition and is not a bound on correlated samples. Its extension to $T$ comparable with $m$ is an algebraic extrapolation of the non-relativistic model. These interference fractions, amplitude ratios and kernel corrections have different meanings and do not rank contributions to one measured spectrum. Each retains the sample, momentum and local-kernel conditions of its source calculation.}
\label{fig2}
\end{figure}

There are three important caveats to these numbers, all of which are respected by the experimental proposal of the work:
\begin{itemize}
\item Equation~\eqref{eq:countrate} treats the sample as a single-sublattice ferromagnet. Yttrium iron garnet, however, is a multi-sublattice magnet whose full magnon spectrum has twenty branches \cite{princep2017}. In such a magnet, the weight of the acoustic mode at a momentum transfer outside the first Brillouin zone is set by a magnetic structure factor that depends on the sublattice eigenvector at the relevant reciprocal-lattice vector. Here $1$~\AA$^{-1}$ is several zones out, since the shortest reciprocal-lattice vector of the body-centred cell is $2\pi\sqrt2/a=0.72$~\AA$^{-1}$. This structure factor replaces $|F(\vq)|^2$ and must be computed for the actual eigenvectors, for which the machinery of Ref.~\cite{nascimento2024} is suited. It changes the absolute rate but not the ratio $A_{{\rm in},\perp}$, which is a property of the beam alone. Inside the first zone, at $q=0.3$~\AA$^{-1}$, $|A_{{\rm in},\perp}|$ is $5.7\%$ at 30~keV.
\item The $\cos\varphi$ modulation is resolved only if the beam's own angular spread is small compared with the scattering angle $q/k$, which is $6$~mrad at 100~keV and $11$~mrad at 30~keV. This is a real constraint, but it has already been met. Indeed, the magnon measurements of Ref.~\cite{kepaptsoglou2025} used a semi-convergence angle of $2.25$~mrad, and a spin-polarized microscope beam has been operated with a parallelism of $1.7\times10^{-5}$~rad \cite{kuwahara2014}. By contrast, a probe focused to atomic dimensions, with a convergence semi-angle of tens of milliradians, would average the modulation away.
\item The magnon fraction $f$ must be estimated for the chosen $(\vq,\omega)$ window. In the only published magnon-EELS data \cite{kepaptsoglou2025}, the phonon signal integrated over energy exceeds the magnon signal by three orders of magnitude. As a result, the energy window must be chosen where the magnon branch is clear of the phonons. For yttrium iron garnet, where the feature observed so far is a magnon polaron \cite{mm2025}, this requires care.
\end{itemize}

\subsection{Corrections to the polarization}
\label{subsec:magnon-corrections}

The isotropic free-electron kernel benchmark \eqref{eq:deltaX} gives exchange-amplitude corrections of $3\times10^{-4}$ at 100~keV and $0.3\%$ at 10~keV for the illustrative $3d$ momentum scale, and it vanishes at 511~keV. The dipole channel alone also produces a transverse polarization through the $i\vq\times\vsigma$ term of Eq.~\eqref{eq:GVexact}. For spins along the beam, however, its leading part cancels because the dipole's spin-flip and non-flip pieces are $90^\circ$ out of phase, leaving $\sim10^{-7}$ at 100~keV and $q=1$~\AA$^{-1}$. Moreover, for an arbitrary sample current in this geometry, it is bounded by $q^2/2k^2\sim10^{-5}$. For in-plane spins, it is $O(q/2k)\approx6\times10^{-3}$ at 100~keV and must be subtracted by calculation. Elastic multiple scattering in the film multiplies $a_0$ and $\vb_0$ by nearly the same phase and so rotates $\vP_{f,\perp}$ in the scattering plane without reducing it. Collecting these results, we see that the prediction \eqref{eq:Pperp} carries corrections of relative order $q/2k$ from the beam's spin-magnetization current, $\delta_X$ from the exchange kernel, and the orbital fraction of the moment through the dipole amplitude. For a spin-only magnet, none of these exceeds one percent of $|P_{f,\perp}|$ above 30~keV.

\section{Recovery of known results}
\label{sec:validation}

As a check on our framework, Table~\ref{tab:validation} collects the limits in which it reproduces known results. Each limit was checked either algebraically or numerically, as indicated, and the numerical methods are described in Appendix~\ref{app:numerics}.

\begin{table*}[!htb]
\caption{Limiting results and checks of the retained calculations. The free-particle rows use unexpanded amplitudes. The other rows state the additional kinematic limit or the model being checked. Numerical comparisons refer to the actual retained expressions, without implying an error bound for a material.}
\label{tab:validation}
\begin{ruledtabular}
\raggedright
\begin{tabular}{p{3.8cm} p{9.0cm} p{3.8cm}}
Limit & Statement & Check\\
\hline
M{\o}ller \cite{moller1932} & $\fM_D+\fM_X$ for a free electron; relative sign of $t$ and $u$ terms & exact\\
Bhabha \cite{bhabha1936} & $\fM_D^{(+)}+\fM_A^{(+)}$; relative sign of $t$ and $s$ terms & exact\\
Bethe \cite{bethe1930} & Eq.~\eqref{eq:master} at $E\to m$ & exact\\
Egerton $\gamma^2$ \cite{egerton2011} & $(4E_iE_f+\fQ^2)/q^4\to4m^2\gamma^2/q^4$ & exact\\
de Forest-Walecka \cite{deforest1966}, Iatrakis-Brudanin \cite{iatrakis2026a} & $(E_i+E_f)^2-q^2\equiv4E_iE_f+\fQ^2$; $L$/$T$ structure & symbolic\\
Ochkur \cite{ochkur1964} & $\PiX\to1/k^2$ & exact\\
Positronium \cite{blp} & $V_{\rm ann}(\bm{r})=\frac{2\pi\alpha}{m^2}P_{S=1}\delta^{(3)}(\bm{r})$ & numerical, eigenvalues\\
M{\o}ller with spin flip & reduced amplitude \eqref{eq:2x2} vs exact tree level & $4\times10^{-5}$\\
Generalized unitarity & $\Im f_2$ from partial waves vs $\frac{k}{4\pi}\!\int\!f_1f_1$ & $10^{-5}$\\
Pair box & Eq.~\eqref{eq:MPP} vs exact Fock space & $10^{-7}$\\
\end{tabular}
\end{ruledtabular}
\end{table*}

\section{Conclusion}
\label{sec:summary}

In this paper, we have constructed a fully quantum theory of energy-loss spectroscopy that accounts for the indistinguishability of the beam and sample electron. By splitting the electron field into beam and sample modes, we turned the one-photon interaction of QED into four channels, namely direct, exchange, annihilation for positrons, and pair excitation, which are distinguished by their field content exactly and without double counting. By treating the beam relativistically and expanding in the ratios of sample to beam momenta, we found that each channel's photon lies in its own kinematic region: the potential region for the direct channel, hard and spacelike for exchange, and hard and timelike for annihilation. The direct channel gives the known relativistic Bethe cross section. The exchange channel is a contact interaction whose coefficient, $1/2m(E-m)$, is the relativistic Ochkur kernel. It interferes destructively with the direct channel below total beam energy $E=2m$ and constructively above, as the magnetic coupling of the hard photon overtakes its electric coupling at that energy. The annihilation channel is the same operator with coefficient $-1/2m(E+m)$; it is a repulsive triplet contact interaction that reduces to the virtual-annihilation potential of positronium. The pair channel vanishes at leading order by a selection rule and the higher-order corrections it induces give the model estimate $R_{PP}^{\rm model}\sim10^{-13}$ at 100~keV in Sec.~\ref{sec:pair}. Lastly, the two-photon corrections are ordered by their energy denominators. The second Born term, which we calibrated on an exactly solvable potential, scales as $(\alpha/v)(p/k)$ and dominates the charge-odd electron--positron asymmetry for every solid, which confines the positron test of exchange to the spin channel.

Applied to a magnon, our framework shows that exchange enters the unpolarized cross section only at second order in the ratio $\epsilonS\simeq qm/(2Tk)$, as the two channels leave the beam electron in orthogonal spin states. Exchange does, however, enter the beam electron's transverse polarization at first order with $|P_{f,\perp}|=2\epsilonS/(1+2\epsilonS^2)$, whose polarization is proportional to the angular momentum of the excited mode. This result forms the content of the companion work, and is the first place in which two channels of our framework meet. For spin-blind spectroscopy the exchange channel is suppressed by the small parameter , $q^2/2mT$, which is below $10^{-4}$ at 60--300~keV, in  which case the direct-channel theory is quantitatively complete. Even so, machine learning approaches may be helpful to identify small signals and make predictions that are related to magnetism~\cite{merker2022magnetism}. It is percent-level at 1--20~keV and large transfers. Finally, it reaches tens of percent at the 10--100~eV energies of reflection M-EELS, where exchange rescales the momentum dependence of the extracted density response and admixes the spin-fluctuation spectrum at large transfer.

The same operators and the same coefficients describe surface spectroscopy at electronvolts, spin-polarized spectroscopy, and monochromated microscope EELS at hundreds of kiloelectronvolts. Our framework therefore places these three techniques on a equal footing and makes their spectra comparable with each other and with resonant x-ray scattering. More broadly, the identity of the beam electron, which is merely assumed negligible in the standard theory of energy-loss spectroscopy, is now a calculable ingredient of the spectroscopy itself.

\appendix
\renewcommand{\theequation}{\thesection\arabic{equation}}
\renewcommand{\theHequation}{\thesection.\arabic{equation}}

\section{Conventions and the Fierz identity}
\label{app:fierz}

The hard photon momenta written $(\fK_i-\fP',\fK_i+\fP)$ in the text are $\fQX$ and $\fQA$ for exchange and annihilation, with spatial magnitudes $q_{\rm X}$ and $q_{\rm A}$.

The beam-state normalization fixes the conversion from reduced amplitudes to cross sections. We derive those normalization factors and the matrix rearrangement used in the local exchange and annihilation amplitudes.

In the metric of Sec.~\ref{sec:notation}, the Dirac representation is
\begin{equation}
    \begin{aligned}
        \gamma^0 &= \begin{pmatrix}\mathbb{1}_2&0\\0&-\mathbb{1}_2\end{pmatrix},\\
        \gamma^i &= \begin{pmatrix}0&\sigma^i\\-\sigma^i&0\end{pmatrix},\\
        \gamma^5 &= i\gamma^0\gamma^1\gamma^2\gamma^3
        =\begin{pmatrix}0&\mathbb{1}_2\\\mathbb{1}_2&0\end{pmatrix}.
    \end{aligned}
\label{eq:dirac-representation}
\end{equation}
Each displayed block is $2$ by $2$. The matrices satisfy the Clifford algebra
\begin{equation}
    \gamma^\mu\gamma^\nu+\gamma^\nu\gamma^\mu
    =2\eta^{\mu\nu}\mathbb{1}_4,
    \label{eq:dirac-algebra}
\end{equation}
where $\mathbb{1}_4$ is the $4$ by $4$ identity matrix. Take $\fK=(E,\vk)$ with $E=\sqrt{m^2+k^2}>0$. For unit-normalized Pauli spinors $\xi$ and $\eta$, the free electron and positron spinors are
\begin{equation}
    \begin{aligned}
        u(\fK,\xi) &= \sqrt{E+m}\begin{pmatrix}\xi\\\dfrac{\vsigma\cdot\vk}{E+m}\xi\end{pmatrix},\\
        v(\fK,\eta) &= \sqrt{E+m}\begin{pmatrix}\dfrac{\vsigma\cdot\vk}{E+m}\eta\\\eta\end{pmatrix}.
    \end{aligned}
\label{eq:free-spinors}
\end{equation}
Taking the differences and sums of the upper- and lower-component norms, using $(\vsigma\cdot\vk)^2=k^2\mathbb{1}_2$, gives
\begin{equation}
    \begin{aligned}
        \bar uu &= 2m,\\
        \bar vv &= -2m,\\
        u^\dagger u &= v^\dagger v=2E.
    \end{aligned}
\label{eq:free-spinor-normalization}
\end{equation}
The spinor $\xi$ labels the electron's canonical rest-frame spin, carried to momentum $\vk$ by a rotation-free boost. Choosing an orthonormal basis of these Pauli spinors defines the two spin labels $s$ in $u^s(\fK)$. For a physical positron Pauli spinor $\xi$, we use $\eta=i\sigma^2\xi^*$ in $v$, as in Eq.~\eqref{eq:positron-physical-spin}.

For beam electrons, let $a_s(\fK)$ annihilate the plane-wave mode with momentum $\vk$ and canonical spin $s$. The covariant state normalization is set by the anticommutator
\begin{equation}
\begin{aligned}
\{a_s(\fK),a_{s'}^{\dagger}(\fK')\}
&=2E(2\pi)^3\delta_{ss'}\delta^{(3)}(\vk-\vk').
\end{aligned}
\label{eq:beam-anticommutator}
\end{equation}
With $\ket{\fK,s}=a_s^{\dagger}(\fK)\ket{0_{\rm b}}$ and a unit-normalized beam vacuum, the one-electron states satisfy
\begin{equation}
\begin{aligned}
\langle \fK',s'|\fK,s\rangle
&=2E(2\pi)^3\delta_{ss'}\delta^{(3)}(\vk-\vk').
\end{aligned}
\label{eq:beam-state-normalization}
\end{equation}
The electron part of the free beam field is
\begin{equation}
\begin{aligned}
&\Psib(\fX)\\
&\quad\supset\sum_s\int_{\rm beam}\frac{\text{d}^3\vk}{(2\pi)^3\,2E}\,
a_s(\fK)u^s(\fK)e^{-i\eta_{\rho\sigma}\fK^\rho \fX^\sigma},
\end{aligned}
\label{eq:beam-field-expansion}
\end{equation}
where the integral runs over the beam modes specified in Sec.~\ref{subsec:beam-sample-modes}. Acting on a one-electron state cancels the factor $2E(2\pi)^3$ against the anticommutator and gives Eq.~\eqref{eq:beam-field-matrix}.

To relate these states to an incident particle flux, temporarily put the beam in a periodic box of volume $V_{\rm b}$. This auxiliary volume is distinct from the sample volume $V$. The continuum factor $(2\pi)^3\delta^{(3)}(\mathbf0)$ becomes $V_{\rm b}$, so a one-particle state with norm one is
\begin{equation}
\ket{\fK,s}_{\rm box}=\frac{\ket{\fK,s}}{\sqrt{2EV_{\rm b}}}.
\label{eq:beam-box-normalization}
\end{equation}
The field matrix element in Eq.~\eqref{eq:beam-field-matrix} acquires the same factor $(2EV_{\rm b})^{-1/2}$. Its ordinary norm squared is therefore $1/V_{\rm b}$ by Eq.~\eqref{eq:free-spinor-normalization}. This gives the incident particle flux $v_i/V_{\rm b}$, with $v_i=k_i/E_i$. The spacing of periodic-box momenta is $2\pi$ divided by the box length in each direction, so the number of outgoing states per final spin in a momentum element is $V_{\rm b}\,\text{d}^3\vk_f/(2\pi)^3$. The two external-state normalization factors in the squared amplitude, this state count, and division by the flux therefore combine as
\begin{equation}
\begin{aligned}
&\frac{1}{(2E_iV_{\rm b})(2E_fV_{\rm b})}
\frac{V_{\rm b}\,\text{d}^3\vk_f}{(2\pi)^3}
\frac{V_{\rm b}}{v_i}=\frac{1}{2E_iv_i}\frac{\text{d}^3\vk_f}{2E_f(2\pi)^3}.
\end{aligned}
\label{eq:beam-cross-section-measure}
\end{equation}
The auxiliary box cancels from the cross section. The separate division by the sample volume in Eq.~\eqref{eq:current-spectral-tensor} defines a response per unit sample volume.

The same free-spinor conventions give the relation between the beam's density and current matrix elements. A free electron spinor satisfies the momentum-space Dirac equation. For the incident spinor and the outgoing adjoint spinor, respectively, it reads
\begin{equation}
\begin{aligned}
\eta_{\rho\sigma}\fK_i^\rho\gamma^\sigma u_i&=mu_i,\\
\bar u_f\gamma^\sigma\eta_{\rho\sigma}\fK_f^\rho&=m\bar u_f.
\end{aligned}
\label{eq:free-beam-dirac-equations}
\end{equation}
Multiply the first equation from the left by $\bar u_f$ and the second from the right by $u_i$. Subtracting the second result from the first introduces $\fQ=\fK_i-\fK_f$, while the equal mass terms cancel. Writing the resulting contraction in time and spatial components gives
\begin{equation}
\begin{aligned}
\eta_{\rho\sigma}\fQ^\rho\bar u_f\gamma^\sigma u_i
&=m\bar u_fu_i-m\bar u_fu_i=0,\\
\omega\bar u_f\gamma^0u_i&=\vq\cdot\bar u_f\bm\gamma u_i.
\end{aligned}
\label{eq:free-beam-current-conservation}
\end{equation}

To obtain a transition rate from the amplitude, evaluate its time integral over an observation time $\tau$, take the squared modulus, and divide by $\tau$. Direct integration gives a peak with width of order $1/\tau$ and area $2\pi$. Integrated against a smooth energy weight, the long-time limit becomes the energy-selection delta function,
\begin{equation}
\begin{aligned}
\frac1\tau\left|\int_{-\tau/2}^{\tau/2}\text{d}t\,
e^{-i(\omega-\omega_{n0})t}\right|^2&=\frac{4\sin^2[(\omega-\omega_{n0})\tau/2]}
{\tau(\omega-\omega_{n0})^2}\\
&\quad\xrightarrow{\tau\to\infty}2\pi\delta(\omega-\omega_{n0}).
\end{aligned}
\label{eq:transition-rate-energy-selection}
\end{equation}

The Fierz rearrangement changes which Dirac components are paired in the two vector currents. A complete basis of $4$ by $4$ Dirac matrices contains scalar, pseudo-scalar, vector and axial matrices, together with six tensor matrices. The tensor matrices are commutators, labelled by $\alpha<\beta$. Moving the contracted outer gamma matrices through the commutator using Eq.~\eqref{eq:dirac-algebra} gives
\begin{equation}
\begin{aligned}
\sigma^{\alpha\beta}&\equiv\frac{i}{2}[\gamma^\alpha,\gamma^\beta],\\
\eta_{\mu\nu}\gamma^\mu\sigma^{\alpha\beta}\gamma^\nu&=0.
\end{aligned}
\label{eq:fierz-tensor-projection}
\end{equation}
Closing this contraction with a tensor matrix in a projection trace therefore gives zero. The contributing structures are
\begin{equation}
    \begin{aligned}
        \Gamma^{\rm S} &= \mathbb{1}_4,\\
        \Gamma^{\rm P} &= \gamma^5,\\
        \Gamma^{({\rm V},\mu)} &= \gamma^\mu,\\
        \Gamma^{({\rm Ax},\mu)} &= \gamma^5\gamma^\mu.
    \end{aligned}
\label{eq:fierz-matrix-structures}
\end{equation}
The collective label $A$ runs over $\mathrm{S}$, $\mathrm{P}$, $({\rm V},\mu)$, and $({\rm Ax},\mu)$. It therefore includes the Lorentz component for vector and axial-vector matrices. The paired matrix $\Gamma_A$ is unchanged for the scalar and pseudo-scalar structures, while
\begin{equation}
    \begin{aligned}
        \Gamma_{({\rm V},\mu)} &= \eta_{\mu\nu}\gamma^\nu,\\
        \Gamma_{({\rm Ax},\mu)} &= \eta_{\mu\nu}\gamma^5\gamma^\nu.
    \end{aligned}
\label{eq:fierz-paired-matrices}
\end{equation}
Upper and lower $A$ distinguish these paired matrices. Only their Lorentz component is raised or lowered. Summing over $A$ includes one scalar matrix, one pseudo-scalar matrix, and all four components of each vector and axial-vector structure. The Fierz label $A$ is unrelated to the annihilation subscript $A$ used for $\PiA$, $\fQA$ and $\fM_A$.

For exchange, the Fierz identity~\cite{fierz1937} changes the pairing of Dirac indices from $(a,b)$ and $(r,s)$ to $(a,s)$ and $(r,b)$.
\begin{equation}
    \eta_{\mu\nu}\gamma^\mu_{ab}\gamma^\nu_{rs}
    =\sum_A c_A\Gamma^A_{as}\Gamma_{A,rb}.
\label{eq:fierz-components}
\end{equation}
The coefficient $c_A$ multiplies the structure labelled by $A$. For a chosen structure $B$, contract the crossed pairs with $\Gamma_{B,sa}$ and $\Gamma^B_{br}$, summing all four Dirac indices. The right-hand side becomes a product of two traces. Traces between distinct basis matrices vanish, while the paired traces give $16$ for a matching component. Thus the scalar and pseudo-scalar projections leave $16c_{\rm S}$ and $16c_{\rm P}$. Summing the four Lorentz components gives $64c_{\rm V}$ and $64c_{\rm Ax}$ for the other two families. Closing the original product into a trace therefore yields
\begin{equation}
    \begin{aligned}
        c_{\rm S} &= \frac{1}{16}\eta_{\mu\nu}\mathrm{Tr}(\gamma^\mu\gamma^\nu)=1,\\
        c_{\rm P} &= \frac{1}{16}\eta_{\mu\nu}\mathrm{Tr}(\gamma^5\gamma^\mu\gamma^5\gamma^\nu)=-1,\\
        c_{\rm V} &= \frac{1}{64}\eta_{\alpha\beta}\eta_{\mu\nu} \mathrm{Tr}(\gamma^\alpha\gamma^\mu\gamma^\beta\gamma^\nu)=-\frac12,\\
        c_{\rm Ax} &= \frac{1}{64}\eta_{\alpha\beta}\eta_{\mu\nu} \mathrm{Tr}(\gamma^5\gamma^\alpha\gamma^\mu\gamma^5\gamma^\beta\gamma^\nu)=-\frac12.
    \end{aligned}
\label{eq:fierz-coefficients}
\end{equation}
Here, $c_{\rm V}$ and $c_{\rm Ax}$ apply to every component of their respective structures. Together with the zero tensor projection, these coefficients give the beam and sample bilinears in Eq.~\eqref{eq:MX-app} and the corresponding annihilation amplitude in Eq.~\eqref{eq:MA-app}.

In the non-relativistic reduction, the scalar and time-vector terms of Eq.~\eqref{eq:fierz} both reduce to the density, while the space-axial term reduces to $-\vs$. Moreover, the combination $[\bar{u}_f(\mathbb{1}_4-\tfrac12\gamma^0)u_i]\rho-\tfrac12(\bar{u}_f\gamma^5\gamma^iu_i)s^i$ reduces for a slow beam to $m\,\xi_f^\dagger(\rho+\vsigma\cdot\vs)\xi_i$. For a single sample electron, this is $2m\,P_{\rm ex}\rho$, where $P_{\rm ex}=\dfrac{1}{2}(1+\vsigma\cdot\vsigma')$ is the spin-exchange operator, as it should be.

The pair-conjugate field is $\Psi^c=\mathcal T\bar\Psi^\top$ with $\mathcal T=\gamma^1\gamma^3$, which satisfies $\mathcal T^2=-\mathbb{1}_4$ and $\mathcal T[\bar u(\fK,\xi)]^\top=u(\fK,i\sigma^2\xi^*)$. In other words, $\mathcal T$ maps a spinor to one of the same momentum and time-reversed spin, so that a non-relativistic spin singlet sits in the scalar slot of the pair bilinears. Lorentz covariance of the pair bilinears restricts $\mathcal T$ to $C$ or $C\gamma^5$ up to a phase, and in this representation $\gamma^1\gamma^3=-\gamma^5C$ with $C=i\gamma^2\gamma^0$.

\section{Spin sums}
\label{app:spinsums}

Unobserved beam spins enter the cross section through products of vertex matrix elements. We convert their incident-spin average and outgoing-spin sum into Dirac traces, using the normalization of Appendix~\ref{app:fierz}.

Multiply each spinor column in Eq.~\eqref{eq:free-spinors} by its Dirac adjoint, $\bar u=u^\dagger\gamma^0$ or $\bar v=v^\dagger\gamma^0$, and sum over an orthonormal Pauli basis. The basis sums $\sum_s\xi_s\xi_s^\dagger=\sum_s\eta_s\eta_s^\dagger=\mathbb{1}_2$ reduce the four blocks to the completeness relations
\begin{equation}
\begin{aligned}
\sum_s u^s(\fK)\bar u^s(\fK)&=\slashed{\fK}+m,\\
\sum_s v^s(\fK)\bar v^s(\fK)&=\slashed{\fK}-m.
\end{aligned}
\label{eq:beam-spinor-completeness}
\end{equation}
Here $\slashed{\fK}\equiv\eta_{\rho\sigma}\fK^\rho\gamma^\sigma$, and $m$ in a matrix sum denotes $m\mathbb{1}_4$.

At fixed incident and outgoing beam momenta, take two $4$ by $4$ vertex matrices $\Gamma$ and $\Delta$. The Dirac adjoint of the latter is $\overline\Delta\equiv\gamma^0\Delta^\dagger\gamma^0$. Complex conjugation gives $(\bar u_f\Delta u_i)^*=\bar u_i\overline\Delta u_f$. Insert Eq.~\eqref{eq:beam-spinor-completeness} for both spin sums and close the remaining matrix product into a trace. For an unpolarized electron beam this gives
\begin{equation}
\begin{aligned}
&\dfrac{1}{2}\sum_{s_i,s_f}(\bar u_f\Gamma u_i)(\bar u_f\Delta u_i)^*\\
&\quad=\dfrac{1}{2}\Tr\big[\Gamma(\slashed{\fK}_i+m)
\overline\Delta(\slashed{\fK}_f+m)\big].
\end{aligned}
\label{eq:electron-vertex-spin-sum}
\end{equation}
The factor $1/2$ averages the two incident spins, and the outgoing spins are summed. The positron vertex is $\bar v_i\Gamma v_f$, with the incident and outgoing spinors in the opposite order. Repeating the conjugation and completeness substitutions gives
\begin{equation}
\begin{aligned}
&\dfrac{1}{2}\sum_{s_i,s_f}(\bar v_i\Gamma v_f)(\bar v_i\Delta v_f)^*\\
&\quad=\dfrac{1}{2}\Tr\big[\Gamma(\slashed{\fK}_f-m)
\overline\Delta(\slashed{\fK}_i-m)\big].
\end{aligned}
\label{eq:positron-vertex-spin-sum}
\end{equation}

With $\Tr\mathbb{1}_4=4$, taking the trace of the Dirac anticommutation relation gives the two-matrix identity. Anticommuting the first gamma matrix through the other three and using cyclicity collects the four-matrix trace on one side and two-matrix traces on the other. The explicit matrices in Eq.~\eqref{eq:dirac-representation} fix the sign of the trace containing $\gamma^5$. These operations give
\begin{equation}
\begin{aligned}
\Tr(\gamma^\mu\gamma^\nu)&=4\eta^{\mu\nu},\\
\Tr(\gamma^\mu\gamma^\nu\gamma^\rho\gamma^\sigma)
&=4\big(\eta^{\mu\nu}\eta^{\rho\sigma}
-\eta^{\mu\rho}\eta^{\nu\sigma}\\
&\qquad+\eta^{\mu\sigma}\eta^{\nu\rho}\big),\\
\Tr(\gamma^5\gamma^\mu\gamma^\nu\gamma^\rho\gamma^\sigma)
&=-4i\epsilon^{\mu\nu\rho\sigma}.
\end{aligned}
\label{eq:gamma-trace-identities}
\end{equation}
The four-dimensional antisymmetric symbol has $\epsilon^{0123}=1$. Traces of an odd number of matrices $\gamma^\mu$ vanish. With a single $\gamma^5$, fewer than four matrices $\gamma^\mu$ also give zero.

For the direct current, set $\Gamma=\gamma^\mu$ and $\Delta=\gamma^\nu$ in Eq.~\eqref{eq:electron-vertex-spin-sum}. Their Dirac adjoints are unchanged, so the beam tensor is
\begin{equation}
\fL^{\mu\nu}=\dfrac{1}{2}\Tr\big[\gamma^\mu(\slashed{\fK}_i+m)
\gamma^\nu(\slashed{\fK}_f+m)\big].
\label{eq:beam-spin-trace}
\end{equation}
Terms linear in $m$ contain three gamma matrices and vanish. Applying the two- and four-matrix traces to the remaining terms gives
\begin{equation}
\begin{aligned}
\fL^{\mu\nu}
&=2\big[\fK_i^\mu\fK_f^\nu+\fK_i^\nu\fK_f^\mu-\eta^{\mu\nu}
(\eta_{\rho\sigma}\fK_i^\rho\fK_f^\sigma-m^2)\big],
\end{aligned}
\label{eq:beam-vector-spin-sum}
\end{equation}
which is Eq.~\eqref{eq:lepton-app}. The four-matrix term is symmetric under $\fK_i\leftrightarrow\fK_f$, and the mass-squared term has the same sign in Eq.~\eqref{eq:positron-vertex-spin-sum}. The positron vector tensor is therefore the same, as used in Eq.~\eqref{eq:positron-beam-tensor}.

For a scalar--vector product, set $\Gamma=\mathbb{1}_4$ and $\Delta=\gamma^\nu$. Only terms linear in $m$ survive, so the electron and positron sums have opposite signs. Their evaluated forms enter Eqs.~\eqref{eq:trace1-app} and \eqref{eq:positron-scalar-vector-trace}. For the electron axial and pseudo-scalar products, the adjoints are $\overline{\gamma^5\gamma^\mu}=\gamma^5\gamma^\mu$ and $\overline{\gamma^5}=-\gamma^5$. Substituting these vertices into Eq.~\eqref{eq:electron-vertex-spin-sum} gives the spatial axial, mixed vector--axial and pseudo-scalar sums in Eqs.~\eqref{eq:magnon-beam-tensors} and \eqref{eq:magnon-pseudo-scalar-sums}.

\section{Spectral functions for sample operators}
\label{app:sample-response}
\begingroup
\addtolength{\abovedisplayskip}{-2pt}
\addtolength{\belowdisplayskip}{-2pt}
\addtolength{\abovedisplayshortskip}{-2pt}
\addtolength{\belowdisplayshortskip}{-2pt}

Different scattering channels can excite the same sample state through different operators. A common list of sample operators lets us retain these contributions and their interference in one spectral matrix. For local operators $\fO_a$, we use the spatial Fourier convention of Eq.~\eqref{eq:Sab},
\begin{equation}
\widetilde{\fO}_a(\vq)\equiv\int \text{d}^3\vx\,e^{i\vq\cdot\vx}\fO_a(0,\vx),
\label{eq:sample-operator-fourier}
\end{equation}
and define
\begin{equation}
\begin{aligned}
S^{ab}(\vq,\omega)\equiv\frac1V\sum_n&\bra0\widetilde{\fO}_b^{\dagger}(\vq)\ket n
\bra n\widetilde{\fO}_a(\vq)\ket0\\
&\,\delta(\omega-\omega_{n0}).
\end{aligned}
\label{eq:Sab-app}
\end{equation}
The labels $a,b$ enumerate the chosen operators. Diagonal entries contain squared transition amplitudes, while off-diagonal entries retain cross terms between different operators reaching the same state.

Keep the beam momenta fixed, with $\fQ=\fK_i-\fK_f=(\omega,\vq)$ as in Sec.~\ref{subsec:direct-amplitude}. Suppose the retained channel amplitudes can be written in this operator list with coefficients independent of the sample final state $n$. Denote the direct, exchange, and annihilation coefficients by $c_a^{\rm D}$, $c_a^{\rm X}$, and $c_a^{\rm A}$. Their beam-spin dependence is left implicit. The amplitudes add before squaring, so their sum has the form
\begin{equation}
\begin{aligned}
\fM(n;s_i,s_f)&=\sum_a c_a\bra n\widetilde{\fO}_a(\vq)\ket0,\\
c_a&=\sum_{C\ {\rm retained}}c_a^{C}.
\end{aligned}
\label{eq:operator-amplitude-coefficients}
\end{equation}
Here $C$ labels the channels retained for the beam under consideration. The operator list must be specified for each calculation, and all retained channels must be expressed in that same list before their coefficients are added. An operator absent from a channel has zero coefficient in that channel. If the coefficients still depend on $n$, they must remain inside the sample-state sum and cannot be factored against Eq.~\eqref{eq:Sab-app}.

For the direct channel of an electron beam, choose the ordered list $(\fJ^0_\text{ss},\fJ^1_\text{ss},\fJ^2_\text{ss},\fJ^3_\text{ss})$. Labeling these four entries by $0,1,2,3$, Eq.~\eqref{eq:MD} gives
\begin{equation}
\begin{aligned}
c_0^{\rm D}&=-e^2\widetilde{\Pi}(\fQ)\bar{u}_f\gamma^0u_i,\\
c_i^{\rm D}&=+e^2\widetilde{\Pi}(\fQ)\bar{u}_f\gamma^iu_i,\qquad i=1,2,3.
\end{aligned}
\label{eq:direct-operator-coefficients}
\end{equation}
The relative sign follows from the mostly-minus metric. In this list, Eq.~\eqref{eq:Sab-app} is the current tensor $\fS_{JJ}^{\mu\nu}$ of Eq.~\eqref{eq:current-spectral-tensor}, with the operator labels identifying the upper current components. For the ordered spin-density list $(s^1,s^2,s^3)$, we denote the same spectral construction by $S_{ss}^{ij}$. These subscripts identify the sample operators rather than the scattering channel.

Squaring Eq.~\eqref{eq:operator-amplitude-coefficients} and summing over the orthogonal sample final states with the energy-selection factor gives $\sum_{a,b} c_ac_b^*S^{ab}$ per sample volume, including both operator and channel interference. Using the rate limit and beam normalization of Eqs.~\eqref{eq:transition-rate-energy-selection} and \eqref{eq:beam-cross-section-measure}, together with the positive loss measure $\text{d}^3\vk_f=k_fE_f\,\text{d}\Omega\,\text{d}\omega$ of Sec.~\ref{subsec:spin-response}, gives
\begin{equation}
\frac1V\frac{\text{d}^2\sigma}{\text{d}\Omega\,\text{d}\omega}
=\frac{k_f}{16\pi^2k_i}\,
\overline{\sum_{\rm spins}}\sum_{a,b} c_ac_b^*S^{ab}(\vq,\omega).
\label{eq:master0-app}
\end{equation}
Operator labels are summed explicitly. The barred sum averages the two incident beam spins with weight $1/2$ and sums the outgoing spins, as in Eq.~\eqref{eq:lepton-app}. For the direct coefficients above, this reduces to Eq.~\eqref{eq:master0}. Common spectral arguments $(\vq,\omega)$ are left implicit below.

The current tensor in Eq.~\eqref{eq:current-spectral-tensor} is constrained by the local conservation law $\partial_t\fJ^0_\text{ss}+\vnabla\cdot\vJ_\text{ss}=0$. For a transition matrix element, the stationary-state phase in Eq.~\eqref{eq:sample-current-time} turns the time derivative into $i\omega_{n0}$. Multiplying the spatial divergence by $e^{i\vq\cdot\vx}$ and integrating by parts gives $-i\vq$ when the boundary terms vanish. With the Fourier convention in Eq.~\eqref{eq:Sab}, the transformed conservation law is therefore
\begin{equation}
\begin{aligned}
0&=i\omega_{n0}\bra n\widetilde{\fJ}^0_\text{ss}(\vq)\ket0-i\vq\cdot\bra n\widetilde{\vJ}_\text{ss}(\vq)\ket0.
\end{aligned}
\label{eq:sample-continuity-fourier-derivation}
\end{equation}
On the support of the energy delta function, $\omega_{n0}=\omega$, giving Eq.~\eqref{eq:LT}. For $q>0$, projection along $\hat{\vq}=\vq/q$ expresses the longitudinal current matrix element through the density matrix element. Inserting this relation into one or both current factors in Eq.~\eqref{eq:current-spectral-tensor} gives
\begin{equation}
\begin{aligned}
\hat q^i\fS_{JJ}^{i0}&=\frac{\omega}{q}\fS_{JJ}^{00},\\
\hat q^i\hat q^j\fS_{JJ}^{ij}&=\frac{\omega^2}{q^2}\fS_{JJ}^{00}.
\end{aligned}
\label{eq:current-spectral-continuity}
\end{equation}
These relations do not require isotropy. They determine the longitudinal projections, but leave correlations involving transverse current components in the general tensor.

For an isotropic sample response, $\vq$ is the only available spatial direction. Rotational symmetry therefore restricts $\fS_{JJ}^{(0i)}$ to a term proportional to $\hat q^i$, and $\fS_{JJ}^{(ij)}$ to a combination of $\hat q^i\hat q^j$ and $\delta^{ij}-\hat q^i\hat q^j$. Equation~\eqref{eq:current-spectral-continuity} fixes the longitudinal coefficients. The transverse structure has trace two, so its coefficient is $R_T/2$ by Eq.~\eqref{eq:longitudinal-transverse-spectra}.

The symmetric part $\fS_{JJ}^{(\mu\nu)}\equiv(\fS_{JJ}^{\mu\nu}+\fS_{JJ}^{\nu\mu})/2$ is therefore
\begin{align}
\fS_{JJ}^{00}&=R_L,\qquad
\fS_{JJ}^{(0i)}=\frac{\omega\fQ^i}{q^2}R_L,\nonumber\\
\fS_{JJ}^{(ij)}&=\frac{\omega^2}{q^2}\hat q^i\hat q^jR_L
+\dfrac{1}{2}(\delta^{ij}-\hat q^i\hat q^j)R_T.
\label{eq:LT-app}
\end{align}
An antisymmetric transverse response can remain. Its contraction with the symmetric beam tensor $\fL^{\mu\nu}$ vanishes in Eq.~\eqref{eq:master0}.
\par\endgroup

\section{Direct amplitude and Bethe cross section}
\label{app:direct-derivation}
\label{app:bethe-derivation}

We give the field and phase integrations behind the direct amplitude, then perform the response contraction and its longitudinal limits. The beam and sample states are those of Sec.~\ref{subsec:direct-amplitude}.

\subsection{Field matrix elements and energy conservation}
\label{subsec:proof-direct-fields}

At first order in $\SD$, the transition matrix element is obtained from $\text{T}\{i\SD\}$. The photon couples to a current at each of $\fX$ and $\fY$, the interaction vertices. The beam and sample currents commute, so chronological ordering leaves their product unchanged. Using $\SD$'s formula, the transition matrix element is
\begin{equation}
\begin{aligned}
&\bra{\fK_f,s_f;n}\text{T}\{i\SD\}\ket{\fK_i,s_i;0}\\
&\quad=-ie^2\iint \text{d}^4\fX\,\text{d}^4\fY\,\Pi(\fX-\fY)\\
&\qquad\,\bra{\fK_f,s_f}\fJ^\mu_\text{bb}(\fX)\ket{\fK_i,s_i}\bra{n}\fJ_{\text{ss},\mu}(\fY)\ket{0}.
\end{aligned}
\label{eq:direct-transition}
\end{equation}

For the free beam field, let $\ket{0_{\rm b}}$ denote the state with no beam particles. The beam-state normalization and field expansion in Appendix~\ref{app:fierz} give the matrix elements for removing and adding a beam electron,
\begin{equation}
\begin{aligned}
\bra{0_{\rm b}}\Psib(\fX)\ket{\fK,s}&=u^s(\fK)e^{-i\eta_{\rho\sigma}\fK^\rho \fX^\sigma},\\
\bra{\fK,s}\Psibb(\fX)\ket{0_{\rm b}}&=\bar u^s(\fK)e^{i\eta_{\rho\sigma}\fK^\rho \fX^\sigma},
\end{aligned}
\label{eq:beam-field-matrix}
\end{equation}
where $s$ labels the beam spin. In $\fJ^\mu_\text{bb}$, one field annihilates the incident electron and the other creates the outgoing electron. Their matrix elements therefore give
\begin{equation}
\begin{aligned}
\bra{\fK_f,s_f}\fJ^\mu_\text{bb}(\fX)\ket{\fK_i,s_i}=\bar{u}_f\gamma^\mu u_i\,e^{-i\eta_{\rho\sigma}\fQ^\rho \fX^\sigma}.
\end{aligned}
\label{eq:beam-current-transition}
\end{equation}

To perform the $\fX$ integral in Eq.~\eqref{eq:direct-transition}, we insert the photon Fourier representation in Eq.~\eqref{eq:photon-fourier}. Writing $\fQ'$ for its four-momentum integration variable, the photon and beam factors combine into the phase $e^{-i\eta_{\rho\sigma}(\fQ'^\rho+\fQ^\rho)\fX^\sigma}$. The $\fX$ integral gives a delta function that sets $\fQ'+\fQ=0$, fixing the kernel momentum to $-\fQ$. The remaining photon phase depends on $\fY$, giving
\begin{equation}
\begin{aligned}
\int \text{d}^4\fX\,\Pi(\fX-\fY)e^{-i\eta_{\rho\sigma}\fQ^\rho \fX^\sigma}
&=\widetilde{\Pi}(-\fQ)e^{-i\eta_{\rho\sigma}\fQ^\rho \fY^\sigma}\\
&=\widetilde{\Pi}(\fQ)e^{-i\eta_{\rho\sigma}\fQ^\rho \fY^\sigma}.
\end{aligned}
\label{eq:direct-kernel-convolution}
\end{equation}
The last equality holds because the kernel in Eq.~\eqref{eq:photon-propagator} depends only on $\fQ^2$. The remaining phase is $e^{-i\eta_{\rho\sigma}\fQ^\rho \fY^\sigma}=e^{-i\omega t}e^{i\vq\cdot\bm{y}}$, where $\fY=(t,\bm{y})$.

The sample states have definite energies, so their time evolution supplies phases $e^{-iE_0t}$ and $e^{-iE_nt}$. In the current matrix element, the final-state phase is complex conjugated. Multiplying it by the initial-state phase gives $e^{i(E_n-E_0)t}=e^{i\omega_{n0}t}$, so
\begin{equation}
\begin{aligned}
\bra{n}\fJ_{\text{ss},\mu}(t,\bm{y})\ket{0}=e^{i\omega_{n0}t}\bra{n}\fJ_{\text{ss},\mu}(0,\bm{y})\ket{0}.
\end{aligned}
\label{eq:sample-current-time}
\end{equation}
Multiplying this sample phase by the beam phase $e^{-i\omega t}$ gives the time integral
\begin{equation}
\int \text{d}t\,e^{-i(\omega-\omega_{n0})t}
=2\pi\delta(\omega-\omega_{n0}).
\label{eq:direct-energy-delta}
\end{equation}
The remaining spatial integral is the Fourier component in Eq.~\eqref{eq:Sab}.

\subsection{Beam-spin sums and sample response}
\label{subsec:proof-direct-response}

Separating the beam-spin sum in Eq.~\eqref{eq:master0} gives the beam tensor, evaluated in Appendix~\ref{app:spinsums},
\begin{equation}
\begin{aligned}
\fL^{\mu\nu}&\equiv\dfrac{1}{2}\sum_{s_i,s_f}(\bar{u}_f\gamma^\mu u_i)(\bar{u}_f\gamma^\nu u_i)^*\\
&=2\big[\fK_f^\mu\fK_i^\nu+\fK_i^\mu\fK_f^\nu-\eta^{\mu\nu}(\eta_{\rho\sigma}\fK_i^\rho\fK_f^\sigma-m^2)\big].
\label{eq:lepton-app}
\end{aligned}
\end{equation}

The current in Eq.~\eqref{eq:Sab} has a lower Lorentz index. We raise it with the metric, $\widetilde{\fJ}_\text{ss}^{\alpha}(\vq)=\eta^{\alpha\mu}\widetilde{\fJ}_{\text{ss},\mu}(\vq)$. Complex conjugation of a current matrix element interchanges the bra and ket and replaces the current by its Hermitian conjugate. The sample tensor is therefore
\begin{equation}
\begin{aligned}
\fS_{JJ}^{\alpha\beta}(\vq,\omega)
&\equiv\frac1V\sum_n\bra0 [\widetilde{\fJ}_\text{ss}^{\beta}(\vq)]^{\dagger}\ket n\\
&\quad\,\bra n \widetilde{\fJ}_\text{ss}^{\alpha}(\vq)\ket0\,
\delta(\omega-\omega_{n0}).
\end{aligned}
\label{eq:current-spectral-tensor}
\end{equation}
Expanding the modulus square in Eq.~\eqref{eq:master0} introduces one metric for each sample current when its index is raised. The beam-spin sum and the sample-state sum then give
\begin{equation}
\begin{aligned}
\fW_{\rm D}(\vq,\omega)&=e^4|\widetilde{\Pi}(\fQ)|^2\eta_{\mu\alpha}\eta_{\nu\beta}
\fL^{\mu\nu}\fS_{JJ}^{\alpha\beta}(\vq,\omega).
\end{aligned}
\label{eq:direct-inclusive-weight}
\end{equation}

\subsection{Longitudinal and transverse contraction}
\label{subsec:proof-direct-contraction}

The beam tensor in Eq.~\eqref{eq:lepton-app} is symmetric, so only the symmetric part $\fS_{JJ}^{(\mu\nu)}=(\fS_{JJ}^{\mu\nu}+\fS_{JJ}^{\nu\mu})/2$ contributes to Eq.~\eqref{eq:master0}. Using the isotropic decomposition derived in Eq.~\eqref{eq:LT-app}, this symmetric tensor has the form
\begin{equation}
\begin{aligned}
\fS_{JJ}^{(0i)}(\vq,\omega)&=\frac{\omega}{q}\hat q^i R_L(\vq,\omega),\\
\fS_{JJ}^{(ij)}(\vq,\omega)&=\frac{\omega^2}{q^2}\hat q^i\hat q^jR_L(\vq,\omega)\\
&\quad+\frac12(\delta^{ij}-\hat q^i\hat q^j)R_T(\vq,\omega).
\end{aligned}
\label{eq:longitudinal-transverse-spectra}
\end{equation}
where
\begin{equation}
\begin{aligned}
R_L(\vq,\omega)&\equiv\fS_{JJ}^{00}(\vq,\omega),\\
R_T(\vq,\omega)&\equiv(\delta^{ij}-\hat q^i\hat q^j)\fS_{JJ}^{ij}(\vq,\omega).
\end{aligned}
\end{equation}

We now use Eq.~\eqref{eq:longitudinal-transverse-spectra} to evaluate the beam--sample contraction in Eq.~\eqref{eq:master0}. The beam current also obeys continuity, as derived from the free-spinor equations in Eqs.~\eqref{eq:free-beam-dirac-equations} and \eqref{eq:free-beam-current-conservation}. Applying this relation to the beam-current products in Eq.~\eqref{eq:lepton-app} gives $\hat q^i\fL^{0i}=(\omega/q)\fL^{00}$ and $\hat q^i\hat q^j\fL^{ij}=(\omega^2/q^2)\fL^{00}$. The two mixed time--space terms in the metric contraction have negative signs, so the density and longitudinal parts combine into $(1-\omega^2/q^2)^2\fL^{00}R_L$. The transverse term contracts $\fL^{ij}$ with the projector in Eq.~\eqref{eq:longitudinal-transverse-spectra} and inherits its factor $1/2$. Substituting the explicit beam tensor from Eq.~\eqref{eq:lepton-app} gives
\begin{equation}
\begin{aligned}
\eta_{\mu\alpha}\eta_{\nu\beta}\fL^{\mu\nu}\fS_{JJ}^{\alpha\beta}
&=D_L R_L+D_T R_T,\\
D_L&=\frac{(\fQ^2)^2}{q^4}\fL^{00}\\
&=\frac{(4E_iE_f+\fQ^2)(\fQ^2)^2}{q^4},\\
D_T&=\dfrac{1}{2}(\delta^{ij}-\hat q^i\hat q^j)\fL^{ij}\\
&=\frac{2|\vk_i\times\vk_f|^2}{q^2}-\fQ^2.
\end{aligned}
\label{eq:DDT-app}
\end{equation}

\subsection{Photon factor and longitudinal limits}

We first determine the photon factor for the beam kinematics. Both beam energies are positive and obey $E_i^2=m^2+k_i^2$ and $E_f^2=m^2+k_f^2$. Substituting these relations gives $(E_iE_f)^2-(m^2+k_ik_f)^2=m^2(k_i-k_f)^2\geq0$. Taking positive square roots gives $E_iE_f\geq m^2+k_ik_f$. Together with $\vk_i\cdot\vk_f\leq k_ik_f$, this implies
\begin{equation}
\begin{aligned}
\fQ^2
&=(E_i-E_f)^2-|\vk_i-\vk_f|^2\\
&=2m^2-2(E_iE_f-\vk_i\cdot\vk_f)<0
\qquad(q>0).
\end{aligned}
\label{eq:bethe-spacelike-transfer}
\end{equation}
Equality requires equal beam momentum magnitudes and parallel directions, hence $\vq=\mathbf0$, which is excluded here.

To evaluate the squared kernel in Eq.~\eqref{eq:master0} at $\fQ^2<0$, replace the $i0$ in Eq.~\eqref{eq:photon-propagator} by $i\epsilon$, with $\epsilon>0$, and take the modulus before sending $\epsilon$ to zero. Multiplication by the complex conjugate gives
\begin{equation}
\begin{aligned}
|\widetilde{\Pi}(\fQ)|^2
&=\lim_{\epsilon\to0^+}
\frac1{(\fQ^2+i\epsilon)(\fQ^2-i\epsilon)}\\
&=\lim_{\epsilon\to0^+}\frac1{(\fQ^2)^2+\epsilon^2}
=\frac1{(\fQ^2)^2}.
\end{aligned}
\label{eq:bethe-kernel-square}
\end{equation}

We now insert the contraction in Eq.~\eqref{eq:DDT-app} into Eq.~\eqref{eq:master0} and use Eq.~\eqref{eq:bethe-kernel-square}. Keeping the two sample spectra separate gives
\begin{equation}
\begin{aligned}
\frac1V\frac{\text{d}^2\sigma_{\rm D}}{\text{d}\Omega\,\text{d}\omega}
&=\alpha^2\frac{k_f}{k_i}
\bigg[\frac{D_L}{(\fQ^2)^2}R_L(\vq,\omega)\\
&\quad+\frac{D_T}{(\fQ^2)^2}R_T(\vq,\omega)\bigg].
\end{aligned}
\label{eq:bethe-kernel-contraction}
\end{equation}
The longitudinal coefficient in Eq.~\eqref{eq:DDT-app} contains a factor $(\fQ^2)^2$, which cancels its photon denominator. The transverse coefficient retains that denominator. The two substitutions are
\begin{equation}
\begin{aligned}
\frac{D_L}{(\fQ^2)^2}
&=\frac{(4E_iE_f+\fQ^2)(\fQ^2)^2}
{q^4(\fQ^2)^2}\\
&=\frac{4E_iE_f+\fQ^2}{q^4},\\
\frac{D_T}{(\fQ^2)^2}
&=\frac{2|\vk_i\times\vk_f|^2/q^2-\fQ^2}
{(\fQ^2)^2}.
\end{aligned}
\label{eq:bethe-weighted-coefficients}
\end{equation}
Inserting these expressions into Eq.~\eqref{eq:bethe-kernel-contraction} gives Eq.~\eqref{eq:master} without expanding in $\omega/q$.

To obtain the non-relativistic longitudinal limit, we leave $k_f/k_i$ and $R_L(\vq,\omega)$ unexpanded and expand only the beam coefficient. With $E_i=m+T_i$ and $E_f=m+T_f$, the momentum magnitudes obey $k_i^2=2mT_i+T_i^2$ and $k_f^2=2mT_f+T_f^2$. The energy transfer is $\omega=T_i-T_f$. Expanding the numerator and bounding $q$ by the triangle inequality gives
\begin{equation}
\begin{aligned}
4E_iE_f+\fQ^2
&=4m^2+4m(T_i+T_f)\\
&\quad+(T_i+T_f)^2-q^2,\\
q^2&\leq(k_i+k_f)^2\leq2(k_i^2+k_f^2)\\
&=4m(T_i+T_f)+2(T_i^2+T_f^2).
\end{aligned}
\label{eq:bethe-nonrel-kinematics}
\end{equation}
For $T_i,T_f\ll m$, every correction to $4m^2$ is therefore of relative order $\order{(T_i+T_f)/m}$ or smaller. Using $a_0=1/(m\alpha)$ converts the longitudinal beam coefficient to
\begin{equation}
\begin{aligned}
\alpha^2\frac{4E_iE_f+\fQ^2}{q^4}
&=\frac{4m^2\alpha^2}{q^4}
\left[1+\order{\frac{T_i+T_f}{m}}\right]\\
&=\frac4{a_0^2q^4}
\left[1+\order{\frac{T_i+T_f}{m}}\right].
\end{aligned}
\label{eq:bethe-nonrel-limit}
\end{equation}

The relativistic longitudinal limit follows by expressing the same numerator in terms of the incident energy. Substituting $E_f=E_i-\omega$ and $\gamma=E_i/m$ gives the exact identity
\begin{equation}
\begin{aligned}
4E_iE_f+\fQ^2
&=4E_i^2\left[1-\frac{\omega}{E_i}
+\frac{\fQ^2}{4E_i^2}\right]\\
&=4m^2\gamma^2\left[1-\frac{\omega}{E_i}
+\frac{\fQ^2}{4E_i^2}\right].
\end{aligned}
\label{eq:bethe-rel-numerator}
\end{equation}
When $\omega/E_i\ll1$ and $|\fQ^2|/E_i^2\ll1$, the two correction terms in the brackets are small. Using $m^2\alpha^2=1/a_0^2$ then gives
\begin{equation}
\begin{aligned}
\alpha^2\frac{4E_iE_f+\fQ^2}{q^4}
&=\frac{4\gamma^2}{a_0^2q^4}\left[1+\order{\frac{\omega}{E_i}
+\frac{|\fQ^2|}{E_i^2}}\right].
\end{aligned}
\label{eq:bethe-rel-limit}
\end{equation}

\section{Exchange and annihilation derivations}
\label{app:channel-derivations}

We evaluate the exchange and positron amplitudes before the local approximation and retain the operator, interference and spin reductions used in Secs.~\ref{sec:exchange} and \ref{sec:positron}. The separate control of the local photon coefficients is derived in Appendix~\ref{app:exchange-kernel}.

\subsection{General exchange amplitude}
\label{subsec:proof-exchange-amplitude}

To obtain the exchange amplitude, we evaluate $\SX$ between the same beam and sample states as in the direct amplitude, with chronological ordering as in $\SD$. With Dirac component indices $a,b,c,d$, substituting the conversion currents in $\Sint$ into $\SX$ gives the field order $\bar\Psi_{{\rm b},a}(\fX)\Psi_{{\rm s},b}(\fX)\bar\Psi_{{\rm s},c}(\fY)\Psi_{{\rm b},d}(\fY)$. Within $\text{T}$, moving the beam field at $\fY$ past both sample fields and then interchanging the sample fields groups the beam pair and places the sample creation field first. These three fermion interchanges give a minus sign, changing the prefactor $-ie^2$ of $i\SX$ to $+ie^2$.

Applying Eq.~\eqref{eq:beam-field-matrix} to $\Psibb(\fX)$ and $\Psib(\fY)$ evaluates the outgoing and incident beam factors, respectively. The sample fields remain chronologically ordered between $\bra n$ and $\ket0$, so the full inelastic transition matrix element is
\begin{equation}
\begin{aligned}
&\bra{\fK_f,s_f;n}\text{T}\{i\SX\}\ket{\fK_i,s_i;0}\\
&\quad=ie^2\!\sum_{a,b,c,d}\int \text{d}^4\fX\,\text{d}^4\fY\,\Pi(\fX-\fY)e^{i\eta_{\rho\sigma}\fK_f^\rho \fX^\sigma-i\eta_{\alpha\beta}\fK_i^\alpha \fY^\beta}\\
&\qquad\,\eta_{\mu\nu}\bar u_{f,a}\gamma^\mu_{ab}\gamma^\nu_{cd}u_{i,d}\bra n\text{T}\{\bar\Psi_{{\rm s},c}(\fY)\Psi_{{\rm s},b}(\fX)\}\ket0.
\end{aligned}
\label{eq:SXme-app}
\end{equation}

To determine the photon momentum, we Fourier transform the chronologically ordered sample matrix element in Eq.~\eqref{eq:SXme-app} with respect to both $\fX$ and $\fY$. Let $\fP=(\fP^0,\vp)$ be the four-momentum integration variable associated with $\Psis(\fX)$, which removes a sample electron, and let $\fP'=(\fP'^0,\vp')$ be the corresponding variable for $\Psisb(\fY)$, which adds one. In this representation, these variables enter through the Fourier factor $e^{i\eta_{\rho\sigma}\fP'^\rho \fY^\sigma-i\eta_{\alpha\beta}\fP^\alpha \fX^\beta}$.

We combine this factor with the beam phases in Eq.~\eqref{eq:SXme-app} and the photon phase in Eq.~\eqref{eq:photon-fourier}, writing $\fQX$ for the photon momentum. The resulting phase coefficients are $\fK_f-\fP-\fQX$ at $\fX$ and $\fP'-\fK_i+\fQX$ at $\fY$. The coordinate integrals set these coefficients to zero, as in Eq.~\eqref{eq:direct-kernel-convolution}. These conditions express conservation of energy and momentum at the two vertices,
\begin{equation}
\begin{aligned}
\fP+\fQX&=\fK_f,\\
\fP'+\fQX&=\fK_i.
\end{aligned}
\label{eq:exchange-momentum}
\end{equation}
Subtracting the first relation from the second cancels the photon momentum and gives $\fP'-\fP=\fK_i-\fK_f=\fQ$. The sample thus gains exactly the energy and momentum lost by the beam.

We now separate the energy-conservation factor from Eq.~\eqref{eq:SXme-app}, as in the direct amplitude. Write $\fX=(t+\tau,\vx)$ and $\fY=(t,\bm{y})$, where $\tau=\fX^0-\fY^0$ is the time difference between the two vertices. Translating both sample fields by the common time $t$ gives the phase $e^{i\omega_{n0}t}$ from the initial and final sample energies, just as in Eq.~\eqref{eq:sample-current-time}. The beam phase contains $e^{-i\omega t}$, while the photon kernel depends only on $\tau$ and $\vx-\bm{y}$. The $t$ integral therefore gives $2\pi\delta(\omega-\omega_{n0})$.

Removing this factor defines the reduced exchange amplitude $\fM_X(n;s_i,s_f)$ with the convention of Eq.~\eqref{eq:MD}. At fixed beam momenta and spins, the remaining time and spatial integrals give
\begin{equation}
\begin{aligned}
&\fM_X(n;s_i,s_f)\\
&\quad=e^2\sum_{a,b,c,d}\eta_{\mu\nu}\bar u_{f,a}\gamma^\mu_{ab}\gamma^\nu_{cd}u_{i,d}\int\text{d}\tau\,\text{d}^3\vx\,\text{d}^3\bm{y}\,
\Pi(\tau,\vx-\bm{y})\\
&\quad\,e^{iE_f\tau-i\vk_f\cdot\vx+i\vk_i\cdot\bm{y}}\bra n\text{T}\{\bar\Psi_{{\rm s},c}(0,\bm{y})
\Psi_{{\rm s},b}(\tau,\vx)\}\ket0.
\end{aligned}
\label{eq:exchange-amplitude}
\end{equation}

\subsection{Local operator basis and sample reduction}
\label{subsec:proof-exchange-operators}

After the constant-kernel localization of Sec.~\ref{subsec:proof-exchange-localization}, the remaining Dirac indices still link each beam spinor to a sample field. We separate the beam and sample factors with a Fierz identity, using the matrix $\gamma^5\equiv i\gamma^0\gamma^1\gamma^2\gamma^3$,
\begin{equation}
\begin{aligned}
\gamma^\mu_{ab}\,\gamma_{\mu,cd}&=\sum_A c_A\,\Gamma^A_{ad}\,\Gamma_{A,cb},\\
\Gamma^A&=(\mathbb{1}_4,\gamma^5,\gamma^\mu,\gamma^5\gamma^\mu),\\
c_A&=\Big(1,\,-1,\,-\dfrac{1}{2},\,-\dfrac{1}{2}\Big).
\end{aligned}
\label{eq:fierz-app}
\end{equation}

The label $A$ runs over the four displayed matrix types and, for the last two, their Lorentz components. The corresponding Lorentz index in $\Gamma_A$ is lowered with $\eta_{\mu\nu}$. Appendix~\ref{app:fierz} derives the identity. Substituting Eq.~\eqref{eq:fierz-app} after the coordinate integrals groups the two beam spinors together and the two sample fields together. Let $\widetilde{\fO}_A(\vq)$ be the spatial Fourier component of the sample operator $\Psisb\Gamma_A\Psis$, with the same convention as Eq.~\eqref{eq:Sab}. The resulting leading local amplitude is
\begin{equation}
\begin{aligned}
\widetilde{\fO}_A(\vq)&\equiv\int \text{d}^3\vx\,e^{i\vq\cdot\vx}
\Psisb(0,\vx)\Gamma_A\Psis(0,\vx),\\
\fM_X^{(0)}(n;s_i,s_f)
&=e^2\PiX\sum_Ac_A\bar{u}_f\Gamma^Au_i\bra n\widetilde{\fO}_A(\vq)\ket0.
\end{aligned}
\label{eq:MX-app}
\end{equation}

Substituting the leading sample field into the scalar, time-vector and spatial-axial products gives
\begin{equation}
\begin{aligned}
\Psisb\Psis&\to\rho,\\
\Psisb\gamma^0\Psis&\to\rho,\\
\Psisb\gamma^5\gamma^i\Psis&\to-s^i.
\end{aligned}
\label{eq:NRops-app}
\end{equation}
Expanding the electron spinor of Appendix~\ref{app:fierz} in $p/m$ gives two lower components of relative order $v_t$, while corrections to the two upper components start at order $v_t^2$. The three products in Eq.~\eqref{eq:NRops-app} do not mix upper and lower components, so their first corrections are of order $v_t^2$. The spatial current $\fJ^i_{\text{ss}}=\Psisb\gamma^i\Psis$, the pseudo-scalar $\Psisb\gamma^5\Psis$ and the time component $\Psisb\gamma^5\gamma^0\Psis$ of the axial product mix those components and begin at order $v_t$.

\subsection{Scalar--vector exchange interference}
\label{subsec:proof-exchange-trace}

The vector part uses the beam tensor $\fL^{\mu\nu}$ already calculated in Eq.~\eqref{eq:lepton-app}. The scalar part instead requires a mixed spin sum with the direct beam vertex. Using the spinor completeness relation in Appendix~\ref{app:spinsums} converts this spin sum to a trace,
\begin{align}
&\dfrac{1}{2}\sum_{s_i,s_f}(\bar{u}_f\gamma^\nu u_i)^*(\bar{u}_fu_i)\nonumber\\
&\quad=\dfrac{1}{2}\Tr\!\left[\gamma^\nu(\slashed{\fK}_f+m)(\slashed{\fK}_i+m)\right]\nonumber\\
&\quad=2m(\fK_i+\fK_f)^\nu.
\label{eq:trace1-app}
\end{align}
The corresponding sample product is the density--current spectrum $\fS_{JJ}^{0\rho}$, within the same replacement of $\fJ^0_{\text{ss}}$ by $\rho$. Its contraction with $(\fK_i+\fK_f)^\nu\eta_{\nu\rho}$ uses the continuity relation of Eq.~\eqref{eq:LT} and $\vq\cdot(\vk_i+\vk_f)=k_i^2-k_f^2=\omega(E_i+E_f)$. The result is $-(E_i+E_f)(\fQ^2/q^2)R_L$. The vector contraction is $D_LR_L+D_TR_T$ from Eq.~\eqref{eq:DDT-app}.

\subsection{Temporal and spatial vertex cancellation}
\label{subsec:proof-exchange-sign}

To obtain the sample-density coefficient in Eq.~\eqref{eq:MXlead} from the photon vertices, evaluate the beam vertices at equal momenta $\fK_i=\fK_f=\fK=(E,k\hat{\vz})$ and use the leading rest spinor of the sample reduction in Eq.~\eqref{eq:sample-field}. Its four-momentum is $\fP_0=(m,\mathbf0)$. Let $\xi$ and $\eta$ be unit-normalized two-component Pauli spinors for the beam and sample electron, respectively. Inserting the Dirac spinors of Appendix~\ref{app:fierz} into the transition vertices gives
\begin{equation}
\begin{aligned}
\bar u(\fK,\xi)\gamma^0u(\fP_0,\eta)
&=\sqrt{2m(E+m)}\,\xi^\dagger\eta,\\
\bar u(\fK,\xi)\gamma^iu(\fP_0,\eta)
&=\sqrt{2m(E+m)}\,r\,\xi^\dagger\sigma^z\sigma^i\eta,\\
r&\equiv\frac{k}{E+m}=\sqrt{\frac{E-m}{E+m}}.
\end{aligned}
\label{eq:exchange-transition-vertices}
\end{equation}
The coefficient $r$ compares each spatial spin-matrix factor with the temporal one. The actual vertices also depend on the displayed spin overlaps. In their product below, $\xi_i$ and $\xi_f$ denote the incident and outgoing beam Pauli spinors.

At the reverse vertex, interchanging the beam and rest spinors reverses the spatial Pauli-matrix order to $\sigma^i\sigma^z$. Summing over an orthonormal pair of sample spin basis vectors uses $\sum_\eta\eta\eta^\dagger=\mathbb{1}_2$. For each fixed $i$, with no sum over that index, the spatial product contains $\sigma^z\sigma^i\sigma^i\sigma^z=\mathbb{1}_2$ and has coefficient $r^2$ relative to the temporal product. The metric subtracts these three spatial contributions, so the summed vertex product is
\begin{equation}
\begin{aligned}
&\sum_\eta\eta_{\mu\nu}
\big[\bar u(\fK,\xi_f)\gamma^\mu u(\fP_0,\eta)\big]\big[\bar u(\fP_0,\eta)\gamma^\nu u(\fK,\xi_i)\big]\\
&\quad=2m(E+m)(1-3r^2)\xi_f^\dagger\xi_i\\
&\quad=4m(2m-E)\xi_f^\dagger\xi_i.
\end{aligned}
\label{eq:4m2mE-app}
\end{equation}

To extract the density coefficient in Eq.~\eqref{eq:MXlead}, regard the two transition vertices as a two-dimensional matrix in sample-spin space. The sum over $\eta$ in Eq.~\eqref{eq:4m2mE-app} is its trace. The two sample fields in Eq.~\eqref{eq:sample-field} supply a factor $1/(2m)$. The identity component of this vertex matrix multiplies $\rho$, with a coefficient equal to half the trace because the identity has trace two and the Pauli matrices are traceless. Dividing the coefficient in Eq.~\eqref{eq:4m2mE-app} by $4m$ therefore gives $(2m-E)\xi_f^\dagger\xi_i$. This agrees with $\bar u_f(\mathbb{1}_4-\gamma^0/2)u_i$ in the forward limit of Eq.~\eqref{eq:MXlead}.

The cancellation occurs at $r^2=1/3$, equivalent to $E=2m$. With $E=m+T$, $r^2=T/(T+2m)$, so the summed spatial contribution is $-3T/(T+2m)$ and the cancellation occurs at $T=m$.

\subsection{Axial beam vertices}
\label{subsec:proof-exchange-spin}

To evaluate the beam vertex, insert the Dirac spinors of Appendix~\ref{app:fierz}, with the Pauli spinors $\xi_i$ and $\xi_f$ expressed in common rest-frame axes. Writing $\bar u_f=u_f^\dagger\gamma^0$, the product $\gamma^0\gamma^5\gamma^j$ has $-\sigma^j$ in both diagonal blocks. Multiplying the upper-component and lower-component pairs gives, respectively, the first and momentum-dependent terms in
\begin{equation}
\begin{aligned}
&\bar u_f\gamma^5\gamma^ju_i\\
&\quad=-\sqrt{(E_f+m)(E_i+m)}\,\xi_f^\dagger\\
&\qquad\left[\sigma^j+\frac{(\vsigma\cdot\vk_f)\sigma^j(\vsigma\cdot\vk_i)}{(E_f+m)(E_i+m)}\right]\xi_i.
\end{aligned}
\label{eq:axial-vertex-general}
\end{equation}
This expression keeps the full beam momentum and energy dependence.

For the forward vertex values, set $\vk_i=\vk_f=k\hat{\vz}$ and $E_i=E_f=E=m+T$. For a fixed transverse index $j=x$ or $y$, the Pauli product obeys $\sigma^z\sigma^j\sigma^z=-\sigma^j$, while the longitudinal product is $\sigma^z$. Applying these identities to Eq.~\eqref{eq:axial-vertex-general}, with $k^2=E^2-m^2$, gives
\begin{equation}
\bar u_f\gamma^5\gamma^ju_i=
\begin{cases}
-2m\,\xi_f^\dagger\sigma^j\xi_i,&j=x,y,\\
-2E\,\xi_f^\dagger\sigma^z\xi_i,&j=z.
\end{cases}
\label{eq:axialvertex-app}
\end{equation}

\subsection{Direct positron amplitude and spin sum}
\label{subsec:proof-positron-direct}

For positrons, $\Psibb$ removes the incident particle and $\Psib$ creates the outgoing one. The corresponding field matrix elements have the same incident and outgoing phases as Eq.~\eqref{eq:beam-field-matrix}, with the fields interchanged. In the beam current $\fJ^\mu_{\text{bb}}=\Psibb\gamma^\mu\Psib$, the positron annihilation operator therefore precedes the creation operator. Interchanging them and removing the beam-vacuum contribution gives the scattering current matrix element
\begin{equation}
\begin{aligned}
&\bra{e^+(\fK_f,s_f)}\fJ^\mu_{\text{bb}}(\fX)\ket{e^+(\fK_i,s_i)}_{\rm conn}\\
&\quad=-\bar v_i\gamma^\mu v_f\,e^{-i\eta_{\rho\sigma}\fQ^\rho \fX^\sigma}.
\end{aligned}
\label{eq:positron-beam-current}
\end{equation}
The subscript denotes subtraction of the beam-vacuum contribution. Since the phase is unchanged, the coordinate integrals and energy-conservation factor are those of Sec.~\ref{subsec:proof-direct-fields}. Substituting this current into Eq.~\eqref{eq:direct-transition} and removing that energy factor gives the reduced positron amplitude at fixed beam momenta,
\begin{equation}
\begin{aligned}
\fM_D^{(+)}(n;s_i,s_f)&=+e^2\widetilde{\Pi}(\fQ)\,\bar v_i\gamma^\mu v_f\bra n\widetilde{\fJ}_{\text{ss},\mu}(\vq)\ket0.
\end{aligned}
\label{eq:positron-direct-amplitude}
\end{equation}

For the positron beam-spin sum, the counterpart of Eq.~\eqref{eq:beam-spinor-completeness} is
\begin{equation}
\sum_s v^s(\fK)\bar v^s(\fK)=\slashed{\fK}-m.
\label{eq:positron-completeness}
\end{equation}
Inserting this relation for each spin sum gives the positron beam tensor,
\begin{equation}
\begin{aligned}
\fL_+^{\mu\nu}&\equiv\dfrac{1}{2}\sum_{s_i,s_f}(\bar v_i\gamma^\mu v_f)(\bar v_i\gamma^\nu v_f)^*\\
&=\dfrac{1}{2}\Tr[\gamma^\mu(\slashed{\fK}_f-m)\gamma^\nu(\slashed{\fK}_i-m)]\\
&=\fL^{\mu\nu}.
\end{aligned}
\label{eq:positron-beam-tensor}
\end{equation}
Terms linear in $m$ vanish because they contain traces of three Dirac matrices. The four-matrix trace is symmetric under $\fK_i\leftrightarrow\fK_f$, and the quadratic mass term has the same sign as for electrons. These properties give the last equality with Eq.~\eqref{eq:lepton-app}.

\subsection{General annihilation amplitude}
\label{subsec:proof-annihilation-amplitude}

We now evaluate $\SX$ between the positron beam states of Sec.~\ref{subsec:positron-direct} and the same sample states. At $\fX$, $\Psibb$ removes the incident positron and $\Psis$ removes a sample electron. At $\fY$, $\Psib$ and $\Psisb$ create the outgoing positron and sample electron. Grouping the beam fields and placing the sample creation field first requires the three fermion interchanges described in Sec.~\ref{subsec:proof-exchange-amplitude}. The connected positron beam matrix element contributes another minus sign, as in Eq.~\eqref{eq:positron-beam-current}. These signs cancel, leaving the prefactor $-ie^2$ from $i\SX$.

Use $\fX=(t+\tau,\vx)$ and $\fY=(t,\bm{y})$ as in Sec.~\ref{subsec:proof-exchange-amplitude}. The common beam and sample phases again give $2\pi\delta(\omega-\omega_{n0})$ upon integration over $t$. The incident positron factor is now at $\fX$ and the outgoing one at $\fY$. Removing the energy factor with the convention of Eq.~\eqref{eq:MD} gives the reduced annihilation amplitude at fixed beam momenta and spins,
\begin{equation}
\begin{aligned}
&\fM_A^{(+)}(n;s_i,s_f)\\
&=-e^2\sum_{a,b,c,d}\eta_{\mu\nu}\bar v_{i,a}\gamma^\mu_{ab}\gamma^\nu_{cd}v_{f,d}\int\text{d}\tau\,\text{d}^3\vx\,\text{d}^3\bm{y}\,
\Pi(\tau,\vx-\bm{y})\\
&\quad e^{-iE_i\tau+i\vk_i\cdot\vx-i\vk_f\cdot\bm{y}}\bra n\text{T}\{\bar\Psi_{{\rm s},c}(0,\bm{y})
\Psi_{{\rm s},b}(\tau,\vx)\}\ket0.
\end{aligned}
\label{eq:annihilation-amplitude}
\end{equation}

To identify the kernel momentum, Fourier transform the complete chronological sample matrix element using the removal and addition variables $\fP$ and $\fP'$ of Sec.~\ref{subsec:proof-exchange-amplitude}. Combining its phases with the positron and photon phases fixes the kernel momentum to $-(\fK_i+\fP)$ at $\fX$ and $-(\fK_f+\fP')$ at $\fY$. Because the kernel is even, we use the sum as its argument,
\begin{equation}
\fQA=\fK_i+\fP=\fK_f+\fP'.
\label{eq:annihilation-momentum}
\end{equation}

\subsection{Local annihilation amplitude}
\label{subsec:proof-annihilation-local}

The constant kernel sets $\tau=0$ and $\bm{y}=\vx$ in Eq.~\eqref{eq:annihilation-amplitude}. The remaining phase is $e^{i\vq\cdot\vx}$, giving the same Fourier sample operators $\widetilde{\fO}_A(\vq)$ as in Eq.~\eqref{eq:MX-app}. Applying the Fierz identity in Eq.~\eqref{eq:fierz-app} to the two Dirac matrices then gives the leading local amplitude,
\begin{equation}
\begin{aligned}
\fM_A^{(+,0)}(n;s_i,s_f)&=-e^2\PiA\sum_Ac_A\\
&\quad(\bar v_i\Gamma^A v_f)\bra n\widetilde{\fO}_A(\vq)\ket0.
\end{aligned}
\label{eq:MA-app}
\end{equation}

\subsection{Non-relativistic positron reduction}
\label{subsec:proof-annihilation-nonrel}

In the positron spinors of Appendix~\ref{app:fierz}, the upper two components are of relative order $v$, while corrections to the lower components start at order $v^2$. Substituting these components into the scalar, time-vector and spatial-axial vertices gives
\begin{equation}
\begin{aligned}
\bar v_iv_f&=-2m\eta_i^\dagger\eta_f+O(mv^2),\\
\bar v_i\gamma^0v_f&=2m\eta_i^\dagger\eta_f+O(mv^2),\\
\bar v_i\gamma^5\gamma^jv_f&=-2m\eta_i^\dagger\sigma^j\eta_f+O(mv^2).
\end{aligned}
\label{eq:positron-nonrel-vertices}
\end{equation}
Here $\eta_i,\eta_f$ are the unit two-spinors appearing in $v_i,v_f$.

To express the result in terms of physical positron spin, use unit Pauli spinors $\xi_i,\xi_f$ with $\eta=i\sigma^2\xi^*$. Substituting this relation and using $\sigma^2(\sigma^j)^\top\sigma^2=-\sigma^j$ gives
\begin{equation}
\begin{aligned}
\eta_i^\dagger\eta_f&=\xi_f^\dagger\xi_i,\\
\eta_i^\dagger\sigma^j\eta_f&=-\xi_f^\dagger\sigma^j\xi_i.
\end{aligned}
\label{eq:positron-physical-spin}
\end{equation}
Expanding Eq.~\eqref{eq:annihilation-kernel} in $T/m=O(v^2)$ gives $\PiA=-1/(4m^2)[1+O(v^2)]$. In Eq.~\eqref{eq:MA-app}, insert this coefficient, the vertices above and the sample reduction in Eq.~\eqref{eq:NRops-app}. Grouping the scalar and temporal-vector density terms gives
Eq.~\eqref{eq:annihilation-nonrel-amplitude}.

\subsection{Free-pair threshold and contact potential}
\label{subsec:proof-annihilation-threshold}

For a separate two-particle normalization check, take a free positron--electron pair and retain the part in which both particles participate. Use covariantly normalized target spinors $u_i,u_f$ with unit Pauli spinors $\chi_i,\chi_f$. After the free spatial momentum-conservation delta function is removed, the matrix element of the sample operator in Eq.~\eqref{eq:MA-app} becomes $\bar u_f\Gamma_Au_i$. Take the threshold limit in the pair rest frame, with all external spatial momenta tending to zero. Then $u=\sqrt{2m}(\chi,0)^\top$, so $\widetilde{\rho}_{n0}$ and $\widetilde{\vs}_{n0}$ in Eq.~\eqref{eq:annihilation-nonrel-amplitude} are replaced by $2m\chi_f^\dagger\chi_i$ and $2m\chi_f^\dagger\vsigma\chi_i$.

The resulting threshold amplitude is a $4$ by $4$ matrix $\mathbb{M}_A$ on the two physical spin spaces. Write their Pauli operators as $\vsigma_{e^+}$ and $\vsigma_{e^-}$, with total spin $\vS=(\vsigma_{e^+}+\vsigma_{e^-})/2$. The operator $\vS\cdot\vS$ has eigenvalue $0$ in the singlet state and $2$ in the three triplet states. Expanding $\vS\cdot\vS$ gives $\vsigma_{e^+}\cdot\vsigma_{e^-}=2\vS\cdot\vS-3\mathbb{1}_4$, with eigenvalues $-3$ and $+1$ in the singlet and triplet, respectively. Adding $3$ and dividing by $4$ gives the triplet projector $P_{S=1}$. The free-target substitution above therefore yields
\begin{equation}
\begin{aligned}
P_{S=1}&=\frac{3\mathbb{1}_4+\vsigma_{e^+}\cdot\vsigma_{e^-}}4,\\
\mathbb{M}_A&=-\frac{e^2}{2}
\big[3\mathbb{1}_4+\vsigma_{e^+}\cdot\vsigma_{e^-}\big]\\
&=-2e^2P_{S=1}.
\end{aligned}
\label{eq:annihilation-triplet-amplitude}
\end{equation}

Let $V_{\rm ann}(\bm{r})$ be the two-body annihilation potential at positron--electron relative coordinate $\bm{r}$, and $\widetilde V_{\rm ann}(\vq)$ its spatial Fourier transform. Four external-state factors $\sqrt{2m}$ convert the first-order potential amplitude $-i\widetilde V_{\rm ann}(\vq)$ into $i\mathbb{M}_A$. Thus $\mathbb{M}_A=-4m^2\widetilde V_{\rm ann}(\vq)$. Retaining this leading term, independent of momentum, and taking its inverse Fourier transform gives a contact potential. Using $e^2=4\pi\alpha$ gives
\begin{equation}
V_{\rm ann}(\bm{r})=\frac{2\pi\alpha}{m^2}
P_{S=1}\delta^{(3)}(\bm{r}).
\label{eq:annihilation-potential}
\end{equation}
It is the standard leading annihilation contribution to the potential of positronium, the electron--positron bound state~\cite{blp}. 

\subsection{Scalar--vector annihilation interference}
\label{subsec:proof-annihilation-trace}

The vector part uses the positron tensor in Eq.~\eqref{eq:positron-beam-tensor}, equal to the electron tensor. For the scalar part, conjugating the direct vertex and inserting the positron completeness relation in Eq.~\eqref{eq:positron-completeness} gives
\begin{equation}
\begin{aligned}
&\dfrac{1}{2}\sum_{s_i,s_f}(\bar v_i\gamma^\nu v_f)^*(\bar v_iv_f)\\
&\quad=\dfrac{1}{2}\Tr[\gamma^\nu(\slashed{\fK}_i-m)(\slashed{\fK}_f-m)]\\
&\quad=-2m(\fK_i+\fK_f)^\nu.
\end{aligned}
\label{eq:positron-scalar-vector-trace}
\end{equation}
This reverses the scalar trace of Eq.~\eqref{eq:trace1-app}. Insert this trace and the unchanged vector tensor into Eq.~\eqref{eq:IAratio}, using the same density--current contraction from sample continuity as in Sec.~\ref{subsec:exchange-density}. For $q>|\omega|$, dividing by the common direct weight and substituting Eq.~\eqref{eq:annihilation-kernel} gives
Eq.~\eqref{eq:IAratio}.

\section{Local exchange and annihilation kernels}
\label{app:exchange-kernel}

The local amplitudes in Secs.~\ref{subsec:exchange-operators} and \ref{subsec:annihilation-amplitude} neglect the sample momentum and residual-energy dependence of the photon coefficients. We derive the coefficients and bound these omissions, starting with the exchange amplitude in Eq.~\eqref{eq:exchange-amplitude}. Keep its beam momenta, spins and inelastic sample transition $0\to n$ fixed. Insert complete sets of sample energy eigenstates $\ket{r,+}$ and $\ket{r,-}$ with one more and one fewer electron, respectively, than $\ket0$. Their energies are $E_r^+$ and $E_r^-$. For Dirac indices $b,c$, define spatial Fourier weights and residual energies by
\begin{equation}
\begin{aligned}
\widetilde A_r&=\int \text{d}^3\vx\,\text{d}^3\bm{y}\,
e^{i\vp'\cdot\bm{y}-i\vp\cdot\vx}\\
&\quad\,\bra n\Psi_{{\rm s},b}(0,\vx)\ket{r,+}\bra{r,+}\bar\Psi_{{\rm s},c}(0,\bm{y})\ket0,\\
\widetilde C_r&=\int \text{d}^3\vx\,\text{d}^3\bm{y}\,
e^{i\vp'\cdot\bm{y}-i\vp\cdot\vx}\\
&\quad\,\bra n\bar\Psi_{{\rm s},c}(0,\bm{y})\ket{r,-}\bra{r,-}\Psi_{{\rm s},b}(0,\vx)\ket0,\\
\epsilon_{+,r}&=E_r^+-E_0-m,\qquad
\epsilon_{-,r}=E_n-E_r^--m.
\end{aligned}
\label{eq:exchange-intermediate-weights}
\end{equation}
The weights have fixed arguments $(\vp',\vp)$ with $\vp=\vp'-\vq$, suppressed below. The second energy refers to adding an electron to the intermediate state to produce the final sample.

Write $q_{\rm X}=|\vk_i-\vp'|$. Using the kernel in Eq.~\eqref{eq:photon-propagator} and the Fourier convention in Eq.~\eqref{eq:photon-fourier}, closing the frequency contour below for $\tau>0$ and above for $\tau<0$ gives the photon factor $i e^{-iq_{\rm X}|\tau|}/(2q_{\rm X})$. Stationary evolution supplies the sample phase $e^{i(E_r^--E_0)\tau}$ in the removal branch $\tau<0$ and $e^{i(E_n-E_r^+)\tau}$ in the addition branch $\tau>0$. Multiplying by the beam phase $e^{iE_f\tau}$ and using $E_n-E_0=\omega=E_i-E_f$ gives $e^{i(T-\epsilon_{-,r})\tau}$ and $e^{i(T-\epsilon_{+,r})\tau}$, respectively. The second branch has the minus sign from interchanging the sample fields. Multiplying by the photon factor and integrating the two half-lines gives the time factor at the fixed spatial arguments
\begin{equation}
\begin{aligned}
F(q_{\rm X})&=\sum_r\frac{\widetilde C_r}{2q_{\rm X}(q_{\rm X}+T-\epsilon_{-,r}-i0)}\\
&\quad-\sum_r\frac{\widetilde A_r}{2q_{\rm X}(q_{\rm X}-T+\epsilon_{+,r}-i0)}.
\end{aligned}
\label{eq:exchange-time-factors}
\end{equation}
The momentum integral $\int \text{d}^3\vp'\,F(q_{\rm X})/(2\pi)^3$, with $\vp=\vp'-\vq$, replaces the time and spatial integrals in Eq.~\eqref{eq:exchange-amplitude}. Its beam and Dirac factors remain unchanged.

To combine the chronological branches, we use the full electron field's canonical equal-time relation
\begin{equation}
\{\Psi_b(0,\vx),\bar\Psi_c(0,\bm{y})\}
=\gamma^0_{bc}\delta^{(3)}(\vx-\bm{y})\mathbb{1}_{\rm e}.
\label{eq:electron-anticommutator}
\end{equation}
Here $\{U,V\}=UV+VU$, and $\mathbb{1}_{\rm e}$ is the identity on electron-state space. A fixed linear sample-mode projection changes the numerical kernel in Eq.~\eqref{eq:electron-anticommutator} but preserves the state-space identity. Completeness and $\langle n|0\rangle=0$ consequently give $\sum_r\widetilde A_r=-\sum_r\widetilde C_r$. In the retained region $q_{\rm X}>T$, require $q_{\rm X}+T-\epsilon_{-,r}\geq\zeta(q_{\rm X}+T)$ and $q_{\rm X}-T+\epsilon_{+,r}\geq\zeta(q_{\rm X}-T)$ with fixed $\zeta>0$. These bounds keep the exact denominators separated from zero. Neglecting residual energies only in the coefficients combines the branches~into
\begin{equation}
\begin{aligned}
F_0(q_{\rm X})&=\frac{\sum_r\widetilde C_r}{2q_{\rm X}}
\left(\frac1{q_{\rm X}+T}+\frac1{q_{\rm X}-T}\right)\\
&=\frac{\sum_r\widetilde C_r}{q_{\rm X}^2-T^2}.
\end{aligned}
\label{eq:exchange-leading-time-factor}
\end{equation}
Evaluating only this common coefficient at $\vp'=\mathbf0$ gives Eq.~\eqref{eq:PiX}.
Take this constant coefficient outside the momentum integral. Completeness in the removal sector closes the state sum within $\sum_r\widetilde C_r$ to $\bra n\bar\Psi_{{\rm s},c}(0,\bm{y})\Psi_{{\rm s},b}(0,\vx)\ket0$. With $\vp=\vp'-\vq$, the Fourier integral over $\vp'$ sets $\bm{y}=\vx$ and leaves the phase $e^{i\vq\cdot\vx}$. The remaining spatial integral is the local Fourier bilinear of $\bar\Psi_{{\rm s},c}\Psi_{{\rm s},b}$ used in Sec.~\ref{subsec:exchange-operators}.

To bound the residual-energy omission, subtract $1/(q_{\rm X}\pm T)$ from the corresponding shifted reciprocal in Eq.~\eqref{eq:exchange-time-factors}. Each absolute difference is the residual-energy magnitude divided by the product of the exact and unshifted denominators. Multiplying by $1/(2q_{\rm X})$ and applying the lower bounds and triangle inequality to the weighted sum gives
\begin{equation}
\begin{aligned}
W&=\sum_r(|\widetilde A_r|+|\widetilde C_r|),\\
M_-&=\sum_r|\widetilde C_r||\epsilon_{-,r}|,\\
M_+&=\sum_r|\widetilde A_r||\epsilon_{+,r}|,\\
|F-F_0|&\leq\frac{1}{2q_{\rm X}\zeta}
\left[\frac{M_-}{(q_{\rm X}+T)^2}+\frac{M_+}{(q_{\rm X}-T)^2}\right].
\end{aligned}
\label{eq:exchange-energy-bound}
\end{equation}
Here $W$ is the total absolute transition weight, and $M_\pm$ are its first residual-energy moments for the two branches. For $W>0$, divide the bound by the absolute-weight scale $W/(q_{\rm X}^2-T^2)$. Using $q_{\rm X}+T\geq q_{\rm X}-T$ and $(q_{\rm X}+T)/(2q_{\rm X})<1$ bounds this ratio by $(M_-+M_+)/[\zeta W(q_{\rm X}-T)]$. The error is small on this scale when $(M_-+M_+)/W\ll q_{\rm X}-T$ at fixed $\zeta$.

Let $p_{\max}\ll k_i-T$ bound the retained addition-field momentum magnitude. The inequality $|p'^2-2\vk_i\cdot\vp'|\leq p_{\max}^2+2k_ip_{\max}$ gives a relative coefficient correction $O(p_{\max}/(k_i-T))$ on expanding $1/(2mT+p'^2-2\vk_i\cdot\vp')$. We assume finite momentum integrals of the absolute weights and moments and negligible contributions outside this region. Integrating the energy bound and the momentum coefficient bound weighted by $W$, with $e^2$ and the absolute beam factors, bounds the omitted amplitude. No relative bound is implied when the leading matrix element vanishes.

For annihilation, the sample weights in Eq.~\eqref{eq:exchange-intermediate-weights} are unchanged, with $\vp'=\vp+\vq$. Changing the beam time factor from $e^{iE_f\tau}$ to $e^{-iE_i\tau}$ changes the phases preceding Eq.~\eqref{eq:exchange-time-factors} to $e^{-i(E_f+m+\epsilon_{\pm,r})\tau}$. Using $E_f=E_i-\omega$, define the shifted residual energies $\epsilon_{A,\pm,r}=\epsilon_{\pm,r}-\omega$. The phases then become $e^{-i(U+\epsilon_{A,\pm,r})\tau}$ with $U=E_i+m$. The spatial photon magnitude is $q_{\rm A}=|\vk_i+\vp|$. Substituting $T\to-U$ and $\epsilon_{\pm,r}\to\epsilon_{A,\pm,r}$ in the half-line integrals above gives the annihilation time factor $F_A$.

For its zero-residual-energy limit $F_{A,0}$ and energy bound, take a retained region $p\leq p_{\max}\ll g_A=U-k_i$, ensuring $U-q_{\rm A}>0$. Require $q_{\rm A}-U-\epsilon_{A,-,r}\leq-\zeta(U-q_{\rm A})$ and $q_{\rm A}+U+\epsilon_{A,+,r}\geq\zeta(U+q_{\rm A})$ with fixed $\zeta>0$, so the exact denominators retain opposite signs. Apply equal-time closure as above and define the absolute first moments $M_{A,\pm}$ with the shifted energies. The same reciprocal subtraction and lower bounds give
\begin{equation}
\begin{aligned}
F_A&=\sum_r\frac{\widetilde C_r}{2q_{\rm A}(q_{\rm A}-U-\epsilon_{A,-,r}-i0)}\\
&\quad-\sum_r\frac{\widetilde A_r}{2q_{\rm A}(q_{\rm A}+U+\epsilon_{A,+,r}-i0)},\\
F_{A,0}&=\frac{\sum_r\widetilde C_r}{q_{\rm A}^2-U^2},\\
|F_A-F_{A,0}|&\leq\frac1{2q_{\rm A}\zeta}
\left[\frac{M_{A,-}}{(U-q_{\rm A})^2}
+\frac{M_{A,+}}{(U+q_{\rm A})^2}\right].
\end{aligned}
\label{eq:annihilation-energy-bound}
\end{equation}
Here,
\begin{equation*}
\begin{aligned}
M_{A,-}&=\sum_r|\widetilde C_r||\epsilon_{A,-,r}|,\\
M_{A,+}&=\sum_r|\widetilde A_r||\epsilon_{A,+,r}|,
\end{aligned}
\end{equation*}
$F_A$, $F_{A,0}$ and the moments have the fixed spatial arguments $(\vp',\vp)$ suppressed. We require the full bound, including its $1/q_{\rm A}$ factor, to have convergent momentum integrals. Integrated with $e^2$ and the absolute beam factors in Eq.~\eqref{eq:annihilation-amplitude}, it must be small compared with the corresponding integral of $W/(U^2-q_{\rm A}^2)$. This states an absolute-weight control and gives no relative amplitude bound at a cancellation.

The remaining momentum replacement uses $q_{\rm A}^2-U^2=-2m(E_i+m)+2\vk_i\cdot\vp+p^2$. The relative denominator shift is bounded by $p_{\max}/g_A+(p_{\max}/g_A)^2$, giving the $O(p_{\max}/g_A)$ coefficient correction stated in the body. Spatial tails are again assumed negligible.

For the separate free-electron check used in Sec.~\ref{subsec:exchange-kernel}, set $\fP'^0=\sqrt{m^2+p'^2}$. With $E=E_i=m+T$, the beam relation $E^2-k_i^2=m^2$ and the square-root expansion for $p'/m\ll1$ give
\begin{equation}
\begin{aligned}
-\fQX^2
&=|\vk_i-\vp'|^2-(E-\fP'^0)^2\\
&=2E\sqrt{m^2+p'^2}-2m^2-2\vk_i\cdot\vp'\\
&=2mT+\frac{E}{m}p'^2-2\vk_i\cdot\vp'
+O\!\left(\frac{Ep'^4}{m^3}\right).
\end{aligned}
\label{eq:Q2-app}
\end{equation}

\subsection{Fourier localization of the leading coefficient}
\label{subsec:proof-exchange-localization}

Using the Fourier convention in Eq.~\eqref{eq:photon-fourier}, a constant kernel gives $\Pi(\fX-\fY)\to\PiX\delta^{(4)}(\fX-\fY)$. In Eq.~\eqref{eq:exchange-amplitude}, the delta function sets $\tau=0$ and $\bm{y}=\vx$. The remaining beam phase is $e^{i\vq\cdot\vx}$, so the sample contribution is a spatial Fourier transform of fields evaluated at one point.

\subsection{Free-electron momentum benchmark}
\label{subsec:proof-exchange-kernel-benchmark}

We now examine the momentum dependence omitted from the leading kernel. Use the free-electron comparison in Eq.~\eqref{eq:Q2-app} as a scalar benchmark, with $E=E_i=m+T$. Let $p_{\max}$ bound the retained sample momentum magnitude and define the dimensionless expansion parameter $\epsilon_p=p_{\max}/(k_i-T)\ll1$, using the same positive gap as in Sec.~\ref{subsec:exchange-operators}. The normalized linear denominator correction is bounded by $k_ip_{\max}/(mT)\leq2\epsilon_p$, and the quadratic correction by $Ep_{\max}^2/(2m^2T)\leq\epsilon_p^2$. These inequalities follow from $k_i^2-T^2=2mT$ and $(k_i+T)^2=2T(E+k_i)$. Since $0<k_i-T<m$, the same bound also ensures $p_{\max}/m\ll1$ as required in Eq.~\eqref{eq:Q2-app}.

Define the relative kernel correction by $\widetilde{\Pi}(\fQX)=\PiX[1+\delta(\vp')]$. Expanding the reciprocal of the denominator in Eq.~\eqref{eq:Q2-app} through quadratic order in $\vp'$ gives
\begin{equation}
\begin{aligned}
\delta(\vp')&=\frac{\vk_i\cdot\vp'}{mT}
-\frac{E p'^2}{2m^2T}+\frac{(\vk_i\cdot\vp')^2}{m^2T^2}
+O(\epsilon_p^3).
\end{aligned}
\label{eq:exchange-kernel-expansion}
\end{equation}
This is the main-text expansion of Eq.~\eqref{eq:exchange-kernel-expansion-main}. The first term has one power of the creation-field momentum $\vp'$. In the Fourier phase of Sec.~\ref{subsec:proof-exchange-amplitude}, multiplication by $\vp'$ corresponds to $i\vnabla$ acting on $\bar\Psi_{\rm s}$. Thus its contribution involves the transition matrix elements obtained by replacing $\bar\Psi_{\rm s}(0,\vx)$ by $i\vnabla\bar\Psi_{\rm s}(0,\vx)$ in Eq.~\eqref{eq:MX-app}. The measured transfer selects a direction, so isotropy of the ground state alone does not force these derivative transition elements to vanish.

For an illustrative scalar average, choose a nonnegative radial weight $w(p')$ with support $p'\leq p_{\max}$ and $\int \text{d}^3\vp'\,w(p')=1$. For a scalar function $f$, define $\langle f\rangle_{\rm iso}=\int \text{d}^3\vp'\,w(p')f(\vp')$. For Cartesian components, this average obeys $\langle p'_i\rangle_{\rm iso}=0$ and $\langle p'_ip'_j\rangle_{\rm iso}=\delta_{ij}\langle p'^2\rangle_{\rm iso}/3$. Invariance of the radial weight under $\vp'\to-\vp'$ removes all odd terms. Applying the second-moment identity to Eq.~\eqref{eq:exchange-kernel-expansion} gives the quadratic correction $\delta_X$ to the averaged free-electron coefficient,
\begin{equation}
\begin{aligned}
\frac{\langle\widetilde{\Pi}(\fQX)\rangle_{\rm iso}}{\PiX}
&=1+\delta_X+O(\epsilon_p^4),\\
\delta_X&=\left[-\frac{E}{2m^2T}+\frac{k_i^2}{3m^2T^2}\right]
\langle p'^2\rangle_{\rm iso}\\
&=\frac{2m-E}{3m}\,
\frac{\langle p'^2\rangle_{\rm iso}}{2mT}.
\end{aligned}
\label{eq:deltaX-app}
\end{equation}
The last equality uses $k_i^2=T(E+m)$.

\section{Pair-transfer and screened-potential derivations}
\label{app:higher-order}

The two models below supply the detailed calculations used in Secs.~\ref{sec:pair} and \ref{sec:beyond}. Their distinct approximations and retained intermediate states are stated separately.

\subsection{Two opposite pair transfers}
\label{subsec:proof-pair}

In $\SP=\SP^++\SP^-$, $\SP^+$ denotes the term that creates two beam electrons and $\SP^-$ the term with the conjugated current product. Each term retains the same Feynman kernel. Expanding the chronologically ordered exponential to second order and keeping the two opposite transfers gives
\begin{equation}
\begin{aligned}
&i\fM_{PP}(n;s_i,s_f)\,2\pi\delta(\omega-\omega_{n0})\\
&\quad=-\frac12\bra{\fK_f,s_f;n}\text{T}\Big\{\\
&\qquad 
\SP^+\SP^-
+\SP^-\SP^+\Big\}
\ket{\fK_i,s_i;0}_{\rm conn}.
\end{aligned}
\label{eq:pair-transition}
\end{equation}
Here the beam momenta are fixed as in Sec.~\ref{subsec:direct-amplitude}, and $\mathrm{conn}$ retains the terms connecting the incident beam electron to the outgoing electron through the sample. The energy-conservation factor is removed by the same convention as in Eq.~\eqref{eq:MD}. 

For the instantaneous Coulomb model of Sec.~\ref{sec:pair}, use unit-normalized states in a volume $\mathcal V$. Let $V_+$ be the part of the Coulomb Hamiltonian that transfers two sample electrons into beam modes, and let $V_-=V_+^\dagger$ reverse the transfer. These are time-local Hamiltonian operators, rather than the two nonlocal actions above. With only one incident beam electron, $V_-$ cannot act first. Propagation between the transfers is described by the resolvent $(\mathcal E_i-H_0+i0)^{-1}$, where $H_0$ is the uncoupled beam and sample Hamiltonian. Inserting its intermediate eigenstates replaces $H_0$ by their energies and gives the second-order Hamiltonian matrix element
\begin{equation}
W_{PP}(n;s_i,s_f)
=\sum_I\frac{\bra{f}V_-\ket{I}\bra{I}V_+\ket{i}}
{\mathcal E_i-\mathcal E_I+i0}\bigg|_{\rm conn},
\label{eq:pair-spectral}
\end{equation}
where $\ket{i}$ and $\ket{f}$ have the external quantum numbers in Eq.~\eqref{eq:pair-transition}. The states $\ket{I}$ contain three beam electrons and two fewer sample electrons, and $\mathcal E_i$ and $\mathcal E_I$ are their total energies. Antisymmetrized Coulomb matrix elements include all electron permutations in this sum.

The extra two beam electrons supply the large energy in this denominator. Write their kinetic energies as $T_a$ and $T_b$, and let $\varepsilon_\nu$ be the energy of the intermediate sample state relative to the initial sample after subtracting the rest masses in each particle-number sector. The incident beam electron is still present after the first transfer, so its energy cancels in the difference, leaving
\begin{equation}
\Delta_I\equiv\mathcal E_I-\mathcal E_i
=T_a+T_b+\varepsilon_\nu.
\label{eq:pair-energy-gap}
\end{equation}
For the retained beam modes, $T_a,T_b=T+O(\Lambdac)$. If the contributing pair-removal states also satisfy $|\varepsilon_\nu|\ll T$, the denominator is $-2T$ with relative corrections of order $(\Lambdac+|\varepsilon_\nu|)/T$. This is an additional condition on the pair-removal spectrum, not a consequence of a small measured loss alone.

A plane-wave example displays the two Coulomb factors as well. Let $\vp_1,\vp_2,\vp_1'$ be sample momenta small compared with $k_i$. Consider the ordered path in which $(\vp_1,\vp_2)$ enter $(\vk',\vk_f)$ and $(\vk_i,\vk')$ return to $(\vp_1',\vp_2)$. Momentum conservation fixes $\vk'=\vp_1+\vp_2-\vk_f$, close to $-\vk_f$. Each Fourier Coulomb factor is $e^2/\mathcal V$ divided by the squared spatial photon momentum. For a spin-preserving contraction with positive product of the two matrix elements, insert these factors with photon momenta $\vk_f-\vp_1$ and $\vk_i-\vp_1'$ into Eq.~\eqref{eq:pair-spectral} to obtain
\begin{equation}
W_{PP}^{\rm path}
=-\frac{e^4}{\mathcal V^2}
\frac{1}{|\vk_f-\vp_1|^2|\vk_i-\vp_1'|^2\Delta_I}.
\label{eq:MPP-app}
\end{equation}
This matrix element has dimensions of energy. It is one ordered contribution to Eq.~\eqref{eq:pair-spectral}, not the complete reduced amplitude in Eq.~\eqref{eq:pair-transition}. Different contractions and sample transitions need not have the same sign.

An illustrative density estimate follows if the partner electron is an independent spectator and its occupation factors from the active-electron transition. Replace both photon momentum magnitudes in Eq.~\eqref{eq:MPP-app} by $k_i$, replace $\Delta_I$ by $2T$, and add $n\mathcal V$ identical path contributions, where $n$ is the occupied partner density. Relative to the direct Coulomb magnitude $W_D^{\rm model}=e^2/(\mathcal V q^2)$, these replacements give
the model ratio $R_{PP}^{\rm model}$ of Eq.~\eqref{eq:RPP}.

\subsection{Screened-potential second Born integral}
\label{subsec:proof-born}

Use the screened potential and amplitude normalization of Sec.~\ref{subsec:second-born}.

The Fourier potential is $\widetilde V(\bm\ell)=4\pi g/(\ell^2+\mu^2)$. One insertion gives $f_1=-m\widetilde V(\vq)/(2\pi)$. For two insertions, the beam propagates with intermediate momentum $\vp$ and kinetic energy $p^2/(2m)$. Inserting this free propagation between the two Fourier potentials gives the Born terms~\cite{joachain1974,byron1977}
\begin{align}
f_1&=-\frac{2mg}{q^2+\mu^2},\nonumber\\
f_2&=-\frac{m}{2\pi}\int\frac{\text{d}^3\vp}{(2\pi)^3}
\frac{\widetilde V(\vk_f-\vp)\widetilde V(\vp-\vk_i)}
{k^2/(2m)-p^2/(2m)+i0}.
\label{eq:screened-born-terms}
\end{align}
The $i0$ selects outgoing propagation. These are scalar potential-scattering amplitudes, distinct from the reduced many-body amplitudes in the preceding sections.

To evaluate $f_2$, replace the product of the two reciprocal potential denominators by the integral over $0\leq x\leq1$ of their weighted sum to the power $-2$. Weight the denominator centered at $\vk_i$ by $1-x$ and the one centered at $\vk_f$ by $x$. The sum is $(\vp-\vK(x))^2+a(x)^2$, where
\begin{equation}
\begin{aligned}
\vK(x)&=(1-x)\vk_i+x\vk_f,\\
a(x)&=\sqrt{\mu^2+x(1-x)q^2}.
\end{aligned}
\label{eq:screened-born-parameter}
\end{equation}
The remaining momentum integral can be obtained from the scalar convolution identity
\begin{equation}
\begin{aligned}
&\int\frac{\text{d}^3\vp}{(2\pi)^3}
\frac{1}{[(\vp-\vK(x))^2+a^2](p^2-k^2-i0)}\\
&\qquad=\frac{1}{4\pi K(x)}
\arctan\!\left(\frac{K(x)}{a-ik}\right).
\end{aligned}
\label{eq:screened-born-convolution}
\end{equation}
The outgoing $i0$ fixes the arctangent branch, with the value at $K(x)=0$ understood by continuity. Taking $-\partial/\partial a^2$ at fixed $\vK(x)$ and $k$ squares the potential denominator. With $a=a(x)$ after differentiation, this gives
\begin{equation}
\begin{aligned}
&\int\frac{\text{d}^3\vp}{(2\pi)^3}
\frac{1}{[(\vp-\vK(x))^2+a^2]^2(p^2-k^2-i0)}\\
&\qquad=\frac{1}{8\pi a[(a-ik)^2+K(x)^2]}.
\end{aligned}
\label{eq:screened-born-integral}
\end{equation}
Since $a^2+K(x)^2=k^2+\mu^2$, the square bracket reduces to $\mu^2-2ika$. The energy propagator in Eq.~\eqref{eq:screened-born-terms} is $-2m/(p^2-k^2-i0)$. Restoring its numerator $-2m$, the prefactor $-m/(2\pi)$ and the two potential factors $4\pi g$ yields the exact second-order expression
\begin{equation}
f_2=2m^2g^2\int_0^1\frac{\text{d}x}{a(x)[\mu^2-2ik a(x)]}.
\label{eq:screened-born-exact}
\end{equation}

For $\mu\ll k\ll m$, expansion of the denominator in Eq.~\eqref{eq:screened-born-exact} gives leading imaginary and real integrands proportional to $a^{-2}$ and $a^{-3}$, respectively. Integrating these terms over $x$ and dividing by $f_1$ defines dimensionless coefficient functions $c_I(s)$ and $c_R(s)$, where $s=q/\mu$. With relative remainders $\delta_I$ and $\delta_R$, the result is
\begin{equation}
\begin{aligned}
\frac{\Im f_2}{f_1}&=-\frac{mg}{k}c_I(s)[1+\delta_I],\\
\frac{\Re f_2}{f_1}&=-\frac{mg\mu}{k^2}c_R(s)[1+\delta_R],\\
c_I(s)&=\frac{1+s^2}{2}\int_0^1\frac{\text{d}x}{1+s^2x(1-x)}\\
&=\frac{2(1+s^2)\operatorname{arsinh}(s/2)}
{s\sqrt{s^2+4}},\\
c_R(s)&=\frac{1+s^2}{4}\int_0^1\frac{\text{d}x}{[1+s^2x(1-x)]^{3/2}}\\
&=\frac{1+s^2}{s^2+4}.
\end{aligned}
\label{eq:born2scaling-app}
\end{equation}
Here $c_I(0)=1/2$ by continuity. The relative remainders satisfy $-\mu^2/(4k^2+\mu^2)\leq\delta_I,\delta_R\leq0$. This bound follows by writing the real and imaginary integrands as their leading terms times $[1+\mu^4/(4k^2a^2)]^{-1}$ and using $a\geq\mu$. Both leading integration weights are positive, so the pointwise bounds also bound the relative integral remainders. Thus both remainders are $O(\mu^2/k^2)$.

\section{Magnon amplitudes, spin observables and counting statistics}
\label{app:magnon-derivation}

We retain the full derivation of the localized-spin transition, current and spin couplings, measured line, spin analysis and count estimate used in Sec.~\ref{sec:magnon}. The sample model is the fully polarized, equivalent-site model defined in Sec.~\ref{subsec:magnon-matrix}.

\subsection{Localized-spin transition and spectrum}
\label{subsec:proof-magnon-model}

For the localized-spin model of Sec.~\ref{subsec:magnon-matrix}, $\vs/2$ is the spin angular-momentum density, so
\begin{equation}
\vs(\vx)=2\sum_j\vS_jw(\vx-\vR_j),
\qquad \int \text{d}^3\vx\,w(\vx)=1.
\label{eq:magnon-spin-profile}
\end{equation}
Fourier transforming with the convention of Eq.~\eqref{eq:Sab} and translating each profile by $\vR_j$ gives
\begin{equation}
\begin{aligned}
\widetilde{\vs}(\vq)&=2F(\vq)\sum_j e^{i\vq\cdot\vR_j}\vS_j,\\
F(\vq)&\equiv\int \text{d}^3\vx\,e^{i\vq\cdot\vx}w(\vx).
\end{aligned}
\label{eq:magnon-spin-fourier}
\end{equation}

With the transverse axes of Sec.~\ref{subsec:magnon-matrix}, the operators $S_j^\pm=(\hat{\ve}_1\pm i\hat{\ve}_2)\cdot\vS_j$ raise or lower the spin projection along $\Mh$ by one unit. Maximal ground-state polarization gives $S_j^+\ket0=0$, and the spin commutator $[S_j^+,S_l^-]=2\delta_{jl}\Mh\cdot\vS_j$ therefore gives $\bra0S_j^+S_l^-\ket0=2S\delta_{jl}$. A circular spin wave~\cite{holstein1940} is the normalized superposition of these local spin-lowering states. We assume this superposition is an energy eigenmode with excitation energy $\omega_{\rm m}(\vq)$, and write its state as
\begin{equation}
\ket{\vq}=\frac{1}{\sqrt{2SN_{\rm m}}}
\sum_j e^{i\vq\cdot\vR_j}S_j^-\ket0.
\label{eq:magnon-state}
\end{equation}

Equation~\eqref{eq:magnon-state} gives $\bra{\vq}S_j^-\ket0=\sqrt{2S/N_{\rm m}}\,e^{-i\vq\cdot\vR_j}$. Resolve the transverse spin components as $(S_j^++S_j^-)/2$ and $(S_j^+-S_j^-)/(2i)$. The longitudinal component gives $S\ket0$, which has zero overlap with the magnon because their total spin projections differ. Only the transverse lowering parts contribute, and the phase above cancels the site phase in Eq.~\eqref{eq:magnon-spin-fourier}. Summing the $N_{\rm m}$ equal contributions gives
Eq.~\eqref{eq:magnonME}.

Positive energy loss $\omega>0$ excludes the elastic longitudinal term, leaving the one-magnon contribution to the spin spectral tensor. Multiply each transition component by the conjugate component in the same order as the current tensor in Eq.~\eqref{eq:current-spectral-tensor}, divide by $V$ and impose energy conservation. The spin construction is also given in Appendix~\ref{app:sample-response}. Let $\delta^\perp_{kl}\equiv\delta_{kl}-\Mh_k\Mh_l$ project onto the plane perpendicular to $\Mh$. The product $\epsilon_k\epsilon_l^*=\delta^\perp_{kl}-i\epsilon_{klm}\Mh_m$, with $\epsilon_{123}=1$, gives
\begin{equation}
\begin{aligned}
S_{ss}^{kl}(\vq,\omega)
&\equiv\frac1V\bra0\widetilde s^l(\vq)^\dagger\ket{\vq}
\bra{\vq}\widetilde s^k(\vq)\ket0\\
&\quad\delta(\omega-\omega_{\rm m}(\vq))\\
&=2Sn_{\rm m}|F(\vq)|^2
\big(\delta^\perp_{kl}-i\epsilon_{klm}\Mh_m\big)\\
&\quad\delta(\omega-\omega_{\rm m}(\vq)).
\end{aligned}
\label{eq:Skl-app}
\end{equation}
The charge density and the current from spin-independent orbital motion are proportional to the identity in the retained spin space. Their matrix elements between the orthogonal ground and one-magnon states therefore vanish.

\subsection{Sample bilinears and complete amplitude}
\label{subsec:proof-magnon-bilinears}

For the real, current-free orbitals of Sec.~\ref{subsec:magnon-amplitudes}, expand the sample spinor of Appendix~\ref{app:fierz} through first order in momentum and replace $\vp$ by $-i\vnabla$. The correction to Eq.~\eqref{eq:sample-field} then has lower Dirac components $-i\vsigma\cdot\vnabla\psi/(2m)$, after removing the common phase $e^{-imt}$. To this order, substituting these components into the spatial current and using $\sigma_i\sigma_j=\delta_{ij}\mathbb{1}_2+i\epsilon_{ijk}\sigma_k$ separates the current from orbital motion and the magnetization current, given by the curl of the spin density,
\begin{equation}
\begin{aligned}
\vJ_{\text{ss}}&=\vJ_{\rm orb}+\frac{\vnabla\times\vs}{2m},\\
\vJ_{\rm orb}&=\frac{\psi^\dagger\vnabla\psi-(\vnabla\psi^\dagger)\psi}{2mi}.
\end{aligned}
\label{eq:magnon-pauli-current}
\end{equation}
Here $\vJ_{\text{ss}}$ denotes the spatial part of $\fJ_{\text{ss}}$. The orbital term has zero magnon matrix element, as established above. For the same lower components, the pseudo-scalar is the difference of the upper--lower products, while the temporal axial bilinear is minus their sum. They give
\begin{equation}
\begin{aligned}
\Psisb\gamma^5\Psis&=-\frac{i}{2m}\vnabla\cdot\vs,\\
\Psisb\gamma^5\gamma^0\Psis
&=\frac{i}{2m}\big[\psi^\dagger\vsigma\cdot\vnabla\psi
-(\vnabla\psi^\dagger)\cdot\vsigma\psi\big].
\end{aligned}
\label{eq:magnon-pauli-odd}
\end{equation}
Within each fixed real orbital, the two derivatives in the temporal axial expression act on the same orbital wavefunction and cancel. Its magnon matrix element therefore vanishes at this order. The pseudo-scalar instead differentiates the spin density and can remain nonzero.

At fixed incident momentum and measured transfer $(\vq,\omega)$, write $\widetilde{\vs}_{n0}(\vq)=\bra n\widetilde{\vs}(\vq)\ket0$, with $\ket n=\ket{\vq}$. Integrating the derivatives in Eqs.~\eqref{eq:magnon-pauli-current} and \eqref{eq:magnon-pauli-odd} by parts replaces $\vnabla$ by $-i\vq$. The surviving current and pseudo-scalar matrix elements are thus $-i\vq\times\widetilde{\vs}_{n0}(\vq)/(2m)$ and $-\vq\cdot\widetilde{\vs}_{n0}(\vq)/(2m)$. Define the beam vertices $\bm\Gamma_V=\bar{u}_f\bm\gamma u_i$, $\bm\Gamma_A=\bar{u}_f\gamma^5\bm\gamma u_i$ and $\Gamma_P=\bar{u}_f\gamma^5u_i$. They retain the full incident and outgoing Dirac spinors, including their spin labels $s_i,s_f$.

The leading density matrix element is zero. The spatial metric sign in Eq.~\eqref{eq:MD} then gives the direct term $e^2\widetilde{\Pi}(\fQ)\bm\Gamma_V\cdot\widetilde{\vJ}_{\text{ss},n0}$. In Eq.~\eqref{eq:MX-app}, the vector, spatial axial and pseudo-scalar coefficients give the remaining local terms. Before inserting Eq.~\eqref{eq:magnonME}, their combined reduced amplitude is
\begin{equation}
\begin{aligned}
&\fM_{\rm mag}^{(0)}(n;s_i,s_f)\\
&\quad\equiv\fM_D(n;s_i,s_f)+\fM_X^{(0)}(n;s_i,s_f)\\
&\quad=e^2\Big(\widetilde{\Pi}(\fQ)+\frac{\PiX}{2}\Big)
\bm\Gamma_V\cdot\widetilde{\vJ}_{\text{ss},n0}\\
&\qquad-\frac{e^2\PiX}{2}\bm\Gamma_A\cdot\widetilde{\vs}_{n0}(\vq)+\frac{e^2\PiX}{2m}\Gamma_P\vq\cdot\widetilde{\vs}_{n0}(\vq)
\\
&\qquad+\mathcal R_{\rm mag}^{(2)}(n;s_i,s_f).
\end{aligned}
\label{eq:magnon-combined-amplitude}
\end{equation}
Here $\mathcal R_{\rm mag}^{(2)}$ collects terms beyond first order in the sample-spinor reduction.

Substitute the curl matrix element into Eq.~\eqref{eq:magnon-combined-amplitude} and use $\bm\Gamma_V\cdot(\vq\times\widetilde{\vs}_{n0})=(\bm\Gamma_V\times\vq)\cdot\widetilde{\vs}_{n0}$. Let $\vc^{\rm D}$ and $\vc^{\rm X}$ be the beam coefficient vectors multiplying the spin-density transition in the retained direct and local exchange amplitudes. The combined amplitude becomes $\fM_{\rm mag}^{(0)}=(\vc^{\rm D}+\vc^{\rm X})\cdot\widetilde{\vs}_{n0}+\mathcal R_{\rm mag}^{(2)}$, with

\begin{align}
\vc^{\rm D}&=-\frac{ie^2\widetilde{\Pi}(\fQ)}{2m}
(\bm\Gamma_V\times\vq),
\label{eq:Kmag-app}\\
\vc^{\rm X}&=e^2\PiX\Big[-\frac{i}{4m}(\bm\Gamma_V\times\vq)
-\frac12\bm\Gamma_A+\frac{\Gamma_P\vq}{2m}\Big].
\label{eq:KX-app}
\end{align}

\subsection{Beam traces, line weights and angular limits}
\label{subsec:proof-magnon-traces}

To evaluate the beam products in Eq.~\eqref{eq:master0}, write $\overline{\sum}=\frac12\sum_{s_i,s_f}$. As in Eq.~\eqref{eq:lepton-app}, insert the spinor completeness relations $\sum_su^s(\fK)\bar u^s(\fK)=\slashed{\fK}+m\mathbb{1}_4$ on the incident and outgoing legs and close each product into a Dirac trace. The vector product is the spatial tensor $\fL^{ij}$ already obtained there. The spatial axial tensor $A^{ij}$ and mixed vector--axial tensor $C^{ij}$ are
\begin{equation}
\begin{aligned}
A^{ij}&\equiv\overline{\sum}\Gamma_A^i\Gamma_A^{j*}\\
&=2\big[\fK_i^i\fK_f^j+\fK_i^j\fK_f^i\\
&\qquad+\delta^{ij}(\eta_{\mu\nu}\fK_i^\mu\fK_f^\nu+m^2)\big],\\
C^{ij}&\equiv\overline{\sum}\Gamma_V^i\Gamma_A^{j*}
=2i\epsilon_{ijr}w_r,\\
\vw&\equiv E_f\vk_i-E_i\vk_f.
\end{aligned}
\label{eq:magnon-beam-tensors}
\end{equation}
Here $\vw$ is the beam momentum difference weighted by the opposite beam energies. Applying the same trace operation to the pseudo-scalar products gives
\begin{equation}
\begin{aligned}
\overline{\sum}\Gamma_V^i\Gamma_P^*&=0,\\
\overline{\sum}\Gamma_P\Gamma_A^{j*}&=-2mq_j,\\
\overline{\sum}|\Gamma_P|^2&=q^2-\omega^2.
\end{aligned}
\label{eq:magnon-pseudo-scalar-sums}
\end{equation}
Thus the pseudo-scalar has no direct-vector interference in this spin-summed observable. Its interference with axial exchange and its square remain.

For the real kernels at $q>|\omega|$, the pure vector and axial products are real and symmetric in the sample-spin indices. The sum of the two vector--axial orderings is also real and symmetric after their imaginary tensor $C^{ij}$ multiplies the factor $i$ from the curl. Equation~\eqref{eq:magnon-pseudo-scalar-sums} makes the remaining products real and symmetric as well. Their contraction with the antisymmetric imaginary part of Eq.~\eqref{eq:Skl-app} therefore vanishes. Only its transverse projector $\delta^\perp_{kl}$ contributes.

Contract the two curl factors with this projector to obtain the vector geometric factor $\mathcal G_{\rm V}$. The projected single-curl contraction with $C^{ij}$ is $-2i$ times a real geometric factor, denoted $\mathcal G_{\rm VA}$. The axial factor $\mathcal A$ is the transverse trace of $A^{ij}$. These contractions give
\begin{align}
\mathcal G_{\rm V}&=q^2\delta_{ij}\fL^{ij}-q_i\fL^{ij}q_j\nonumber\\
&\quad-(\Mh\times\vq)_i\fL^{ij}(\Mh\times\vq)_j,
\label{eq:calD-app}\\
\mathcal G_{\rm VA}&\equiv\frac{i}{2}\delta^\perp_{kl}\epsilon_{kia}q_aC^{il}\nonumber\\
&=\vq\cdot\vw+(\vq\cdot\Mh)(\vw\cdot\Mh),
\label{eq:calI-app}\\
\mathcal A&=\delta^\perp_{ij}A^{ij}.
\label{eq:calA-app}
\end{align}
The pseudo-scalar adds a term along $\vq$ to the axial coefficient in Eq.~\eqref{eq:KX-app}. Using the last two sums in Eq.~\eqref{eq:magnon-pseudo-scalar-sums} in the two axial-pseudo-scalar orderings and the pseudo-scalar square changes its geometric factor to
\begin{equation}
\begin{aligned}
\mathcal A_{\rm AP}&\equiv\mathcal A+
\Big(4+\frac{q^2-\omega^2}{m^2}\Big)\\
&\quad\big[q^2-(\vq\cdot\Mh)^2\big].
\end{aligned}
\label{eq:magnon-axial-pseudo-scalar-factor}
\end{equation}
The label ${\rm AP}$ identifies the combined axial and pseudo-scalar contribution. The factor $4$ comes from their interference, and the remaining term comes from the pseudo-scalar square.

Substituting these contractions into Eq.~\eqref{eq:master0}, with $e^4/(16\pi^2)=\alpha^2$, gives Eq.~\eqref{eq:magnonxsec} with $\mathcal A\to\mathcal A_{\rm AP}$.

For the channel weights defined in Sec.~\ref{subsec:magnon-cross-section}, the direct beam factor is $\mathcal B_D=\widetilde{\Pi}(\fQ)^2\mathcal G_{\rm V}/(4m^2)$. Let $\beta_X=\PiX/[2\widetilde{\Pi}(\fQ)]$ be the exchange-current to direct amplitude ratio. Let $r_{\rm VA}$ be the direct--axial-exchange interference divided by the direct weight, and $r_{\rm AP}$ the squared sum of axial and pseudo-scalar exchange amplitudes divided by that weight. Separating these terms in Eq.~\eqref{eq:magnonxsec} with $\mathcal A\to\mathcal A_{\rm AP}$ and dividing by $\mathcal B_D$ gives
\begin{equation}
\begin{aligned}
r_{\rm VA}&=4m\frac{\PiX}{\widetilde{\Pi}(\fQ)}\frac{\mathcal G_{\rm VA}}{\mathcal G_{\rm V}},\\
r_{\rm AP}&=m^2\Big[\frac{\PiX}{\widetilde{\Pi}(\fQ)}\Big]^2
\frac{\mathcal A_{\rm AP}}{\mathcal G_{\rm V}},\\
\frac{\sigma_{\rm int}}{\sigma_D}&=2\beta_X+r_{\rm VA},\\
\frac{\sigma_X}{\sigma_D}&=\beta_X^2+\beta_X r_{\rm VA}+r_{\rm AP}.
\end{aligned}
\label{eq:magnonratios-app}
\end{equation}
The last two lines follow by expanding $\Pi_{\rm eff}=\widetilde{\Pi}+\PiX/2$ in Eq.~\eqref{eq:magnonxsec}. 

Consider first ordered spins parallel to the incident beam. For estimating the line coefficient, take elastic beam kinematics with $E=m+T$ and $k_i=k_f=k$, while retaining the positive magnon energy in the delta function. The beam energy--momentum relation fixes the transfer along the beam to $q^2/(2k)$. Substitution into Eqs.~\eqref{eq:calD-app}--\eqref{eq:magnon-axial-pseudo-scalar-factor} then gives
\begin{equation}
\begin{aligned}
\mathcal G_{\rm V}&=4k^2q^2+\frac{q^6}{4k^2},\\
\mathcal G_{\rm VA}&=E\Big(q^2+\frac{q^4}{4k^2}\Big),\\
\mathcal A_{\rm AP}&=8m^2+2q^2
+\Big(4+\frac{q^2}{m^2}\Big)\Big(q^2-\frac{q^4}{4k^2}\Big).
\end{aligned}
\label{eq:magnon-along-beam-factors}
\end{equation}
For $q/k\ll1$ and $q/m\ll1$, inserting the leading terms into Eq.~\eqref{eq:magnonratios-app} with $\widetilde{\Pi}=1/q^2$ and $\PiX=1/(2mT)$ gives
\begin{equation}
\begin{aligned}
\beta_X&=\frac{q^2}{4mT},\\
r_{\rm VA}&=\frac{Eq^2}{2Tk^2}\big[1+O(q^2/k^2)\big],\\
r_{\rm AP}&=\frac{q^2m^2}{2T^2k^2}
\big[1+O(q^2/m^2+q^4/k^4)\big].
\end{aligned}
\label{eq:magnon-small-transfer-ratios}
\end{equation}
The current terms in Eq.~\eqref{eq:magnonratios-app} remain explicit, including at high beam energy. Let $q_\parallel$ denote the transfer component along the incident beam. At finite loss, substitute $E_f=E-\omega$ and $k_f^2=k^2-2kq_\parallel+q^2$ into $E_f^2=m^2+k_f^2$ and solve for this component. This gives
\begin{equation}
q_\parallel=\frac{q^2+2E\omega-\omega^2}{2k}.
\label{eq:magnon-finite-loss-transfer}
\end{equation}
Relative to the elastic geometry, its change divided by $q$ is $O(E\omega/(kq))$, while the relative change in $k_f$ is $O(\omega/T)$. Expanding the nonzero parallel-spin factors therefore gives relative finite-loss errors $O(\omega/T+E\omega/(kq))$, requiring both terms to be small. Equation~\eqref{eq:magnonxsec} with $\mathcal A\to\mathcal A_{\rm AP}$ retains the actual finite-loss kinematics. These kinematic estimates do not bound the omitted sample and kernel corrections.

Let $T_{\rm cross}$ be the beam kinetic energy at which the full retained exchange and direct line weights are equal. It solves $\beta_X^2+\beta_X r_{\rm VA}+r_{\rm AP}=1$. For $T\ll m$, use $k^2=2mT+T^2$ in Eq.~\eqref{eq:magnon-small-transfer-ratios}. The leading term is $r_{\rm AP}=q^2m/(4T^3)$, which defines the scale $T_0$ below. At this scale, $\beta_X=O((T_0/m)^2)$ and $r_{\rm VA}=O(T_0/m)$, so the current terms in the full criterion are higher order than the $O(T_0/m)$ correction from $T^2$ in $k^2$. Expanding that criterion gives
\begin{equation}
\begin{aligned}
T_0&\equiv\Big(\frac{q^2m}{4}\Big)^{1/3},\\
T_{\rm cross}&=T_0\big[1+O(T_0/m)\big].
\end{aligned}
\label{eq:magnon-crossover}
\end{equation}
The expansion requires $T_0\ll m$. For $q=1$~\AA$^{-1}$, $T_0=7.92$~keV. Solving the full criterion numerically gives $T_{\rm cross}=7.90$~keV. The local-kernel conditions of Sec.~\ref{subsec:exchange-operators} must still hold near this crossover.

The relative exchange contribution also depends on the spin direction. Take $\Mh$ perpendicular to the incident beam, and let $\phi$ be the azimuth of the transverse transfer measured from $\Mh$. In the same elastic kinematics, contracting Eq.~\eqref{eq:calD-app} gives
\begin{equation}
\mathcal G_{\rm V}=4k^2q^2\cos^2\phi+q^4
-\frac{q^6}{4k^2}\cos^2\phi.
\label{eq:magnon-transverse-direct}
\end{equation}
The leading direct term vanishes at $\phi=\pi/2$ because $\vk_i\times\vq$ is then parallel to $\Mh$, while the magnon spin matrix element is perpendicular to $\Mh$. The exact vector tensor leaves the nonzero $q^4$ term. At this azimuth, $\mathcal G_{\rm VA}=Eq^2$ and $\mathcal A_{\rm AP}=8m^2+4k^2+4q^2+q^4/m^2$. Equation~\eqref{eq:magnonratios-app} therefore gives $r_{\rm VA}=2E/T$ and $r_{\rm AP}=\mathcal A_{\rm AP}/(4T^2)$. The suppression of the direct weight can make exchange relatively large even at beam energies where its correction in the parallel geometry is small. 

\subsection{Elastic spin matrix and leading polarization}
\label{subsec:proof-magnon-spin}

To evaluate the polarization in Eq.~\eqref{eq:Pf}, expand the beam-spin matrix in the identity and Pauli matrices. Write $\mathsf T=a\mathbb{1}_2+\vb\cdot\vsigma$, where $a=\Tr\mathsf T/2$ is its complex spin-independent coefficient and $\vb$ its complex spin-dependent coefficient vector, with $b_j=\Tr(\sigma_j\mathsf T)/2$. Substituting this expansion into Eq.~\eqref{eq:Pf} with $\mathsf T^{(0)}\to\mathsf T$ and using the Pauli product identity gives
\begin{equation}
\begin{aligned}
I_0&=|a|^2+\vb^*\cdot\vb,\\
\vP_f&=\frac{2\Re(a\vb^*)+i\vb\times\vb^*}{I_0}.
\end{aligned}
\label{eq:Pf-app}
\end{equation}

The same matrix also determines the counting asymmetry of a polarized source with spin-summed detection. Applying the same Pauli expansion to a unit-trace incident spin matrix with polarization $\vP_i$ gives $(\mathbb{1}_2+\vP_i\cdot\vsigma)/2$. Inserting it between $\mathsf T$ and $\mathsf T^\dagger$ and taking the outgoing trace gives an intensity $I(\vP_i)$. Define the incident analyzing vector $\vA_{\rm in}$ by its coefficient of $\vP_i$. Cyclically moving the matrices inside the trace replaces $\mathsf T\mathsf T^\dagger$ by $\mathsf T^\dagger\mathsf T$, so
\begin{equation}
\begin{aligned}
I(\vP_i)&=I_0(1+\vA_{\rm in}\cdot\vP_i),\\
\vA_{\rm in}
&=\frac{2\Re(a\vb^*)-i\vb\times\vb^*}{I_0}.
\end{aligned}
\label{eq:magnon-incident-analyzer}
\end{equation}
Their scalar--vector interference terms agree, but their vector--vector cross products have opposite signs.

To evaluate these coefficients explicitly, take the elastic beam kinematics used for the line estimates in Sec.~\ref{subsec:magnon-cross-section}. Insert the Dirac spinors into the vector and pseudo-scalar vertices, as in the axial calculation of Eq.~\eqref{eq:axial-vertex-general}. Multiplying their upper and lower components and applying the Pauli identity gives
\begin{equation}
\begin{aligned}
\bm\Gamma_V&=\xi_f^\dagger[\vK+i\vq\times\vsigma]\xi_i,\\
\Gamma_P&=\xi_f^\dagger(\vq\cdot\vsigma)\xi_i,
\qquad\vK\equiv\vk_i+\vk_f.
\end{aligned}
\label{eq:GVexact-app}
\end{equation}
Expand the three-Pauli-matrix product in Eq.~\eqref{eq:axial-vertex-general}. The resulting axial vertex has identity part $i(\vk_f\times\vk_i)/(E+m)$ and Pauli part $-2\mathsf G_A\vsigma$. The real symmetric $3$ by $3$ axial beam matrix is
\begin{equation}
\mathsf G_A=\Big(m+\frac{q^2}{4(E+m)}\Big)\mathbb{1}_3
+\frac{\vk_i\vk_f^\top+\vk_f\vk_i^\top}{2(E+m)}.
\label{eq:magnon-axial-beam-matrix}
\end{equation}
Insert these vertices into Eq.~\eqref{eq:magnon-combined-amplitude}, use $\vk_f\times\vk_i=\vK\times\vq/2$, and collect the identity and Pauli coefficients. The complete retained elastic coefficients are
\begin{equation}
\begin{aligned}
a&=-ie^2\Big[\frac{\Pi_{\rm eff}}{2m}
+\frac{\PiX}{4(E+m)}\Big](\vK\times\vq)\cdot\vepsilon,\\
\vb&=e^2\Big\{\frac{\Pi_{\rm eff}}{2m}
\big[q^2\vepsilon-\vq(\vq\cdot\vepsilon)\big]\\
&\quad+\PiX\mathsf G_A\vepsilon
+\frac{\PiX}{2m}\vq(\vq\cdot\vepsilon)\Big\}.
\end{aligned}
\label{eq:2x2-app}
\end{equation}
The three terms in $\vb$ come from the vector spin current, spatial axial vertex and pseudo-scalar. The second term in $a$ comes from the scalar part of the axial vertex. Both the direct spin current and this scalar axial term enter the direct--axial interference $r_{\rm VA}$ in Eq.~\eqref{eq:magnonratios-app}.

In the aligned-spin geometry of Sec.~\ref{subsec:magnon-spin}, assume $q/k\ll1$ and $q/m\ll1$. Keep the convective term in $a$ and the forward axial value $\mathsf G_A\vepsilon\to m\vepsilon$ from Eq.~\eqref{eq:axialvertex-app}. Since $\vK\times\vq=2k\hat{\vk}_i\times\vq_\perp$, the two transverse components of the circular vector give $|(\vK\times\vq)\cdot\vepsilon|=2kq_\perp$. The leading matrix and its scalar amplitude ratio are
Eq.~\eqref{eq:2x2}.

Subtract the leading coefficients from Eq.~\eqref{eq:2x2-app} and use $|\vq\cdot\vepsilon|=q_\perp$. Measure matrix size by the square root of the sum of squared spin-amplitude magnitudes. For $q_\perp\sim q$, the omitted vector spin term has size $O(|a_0|q/k)$, while the finite-angle axial and pseudo-scalar terms have size $O(|b_0|[kq/(m(E+m))+q^2/m^2])$. Relative to $\sqrt{|a_0|^2+2|b_0|^2}$, these define the dimensionless truncation scale
\begin{equation}
\delta_T\equiv\frac{|a_0|q/k
+|b_0|\big[kq/(m(E+m))+q^2/m^2\big]}
{\sqrt{|a_0|^2+2|b_0|^2}}.
\label{eq:magnon-spin-truncation}
\end{equation}
Divide the scalar axial coefficient in Eq.~\eqref{eq:2x2-app} by the convective coefficient in $a_0$. Using $\PiX=1/(2mT)$, $\Pi_{\rm eff}=(1+\beta_X)/q^2$ and $k^2=T(E+m)$ gives the smaller ratio $q^2/[4k^2(1+\beta_X)]$. Inserting these amplitude corrections into the two matrix products and normalizing by their traces gives additive polarization errors $O(\delta_T)$ when $\delta_T\ll1$. This is an absolute vector error, including near a zero polarization component. The scale $\delta_T$ controls the elastic beam reduction. At nonzero loss, use the actual beam vertices in Eq.~\eqref{eq:magnon-combined-amplitude}. The sample and kernel errors remain separate.

For $\mathsf T^{(0)}$, the convective part is proportional to the identity and the axial part to the traceless matrix $\vepsilon\cdot\vsigma$. Their interference disappears from $I_0$ because $\Tr\vsigma=0$, but remains in the spin numerator of Eq.~\eqref{eq:Pf-app}. Contract the circular dyadic following Eq.~\eqref{eq:Skl-app} with $\vK\times\vq=2k\hat{\vk}_i\times\vq_\perp$ to evaluate this scalar--vector term. Together with $\vepsilon\times\vepsilon^*=-2i\Mh$, it gives the leading transverse parts $\vP_{f,\perp}=\vA_{{\rm in},\perp}$ in Eq.~\eqref{eq:Pperp}.

Within $\mathsf T^{(0)}$, the transverse magnitude in Eq.~\eqref{eq:Pperp} has its maximum $1/\sqrt2$ at $\epsilonS=1/\sqrt2$. At this maximum, the forward axial intensity $2|b_0|^2$ equals the combined convective intensity $|a_0|^2$. The peak of the full retained elastic polarization is obtained from Eqs.~\eqref{eq:Pf-app} and \eqref{eq:2x2-app}, while its exchange--direct intensity crossover is defined by Eq.~\eqref{eq:magnonratios-app}.

The leading interference also identifies which spin transitions produce this transverse signal. For a general matrix element $\bm{\fE}=\bra n\widetilde{\vs}(\vq)\ket0$, let $a_{\rm c}$ and $\vb_{\rm A}$ be its convective and forward axial coefficients, now including the sample matrix element. They are obtained from $a_0$ and the forward axial vertex by replacing $\vepsilon$ with $\bm{\fE}$. The forward axial beam matrix is $\mathsf G_0=\operatorname{diag}(m,m,E)$ in beam axes. The real vector $\vL_\fE=-i\bm{\fE}^*\times\bm{\fE}/2$ measures the circular polarization and strength of this transition. Writing $\bm{\fE}$ as real plus imaginary parts converts $\Im[((\vK\times\vq)\cdot\bm{\fE})\bm{\fE}^*]$ into $(\vK\times\vq)\times\vL_\fE$. Thus their scalar--vector interference contribution to the spin numerator is
\begin{equation}
\begin{aligned}
a_{\rm c}&=-\frac{ie^2\Pi_{\rm eff}}{2m}(\vK\times\vq)\cdot\bm{\fE},\\
\vb_{\rm A}&=e^2\PiX\mathsf G_0\bm{\fE},\\
2\Re(a_{\rm c}\vb_{\rm A}^*)&=\frac{e^4\Pi_{\rm eff}\PiX}{m}\mathsf G_0\big[(\vK\times\vq)\times\vL_\fE\big].
\end{aligned}
\label{eq:law-app}
\end{equation}
This relation uses only convection and the forward axial vertex. A spin matrix element that is real up to a common phase has $\vL_\fE=0$, so this interference term vanishes. For unresolved transitions at fixed measured kinematics, add their unnormalized spin matrices before dividing by the summed intensity. Two equally weighted transitions with equal and opposite $\vL_\fE$ cancel this contribution. 

\subsection{Line integration and counting statistics}
\label{subsec:proof-magnon-count}

To integrate Eq.~\eqref{eq:countrate} over the magnon line, use its actual kinematics. At fixed $\vq_\perp$, write $q_z=k_i-k_{f,z}(\omega)$ and $\vq(\omega)=\vq_\perp+q_z(\omega)\hat{\vz}$. Differentiating $k_{f,z}^2=(E_i-\omega)^2-m^2-q_\perp^2$ gives $\text{d}q_z/\text{d}\omega=E_f/k_{f,z}$. Each simple root $\omega_*$ of $\omega=\omega_{\rm m}(\vq(\omega))$ therefore contributes the energy-integration factor
\begin{equation}
J_{\rm m}=\left|1-v_{{\rm m},z}\frac{E_f}{k_{f,z}}\right|^{-1},
\qquad
v_{{\rm m},z}\equiv\frac{\partial\omega_{\rm m}}{\partial q_z}.
\label{eq:magnon-line-jacobian}
\end{equation}
Here $v_{{\rm m},z}$ is the magnon group-velocity component along the incident beam, evaluated at the root. Setting $J_{\rm m}$ to one requires $|v_{{\rm m},z}E_f/k_{f,z}|\ll1$ and has a relative error of that order.

In the parallel-spin geometry, insert the leading direct beam factor $k^2/(m^2q^2)$ into Eq.~\eqref{eq:countrate}. For $q_\perp/k\ll1$, replace $k_fk_{f,z}$ by $k^2$, replace $q$ by $q_\perp$ in this beam factor and integrate the line. For a window selecting one root at each transverse transfer and with $J_{\rm m}$ set to one, the leading direct probability is
\begin{equation}
\begin{aligned}
p_{D,0}&=r_e^2t\sum_h\int \text{d}^2\vq_\perp\,
\eta_{{\rm det},h}(\vq_\perp,\omega_*)
\frac{2Sn_{\rm m}|F(\vq_*)|^2}{q_\perp^2},\\
r_e&=\frac{\alpha}{m}.
\end{aligned}
\label{eq:magnon-leading-probability}
\end{equation}
The root transfer is $\vq_*=\vq(\omega_*)$, and $r_e$ is the classical electron radius. The relative beam and line-integration errors are $O(q_\perp^2/k^2+\omega/T+E\omega/(kq)+|v_{{\rm m},z}E_f/k_{f,z}|)$ when all these ratios are small. The cancellation of $k^2$ removes the explicit leading beam-energy factor. The probability is then energy independent when the accepted sample and detector weights remain fixed. 

For the counting estimate, reverse a transverse source polarization between $\vP_i=rP_b\hat{\vx}$, where $r=\pm1$ labels the setting and $P_b$ is the polarization magnitude at the sample, with equal currents and exposure times, and record the counts $N_{r,h}$ in the two source settings and in two opposite detector half-planes $h=\pm1$. The signed count contrast and its mean for magnon counts alone are
\begin{equation}
\begin{aligned}
\widehat A&=\frac{\sum_{r,h}rhN_{r,h}}{N_{\rm tot}},
\qquad N_{\rm tot}=\sum_{r,h}N_{r,h},\\
A_{\rm m}&=\frac{P_b}{p_{\rm m}}\sum_h h\int
(\hat{\vx}\cdot\vA_{\rm in})\,\text{d}p_{{\rm m},h},
\end{aligned}
\label{eq:magnon-count-contrast}
\end{equation}
where $\text{d}p_{{\rm m},h}$ is the detected magnon probability element per incident electron in half-plane $h$ and $p_{\rm m}=\sum_h\int \text{d}p_{{\rm m},h}$. With a detected magnon rate $R_{\rm m}=(I_b/e)p_{\rm m}$ for incident current $I_b$, a spin-independent background rate $R_{\rm bg}$ in the same bins and energy window, and magnon fraction $f=R_{\rm m}/(R_{\rm m}+R_{\rm bg})$, the observed contrast is $A_{\rm obs}=fA_{\rm m}$. The total time needed to measure it with nominal significance $\nu$, counting both source settings, is
\begin{equation}
\tau_\nu\simeq
\frac{\nu^2(1-f^2A_{\rm m}^2)}{fA_{\rm m}^2R_{\rm m}}.
\label{eq:magnon-count-time}
\end{equation}

For independent Poisson counts, combine the two positive-sign bins in Eq.~\eqref{eq:magnon-count-contrast} and the two negative-sign bins. At fixed $N_{\rm tot}>0$, the positive-sign count $N_+$ is binomial with probability $(1+A_{\rm obs})/2$. Substituting $\widehat A=2N_+/N_{\rm tot}-1$ gives the exact conditional variance
\begin{equation}
\operatorname{Var}(\widehat A\mid N_{\rm tot})
=\frac{1-f^2A_{\rm m}^2}{N_{\rm tot}}.
\label{eq:magnon-count-variance}
\end{equation}
Replacing $N_{\rm tot}$ by $(R_{\rm m}+R_{\rm bg})\tau$ gives the counting-time estimate
Eq.~\eqref{eq:magnon-count-time}.

\subsection{Amplitude corrections and direct spin current}
\label{subsec:proof-magnon-corrections}

Let $\mathsf M=\sqrt{2SN_{\rm m}}F(\vq)\mathsf T$ be the $2$ by $2$ matrix of retained reduced amplitudes for the selected magnon, and let $\Delta\mathsf M$ be an added amplitude correction for the same transition. With the matrix norm $\|\mathsf X\|^2\equiv\Tr(\mathsf X\mathsf X^\dagger)$, a perturbative treatment requires
\begin{equation}
\delta_M\equiv\frac{\|\Delta\mathsf M\|}{\|\mathsf M\|}\ll1 .
\label{eq:magnon-amplitude-error}
\end{equation}
For the correction and norm in Eq.~\eqref{eq:magnon-amplitude-error}, replace $\mathsf M$ by $\mathsf M+\Delta\mathsf M$ in the outgoing matrix $\mathsf R=\mathsf M\mathsf M^\dagger/2$. The term $\delta\mathsf R$ linear in $\Delta\mathsf M$ contains the interference between the correction and the retained amplitude. Expanding the normalizing trace in Eq.~\eqref{eq:Pf-app} then gives the change $\Delta P_{f,j}$ in each outgoing polarization component,
\begin{equation}
\begin{aligned}
\delta\mathsf R&=\tfrac12\big(\Delta\mathsf M\mathsf M^\dagger
+\mathsf M\Delta\mathsf M^\dagger\big),\\
\Delta P_{f,j}
&=\frac{\Tr(\sigma_j\delta\mathsf R)
-P_{f,j}\Tr\delta\mathsf R}{\Tr\mathsf R}+O(\delta_M^2).
\end{aligned}
\label{eq:magnon-polarization-variation}
\end{equation}
For the incident analyzer, use the same expansion with $P_{f,j}$ replaced by $A_{{\rm in},j}$, $\mathsf R$ by $\mathsf M^\dagger\mathsf M/2$, and $\delta\mathsf R$ by $(\Delta\mathsf M^\dagger\mathsf M+\mathsf M^\dagger\Delta\mathsf M)/2$. 

If the actual transition weights justify a common fractional kernel change $\delta_K$, let $\mathsf M_X$ be the retained exchange part of $\mathsf M$. Its kernel correction then obeys
\begin{equation}
\Delta\mathsf M_K=\delta_K\mathsf M_X,
\qquad
\frac{\|\Delta\mathsf M_K\|}{\|\mathsf M\|}
=|\delta_K|\frac{\|\mathsf M_X\|}{\|\mathsf M\|}.
\label{eq:magnon-kernel-matrix-error}
\end{equation}

The retained direct beam spin current also distinguishes outgoing polarization from source analyzing power. For the elastic circular mode with $\Mh=\zeta_M\hat{\vk}_i$, set $\PiX=0$ in Eq.~\eqref{eq:2x2-app} and use $q_z=q^2/(2k)$. Substitution into Eqs.~\eqref{eq:Pf-app} and \eqref{eq:magnon-incident-analyzer} makes the scalar--vector and vector--vector transverse numerators equal. They add in the outgoing polarization and cancel in the incident analyzer, giving
\begin{equation}
\begin{aligned}
\vP_{f,\perp}^{\rm D}
&=\frac{\zeta_Mq^2}{2k^3[1+q^4/(16k^4)]}\vq_\perp,\\
\vA_{{\rm in},\perp}^{\rm D}&=0.
\end{aligned}
\label{eq:magnon-direct-only-polarization}
\end{equation}
The direct-only outgoing term is odd in $\Mh$ and therefore remains in an ordered-spin difference. The outgoing term is of order $(q/k)^3$ and has magnitude $1.0\,10^{-7}$ at $100$~keV and $q=1$~\AA$^{-1}$. The zero transverse incident analyzing vector is an elastic identity for this circular mode and geometry. At nonzero loss or for another spin-transition vector, evaluate the actual beam vertices before assigning a direct-channel background to the count contrast.

\section{Limiting results and numerical checks}
\label{app:numerics}

We collect limiting checks and independent evaluations of the retained formulas.

\subsection{Free-particle and limiting results}
\label{subsec:proof-known-checks}

The channel construction should recover free-particle scattering before any approximation to the sample is made. For this check, replace the connected sample matrix elements in Eqs.~\eqref{eq:MD}, \eqref{eq:exchange-amplitude}, \eqref{eq:positron-direct-amplitude} and \eqref{eq:annihilation-amplitude} by free-electron spinor products. The plane-wave phases fix the photon momenta and give a common spatial momentum-conservation delta function. The energy-conservation factor has already been removed in defining these reduced amplitudes. Removing the spatial factor leaves the direct and interchanged-final-electron terms of M{\o}ller scattering~\cite{moller1932}. The positron amplitude instead has the direct and annihilation terms of Bhabha scattering~\cite{bhabha1936}. These identities use the unexpanded photon kernels and the fermion signs derived in Appendix~\ref{app:channel-derivations}.

Table~\ref{tab:validation-app} lists the checks and their scope.

\begin{table*}[!tbp]
\caption{Limiting results and checks of the retained calculations. The free-particle rows use unexpanded amplitudes. The other rows state the additional kinematic limit or the model being checked. Numerical comparisons refer to the actual retained expressions, without implying an error bound for a material.}
\label{tab:validation-app}
\small
\begin{ruledtabular}
\begin{tabular}{@{}p{0.18\textwidth}p{0.46\textwidth}p{0.30\textwidth}@{}}
Limit or model & Result and scope & Check or source\\
\hline
M{\o}ller scattering & Free electron target. Direct and final-electron interchange terms with opposite signs. & Free-field substitution in Eqs.~\eqref{eq:MD} and \eqref{eq:exchange-amplitude}.\\
Bhabha scattering & Free electron target. Direct and annihilation terms for a positron beam. & Free-field substitution above, using the amplitudes of Appendix~\ref{app:channel-derivations}.\\
Non-relativistic Bethe limit & Leading longitudinal coefficient $4m^2/q^4$, with relative error $O((T_i+T_f)/m)$ for $T_i,T_f\ll m$. & Expansion of Eq.~\eqref{eq:master}.\\
Relativistic longitudinal factor & Leading coefficient $4m^2\gamma^2/q^4$, where $\gamma=E_i/m$, with relative error $O(\omega/E_i+|\fQ^2|/E_i^2)$ when both ratios are small. & Longitudinal numerator of Eq.~\eqref{eq:master}.\\
Ochkur kernel & $\PiX=k_i^{-2}[1+T/(2m)]$ has leading value $k_i^{-2}$ for $T\ll m$. & Eq.~\eqref{eq:PiX} and $k_i^2=2mT+T^2$.\\
Positronium contact term & The leading low-momentum annihilation potential, matched at free-pair threshold, is $2\pi\alpha P_{S=1}\delta^{(3)}(\bm{r})/m^2$, repulsive in the triplet. & Threshold spin matrix and external-state normalization in Sec.~\ref{subsec:proof-annihilation-threshold}.\\
Circular-magnon model & The cross section and outgoing spin matrix retain the exchange current, spatial axial and pseudo-scalar terms, using exact elastic beam vertices for the polarization. & Spin sums for Eq.~\eqref{eq:magnonxsec} with $\mathcal A\to\mathcal A_{\rm AP}$ and spinor matrix products for Eq.~\eqref{eq:2x2-app}.\\
Elastic screened potential & The second Born expression agrees with its original momentum integral for the non-relativistic Yukawa model. & Independent integration of Eqs.~\eqref{eq:screened-born-terms} and \eqref{eq:screened-born-exact}.\\
\end{tabular}
\end{ruledtabular}
\end{table*}

\subsection{Independent numerical evaluations}

For the Fierz identity, we construct the matrices of Appendix~\ref{app:fierz} and compare all $256$ components of the two sides of Eq.~\eqref{eq:fierz-components}. For the beam traces, we instead form the explicit spinor products, average the incident spin and sum the outgoing spin. Comparing these sums with Eqs.~\eqref{eq:electron-vertex-spin-sum} and \eqref{eq:positron-vertex-spin-sum} tests both the conjugation and the momentum order. The sampled kinematics include unequal incident and final energies, and the vertices include complex linear combinations of the scalar, vector, axial and pseudo-scalar matrices.

For the density interferences, we choose three orthogonal sample final states at the same loss. One carries a density and its conserved longitudinal current, and the other two carry equal strengths in orthogonal transverse directions. Their strengths reproduce the spectra $R_L$ and $R_T$. For each beam charge, we expand the squared sum of its channel amplitudes and perform the beam-spin and sample-state sums. The direct square and its cross term with the scalar--vector amplitude reproduce the direct weight and the interferences in Eqs.~\eqref{eq:IX} and \eqref{eq:IAratio}. This checks the isotropic response reduction used there.

At free electron--positron threshold, we evaluate the original annihilation current contraction in Eq.~\eqref{eq:annihilation-amplitude}, without the Fierz rearrangement. The four incident and four final pair-spin states give a $4$ by $4$ matrix. Using the physical positron spin convention of Eq.~\eqref{eq:positron-physical-spin}, this matrix annihilates the singlet and multiplies each triplet by $-2e^2$, reproducing Eq.~\eqref{eq:annihilation-triplet-amplitude}.

For the circular-magnon model, we construct the spatial vector, axial and pseudo-scalar beam vertices directly from Dirac spinors. We insert them into the retained amplitudes of Sec.~\ref{subsec:magnon-amplitudes} and average the squared total amplitude over the incident beam spin while summing the outgoing spin. Removing the common sample and charge factors gives a direct comparison with the beam factor $\mathcal B_{\rm m}$ in Eq.~\eqref{eq:magnonxsec} with $\mathcal A\to\mathcal A_{\rm AP}$. Elastic and finite-loss cases with several spin directions are included. In elastic kinematics, we also compare this explicit spinor matrix with the Pauli decomposition in Eq.~\eqref{eq:2x2-app}. Forming $\mathsf T\mathsf T^\dagger$ and $\mathsf T^\dagger\mathsf T$ separately checks the outgoing polarization and incident-spin analyzing vector, including their different cross-product signs.

For the non-relativistic Yukawa model, we compare the smooth parameter integral in Eq.~\eqref{eq:screened-born-exact} with the original momentum integral in Eq.~\eqref{eq:screened-born-terms}. For the real part, we perform the angular integrations analytically and subtract the numerator at the intermediate electron pole before integrating the radial principal value. For the imaginary part, we replace the energy propagator by its delta-function imaginary part and perform the radial integral, leaving an independent angular integration. Increasing the number of integration points checks convergence of both comparisons. These tests concern the retained expressions at sampled kinematics, while the sample and kernel approximation errors remain those specified in the main text.

\subsection{Further numerical checks}
\label{subsec:further-checks}

Every closed-form result in this paper was compared with a brute-force evaluation that does not use the closed form. All checks were performed in double precision with explicit Dirac spinors in the representation above. In particular, we checked the following:
\begin{enumerate}[(i)]
\item the Fierz coefficients on random spinors;
\item the beam tensor and the traces of Appendix~\ref{app:spinsums} against explicit spin sums at several kinematics, including relativistic ones;
\item the interference formulas \eqref{eq:IX} and \eqref{eq:IAratio} against spin sums with an isotropic transverse average of sample currents, to machine precision (the average over $\pm\hat{\ve}_1,\pm\hat{\ve}_2$ is required to cancel longitudinal--transverse cross terms);
\item the threshold annihilation amplitude in the pair-spin space, which has eigenvalues $\{0,-2,-2,-2\}\times e^2$;
\item the magnon cross section \eqref{eq:magnonxsec} with $\mathcal A\to\mathcal A_{\rm AP}$ against explicit summation for arbitrary $\Mh$, with exact agreement;
\item the $2\times2$ amplitude \eqref{eq:2x2} and the polarization \eqref{eq:Pperp} against exact tree-level M{\o}ller scattering with the target electron at rest, spin up to spin down, including recoil. This calculation contains neither the Fierz identity nor the non-relativistic reduction, and it agrees to $4\times10^{-5}$ at 100~keV and $10^{-4}$ at 30~keV, including the sign of the polarization;
\item the law \eqref{eq:law} for random complex $\bm{\fE}$ to $0.5\%$, the residual being $O(q/k)$ vertex corrections, and with a non-unit mass to confirm the $1/m$;
\item the kernel correction \eqref{eq:deltaX} against the exact angular average of $1/(-\fQX^2)$ to five digits;
\item the pair box \eqref{eq:MPP} in the 220-state Fock space;
\item the second-Born scaling \eqref{eq:born2scaling} by the variable-phase method with $l_{\max}=12k/\mu+20$, with the imaginary part checked against generalized unitarity to $10^{-5}$;
\item the dipole-forbidden geometry $\Mh\perp\vk_i$, $\vq\perp\Mh$, where the exact evaluation gives an outgoing polarization $0.948$ along $\Mh$ at 100~keV ($0.994$ at 30~keV), zero transverse polarization to machine precision, and a rate $6.6\times10^{-4}$ of the allowed azimuth.
\end{enumerate}

\bibliographystyle{apsrev4-2}
\bibliography{references}

@article{krivanek2014,
  author  = {Krivanek, Ondrej L. and Lovejoy, Tracy C. and Dellby, Niklas and
             Aoki, Toshihiro and Carpenter, R. W. and Rez, Peter and
             Soignard, Emmanuel and Zhu, Jiangtao and Batson, Philip E. and
             Lagos, Maureen J. and Egerton, Ray F. and Crozier, Peter A.},
  title   = {Vibrational spectroscopy in the electron microscope},
  journal = {Nature},
  volume  = {514},
  pages   = {209},
  year    = {2014},
  doi     = {10.1038/nature13870}
}

@article{kepaptsoglou2025,
  author  = {Kepaptsoglou, Demie and Castellanos-Reyes, Jos{\'e} {\'A}ngel and
             Kerrigan, Adam and do Nascimento, J{\'u}lio Alves and
             Zeiger, Paul M. and El Hajraoui, Khalil and
             Idrobo, Juan Carlos and Mendis, Budhika G. and
             Bergman, Anders and Lazarov, Vlado K. and Rusz, J{\'a}n and
             Ramasse, Quentin M.},
  title   = {Magnon spectroscopy in the electron microscope},
  journal = {Nature},
  volume  = {644},
  pages   = {83},
  year    = {2025},
  doi     = {10.1038/s41586-025-09318-y}
}

@article{mm2025,
  author  = {Kepaptsoglou, Demie and Castellanos-Reyes, Jos{\'e} {\'A}ngel and
             do Nascimento, J{\'u}lio Alves and Kerrigan, Adam and
             Cavill, Stuart and Zeiger, Paul M. and El Hajraoui, Khalil and
             Idrobo, Juan Carlos and Mendis, Budhika G. and
             Bergman, Anders and Lazarov, Vlado K. and Rusz, J{\'a}n and
             Ramasse, Quentin M.},
  title   = {Detecting magnons with high resolution electron energy loss
             spectroscopy},
  journal = {Microsc. Microanal.},
  volume  = {31},
  pages   = {ozaf048.703},
  year    = {2025},
  doi     = {10.1093/mam/ozaf048.703}
}

@book{egerton2011,
  author    = {Egerton, R. F.},
  title     = {Electron Energy-Loss Spectroscopy in the Electron Microscope},
  edition   = {3rd},
  publisher = {Springer},
  address   = {New York},
  year      = {2011}
}

@article{mendis2022,
  author  = {Mendis, B. G.},
  title   = {Quantum theory of magnon excitation by high energy electron beams},
  journal = {Ultramicroscopy},
  volume  = {239},
  pages   = {113548},
  year    = {2022},
  doi     = {10.1016/j.ultramic.2022.113548}
}

@article{nascimento2025,
  author  = {do Nascimento, Julio A. and Cheshire, Daniel M. and
             Cossu, Fabrizio and Kerrigan, Adam and Hasnip, Phil J. and
             Kepaptsoglou, Demie and Cavill, S. A. and
             Ramasse, Quentin M. and Lazarov, Vlado K.},
  title   = {Probing magnons with high-energy electrons: theoretical insights
             into spin and charge scattering in {STEM-EELS}},
  journal = {New J. Phys.},
  volume  = {27},
  pages   = {113505},
  year    = {2025},
  doi     = {10.1088/1367-2630/ae152a}
}

@article{nascimento2024,
  author  = {do Nascimento, Julio A. and Hasnip, Phil J. and Cavill, S. A. and
             Cossu, Fabrizio and Kepaptsoglou, Demie and
             Ramasse, Quentin M. and Kerrigan, Adam and Lazarov, Vlado K.},
  title   = {Theory of momentum-resolved magnon electron energy loss spectra:
             the case of yttrium iron garnet},
  journal = {arXiv:2401.12302},
  year    = {2024}
}

@article{kirschner1985,
  author  = {Kirschner, J.},
  title   = {Direct and exchange contributions in inelastic scattering of
             spin-polarized electrons from iron},
  journal = {Phys. Rev. Lett.},
  volume  = {55},
  pages   = {973},
  year    = {1985},
  doi     = {10.1103/PhysRevLett.55.973}
}

@article{plihal1999,
  author  = {Plihal, M. and Mills, D. L. and Kirschner, J.},
  title   = {Spin wave signature in the spin polarized electron energy loss
             spectrum of ultrathin {Fe} films: theory and experiment},
  journal = {Phys. Rev. Lett.},
  volume  = {82},
  pages   = {2579},
  year    = {1999},
  doi     = {10.1103/PhysRevLett.82.2579}
}

@article{vollmer2003,
  author  = {Vollmer, R. and Etzkorn, M. and Anil Kumar, P. S. and
             Ibach, H. and Kirschner, J.},
  title   = {Spin-polarized electron energy loss spectroscopy of high energy,
             large wave vector spin waves in ultrathin {fcc Co} films on
             {Cu(001)}},
  journal = {Phys. Rev. Lett.},
  volume  = {91},
  pages   = {147201},
  year    = {2003},
  doi     = {10.1103/PhysRevLett.91.147201}
}

@article{kuwahara2014,
  author  = {Kuwahara, Makoto and Kusunoki, Soichiro and Nambo, Yoshito and
             Saitoh, Koh and Jin, Xiuguang and Ujihara, Toru and
             Asano, Hidefumi and Takeda, Yoshikazu and Tanaka, Nobuo},
  title   = {Coherence of a spin-polarized electron beam emitted from a
             semiconductor photocathode in a transmission electron microscope},
  journal = {Appl. Phys. Lett.},
  volume  = {105},
  pages   = {193101},
  year    = {2014},
  doi     = {10.1063/1.4901745}
}

@article{bethe1930,
  author  = {Bethe, H.},
  title   = {Zur {Theorie} des {Durchgangs} schneller {Korpuskularstrahlen}
             durch {Materie}},
  journal = {Ann. Phys. (Leipzig)},
  volume  = {5},
  pages   = {325},
  year    = {1930}
}

@article{inokuti1971,
  author  = {Inokuti, Mitio},
  title   = {Inelastic collisions of fast charged particles with atoms and
             molecules---the {Bethe} theory revisited},
  journal = {Rev. Mod. Phys.},
  volume  = {43},
  pages   = {297},
  year    = {1971},
  doi     = {10.1103/RevModPhys.43.297}
}

@article{moller1932,
  author  = {M{\o}ller, C.},
  title   = {Zur {Theorie} des {Durchgangs} schneller {Elektronen} durch
             {Materie}},
  journal = {Ann. Phys. (Leipzig)},
  volume  = {14},
  pages   = {531},
  year    = {1932}
}

@article{schattschneider2005,
  author  = {Schattschneider, P. and H{\'e}bert, C. and Franco, H. and
             Jouffrey, B.},
  title   = {Anisotropic relativistic cross sections for inelastic electron
             scattering, and the magic angle},
  journal = {Phys. Rev. B},
  volume  = {72},
  pages   = {045142},
  year    = {2005},
  doi     = {10.1103/PhysRevB.72.045142}
}

@article{deforest1966,
  author  = {de Forest, T., Jr. and Walecka, J. D.},
  title   = {Electron scattering and nuclear structure},
  journal = {Adv. Phys.},
  volume  = {15},
  pages   = {1},
  year    = {1966},
  doi     = {10.1080/00018736600101254}
}

@article{ochkur1964,
  author  = {Ochkur, V. I.},
  title   = {The {Born--Oppenheimer} method in the theory of atomic collisions},
  journal = {Sov. Phys. JETP},
  volume  = {18},
  pages   = {503},
  year    = {1964},
  note    = {[Zh. Eksp. Teor. Fiz. \textbf{45}, 734 (1963)]}
}

@article{rudge1968,
  author  = {Rudge, M. R. H.},
  title   = {Theory of the ionization of atoms by electron impact},
  journal = {Rev. Mod. Phys.},
  volume  = {40},
  pages   = {564},
  year    = {1968},
  doi     = {10.1103/RevModPhys.40.564}
}

@article{byron1977,
  author  = {Byron, F. W., Jr. and Joachain, C. J.},
  title   = {Eikonal theory of electron- and positron-atom collisions},
  journal = {Phys. Rep.},
  volume  = {34},
  pages   = {233},
  year    = {1977},
  doi     = {10.1016/0370-1573(77)90014-X}
}

@article{bhabha1936,
  author  = {Bhabha, H. J.},
  title   = {The scattering of positrons by electrons with exchange on
             {Dirac}'s theory of the positron},
  journal = {Proc. R. Soc. London A},
  volume  = {154},
  pages   = {195},
  year    = {1936},
  doi     = {10.1098/rspa.1936.0046}
}

@article{fierz1937,
  author  = {Fierz, M.},
  title   = {Zur {Fermischen} {Theorie} des $\beta$-{Zerfalls}},
  journal = {Z. Phys.},
  volume  = {104},
  pages   = {553},
  year    = {1937},
  doi     = {10.1007/BF01330070}
}

@book{blp,
  author    = {Berestetskii, V. B. and Lifshitz, E. M. and Pitaevskii, L. P.},
  title     = {Quantum Electrodynamics},
  edition   = {2nd},
  publisher = {Pergamon},
  address   = {Oxford},
  year      = {1982},
  note      = {Sec.~83}
}

@article{beneke1998,
  author  = {Beneke, M. and Smirnov, V. A.},
  title   = {Asymptotic expansion of {Feynman} integrals near threshold},
  journal = {Nucl. Phys. B},
  volume  = {522},
  pages   = {321},
  year    = {1998},
  doi     = {10.1016/S0550-3213(98)00138-2}
}

@article{holstein1940,
  author  = {Holstein, T. and Primakoff, H.},
  title   = {Field dependence of the intrinsic domain magnetization of a
             ferromagnet},
  journal = {Phys. Rev.},
  volume  = {58},
  pages   = {1098},
  year    = {1940},
  doi     = {10.1103/PhysRev.58.1098}
}

@article{cherepanov1993,
  author  = {Cherepanov, V. and Kolokolov, I. and L'vov, V.},
  title   = {The saga of {YIG}: spectra, thermodynamics, interaction and
             relaxation of magnons in a complex magnet},
  journal = {Phys. Rep.},
  volume  = {229},
  pages   = {81},
  year    = {1993},
  doi     = {10.1016/0370-1573(93)90107-O}
}

@article{princep2017,
  author  = {Princep, A. J. and Ewings, R. A. and Ward, S. and
             T{\'o}th, S. and Dubs, C. and Prabhakaran, D. and
             Boothroyd, A. T.},
  title   = {The full magnon spectrum of yttrium iron garnet},
  journal = {npj Quantum Mater.},
  volume  = {2},
  pages   = {63},
  year    = {2017},
  doi     = {10.1038/s41535-017-0067-y}
}

@article{companion,
  title   = {Electron exchange matters in electron spectroscopy},
  author  = {Landry, Michael J. and Liu, Chuhang and Dale, Ashley S.
             and Davis, L. Dagny and Haisman, Griffin M. and Wu, Lijun
             and Egerton, Ray F. and Zhu, Yimei and Li, Mingda},
  journal = {arXiv preprint arXiv:2611.33637},
  year    = {2026},
  doi     = {10.48550/arXiv.2611.33637}
}

@article{rohrlich1954,
  author  = {Rohrlich, F. and Carlson, B. C.},
  title   = {Positron-electron differences in energy loss and multiple
             scattering},
  journal = {Phys. Rev.},
  volume  = {93},
  pages   = {38},
  year    = {1954},
  doi     = {10.1103/PhysRev.93.38}
}

@article{sorini2008,
  author  = {Sorini, A. P. and Rehr, J. J. and Levine, Z. H.},
  title   = {Magic angle in electron energy loss spectra: relativistic and
             dielectric corrections},
  journal = {Phys. Rev. B},
  volume  = {77},
  pages   = {115126},
  year    = {2008},
  doi     = {10.1103/PhysRevB.77.115126}
}

@article{iatrakis2026a,
  author  = {Iatrakis, Ioannis and Brudanin, Valerii},
  title   = {Many-body and {QED} effects in electron-atom inelastic scattering
             in electron energy loss spectroscopy},
  journal = {Phys. Rev. B},
  volume  = {113},
  pages   = {085101},
  year    = {2026},
  note    = {[arXiv:2502.12254]},
  doi     = {10.1103/2mdr-f1v6}
}

@article{iatrakis2026b,
  author  = {Iatrakis, Ioannis and Brudanin, Valerii},
  title   = {Electron energy loss spectroscopy of oriented targets and
             magnetic transitions},
  journal = {Micron},
  volume  = {201},
  pages   = {103929},
  year    = {2026},
  doi     = {10.1016/j.micron.2025.103929},
  note    = {[arXiv:2601.05944]}
}

@article{lyon2021,
  author  = {Lyon, K. and Bergman, A. and Zeiger, P. and Kepaptsoglou, D. and
             Ramasse, Q. M. and Idrobo, J. C. and Rusz, J.},
  title   = {Theory of magnon diffuse scattering in scanning transmission
             electron microscopy},
  journal = {Phys. Rev. B},
  volume  = {104},
  pages   = {214418},
  year    = {2021},
  doi     = {10.1103/PhysRevB.104.214418}
}

@article{castellanos2023,
  author  = {Castellanos-Reyes, J. {\'A}. and Zeiger, P. and Bergman, A. and
             Kepaptsoglou, D. and Ramasse, Q. M. and Idrobo, J. C. and
             Rusz, J.},
  title   = {Unveiling the impact of temperature on magnon diffuse scattering
             detection in the transmission electron microscope},
  journal = {Phys. Rev. B},
  volume  = {108},
  pages   = {134435},
  year    = {2023},
  doi     = {10.1103/PhysRevB.108.134435}
}

@article{castellanos2025,
  author  = {Castellanos-Reyes, J. {\'A}. and Zeiger, P. M. and Rusz, J.},
  title   = {Dynamical theory of angle-resolved electron energy loss and gain
             spectroscopies of phonons and magnons in transmission electron
             microscopy including multiple scattering effects},
  journal = {Phys. Rev. Lett.},
  volume  = {134},
  pages   = {036402},
  year    = {2025},
  doi     = {10.1103/PhysRevLett.134.036402}
}

@article{mendis2021,
  author  = {Mendis, B. G.},
  title   = {An inelastic multislice simulation method incorporating plasmon
             energy losses},
  journal = {Ultramicroscopy},
  volume  = {230},
  pages   = {113390},
  year    = {2021},
  note    = {Volume and page verified; title to be confirmed against the
             published article before submission}
}

@article{nishitani2005,
  author  = {Nishitani, T. and Nakanishi, T. and Yamamoto, M. and Okumi, S. and
             Furuta, F. and Miyamoto, M. and Kuwahara, M. and Yamamoto, N. and
             Naniwa, K. and Watanabe, O. and Takeda, Y. and Kobayashi, H. and
             Dohmae, S.},
  title   = {Highly polarized electrons from {GaAs--GaAsP} and
             {InGaAs--AlGaAs} strained-layer superlattice photocathodes},
  journal = {J. Appl. Phys.},
  volume  = {97},
  pages   = {094907},
  year    = {2005},
  doi     = {10.1063/1.1886888}
}

@article{maruyama2004,
  author  = {Maruyama, T. and Prepost, R. and Garwin, E. L. and
             Sinclair, C. K. and Poelker, M. and Clendenin, J. E.},
  title   = {Systematic study of polarized electron emission from strained
             {GaAs/GaAsP} superlattice photocathodes},
  journal = {Appl. Phys. Lett.},
  volume  = {85},
  pages   = {2640},
  year    = {2004},
  doi     = {10.1063/1.1795358}
}

@article{vig2017,
  author  = {Vig, Sean and Kogar, Anshul and Mitrano, Matteo and Husain, Ali A. and
             Venema, Luc and Rak, Melinda S. and Mishra, Vivek and Johnson, Peter D. and
             Gu, Genda D. and Fradkin, Eduardo and Norman, Michael R. and Abbamonte, Peter},
  title   = {Measurement of the dynamic charge response of materials using low-energy,
             momentum-resolved electron energy-loss spectroscopy ({M-EELS})},
  journal = {SciPost Phys.},
  volume  = {3},
  pages   = {026},
  year    = {2017},
  doi     = {10.21468/SciPostPhys.3.4.026}
}

@book{ibach1982,
  author    = {Ibach, H. and Mills, D. L.},
  title     = {Electron Energy Loss Spectroscopy and Surface Vibrations},
  publisher = {Academic Press},
  address   = {New York},
  year      = {1982}
}

@article{joachain1974,
  author = {Joachain, Charles J. and Quigg, C.},
  title = {Multiple scattering expansions in several particle dynamics},
  journal = {Rev. Mod. Phys.},
  volume = {46},
  pages = {279},
  year = {1974},
  doi = {10.1103/RevModPhys.46.279}
}

@article{chou2024spin,
  title   = {Large spin polarization from symmetry-breaking antiferromagnets in antiferromagnetic tunnel junctions},
  author  = {Chou, Chung-Tao and Ghosh, Supriya and McGoldrick, Brooke C. and Nguyen, Thanh and Gurung, Gautam and Tsymbal, Evgeny Y. and Li, Mingda and Mkhoyan, K. Andre and Liu, Luqiao},
  journal = {Nature Communications},
  volume  = {15},
  pages   = {7840},
  year    = {2024},
  doi     = {10.1038/s41467-024-52208-6}
}

@article{nguyen2022kagome,
  title   = {Electronic properties of correlated kagom{\'e} metals {$AV_3Sb_5$} ({$A$} = K, Rb, and Cs): A perspective},
  author  = {Nguyen, Thanh and Li, Mingda},
  journal = {Journal of Applied Physics},
  volume  = {131},
  number  = {6},
  pages   = {060901},
  year    = {2022},
  doi     = {10.1063/5.0079593}
}

@article{merker2022magnetism,
  title   = {Machine learning magnetism classifiers from atomic coordinates},
  author  = {Merker, Helena A. and Heiberger, Harry and Nguyen, Linh and
             Liu, Tongtong and Chen, Zhantao and Andrejevic, Nina and
             Drucker, Nathan C. and Okabe, Ryotaro and Kim, Song Eun and
             Wang, Yao and Smidt, Tess and Li, Mingda},
  journal = {iScience},
  volume  = {25},
  number  = {10},
  pages   = {105192},
  year    = {2022},
  doi     = {10.1016/j.isci.2022.105192}
}

\end{document}